\documentclass[aps,twocolumn,superscriptaddress,floatfix,nofootinbib,showpacs,altaffilletter]{revtex4-2}

\usepackage{epsfig,graphicx,amssymb,amsmath,wasysym,graphicx}

\usepackage{amsmath}
\usepackage{amsfonts}
\usepackage{amssymb}
\usepackage{graphicx}
\usepackage{bm}
\usepackage{subfigure}
\usepackage{url}
\usepackage[hyperindex]{hyperref}
\usepackage{color}
\usepackage[ddmmyy,24hr]{datetime}
\usepackage{bigdelim}
\usepackage{booktabs}
\usepackage{dcolumn}
\usepackage{multirow}
\usepackage{subfigure}
\usepackage{cancel}
\usepackage{stackrel}
\usepackage{paralist}
\usepackage{xspace}
\usepackage{slashed}
\usepackage{cancel}
\usepackage{todonotes}
\usepackage{enumerate}
\usepackage{float}
\usepackage{fullpage}
\usepackage{ulem}
\usepackage{physics}
\usepackage{orcidlink}

\newcommand{\nua}[1]{\ensuremath{\rlap{\kern-2.5pt\ensuremath{\overset{\scriptscriptstyle(-)}{\phantom{\nu}}}}{\ensuremath{{\nu}_{#1}}}}}
\newcommand{\vet}[1]{\ensuremath{\hskip-1pt\vec{\hskip1pt#1}}}
\usepackage{enumitem}

\newcommand{\cenns}{CE$\nu$NS\xspace}
\newcommand{\cawo}{$\mathrm{CaWO_4}$\xspace}

\newcommand{\be}{\begin{equation}}
\newcommand{\ee}{\end{equation}}
\newcommand{\ba}{\begin{array}}
\newcommand{\ea}{\end{array}}

\newcommand{\HEPHY}{Marietta-Blau-Institut f{\"u}r Teilchenphysik der {\"O}sterreichischen Akademie der Wissenschaften, Dominikanerbastei~16, Wien, A-1010, Austria}
\newcommand{\TUW}{Atominstitut, Technische Universit\"at Wien, Stadionallee~2, Wien, A-1020, Austria}
\newcommand{\CEA}{IRFU, CEA, Universit\'e Paris-Saclay, B\^{a}timent 141, Gif-sur-Yvette, F-91191, France}

\newcommand{\MPIK}{Max-Planck-Institut f\"ur Kernphysik, Saupfercheckweg 1, Heidelberg, D-69117, Germany}
\newcommand{\MPP}{Max-Planck-Institut f\"ur Physik, Boltzmannstra{\ss}e~8, Garching, D-85748, Germany}
\newcommand{\TUM}{Physik-Department, TUM School of Natural Sciences, Technische Universit\"at M\"unchen, James-Franck-Straße 1, Garching, D-85748, Germany}
\newcommand{\INFNRoma}{Istituto Nazionale di Fisica Nucleare -- Sezione di Roma, Piazzale Aldo Moro 2, Roma, I-00185, Italy}
\newcommand{\Sapienza}{Dipartimento di Fisica, Sapienza Universit\`{a} di Roma, Piazzale Aldo Moro 5, Roma, I-00185, Italy}
\newcommand{\INFNTorVergata}{Istituto Nazionale di Fisica Nucleare -- Sezione di Roma ``Tor Vergata'', Via della Ricerca Scientifica 1, Roma, I-00133, Italy}
\newcommand{\TorVergata}{Dipartimento di Fisica, Universit\`{a} di Roma ``Tor Vergata'', Via della Ricerca Scientifica 1, Roma, I-00133, Italy}

\newcommand{\Ferrara}{Dipartimento di Fisica, Universit\`{a} di Ferrara, Via Giuseppe Saragat 1, Ferrara, I-44122, Italy}
\newcommand{\INFNFerrara}{Istituto Nazionale di Fisica Nucleare -- Sezione di Ferrara, Via Giuseppe Saragat 1c, Ferrara, I-44122, Italy}
\newcommand{\Coimbra}{LIBPhys-UC, Departamento de Fisica, Universidade de Coimbra, Rua Larga 3004-516, Coimbra, P3004-516, Portugal}
\newcommand{\INFNLnGS}{Istituto Nazionale di Fisica Nucleare -- Laboratori Nazionali del Gran Sasso, Via Giovanni Acitelli 22, Assergi (L’Aquila), I-67100, Italy}
\newcommand{\Kiutra}{kiutra GmbH, Fl{\"o}{\ss}ergasse 2, D-81369 Munich, Germany}

\begin{document}

\author{H.~Abele\,\orcidlink{0000-0002-6832-9051}} 
\affiliation{\TUW} 
\author{G.~Angloher} 
\affiliation{\MPP}
\author{B.~Arnold} 
\affiliation{\HEPHY}
\author{M.~Atzori~Corona\,\orcidlink{0000-0001-5092-3602}}
\thanks{Corresponding authors: mcorona@roma2.infn.it; elisabetta.bossio@cea.fr}
\affiliation{\INFNTorVergata}
\author{A.~Bento\,\orcidlink{0000-0002-3817-6015}} 
\altaffiliation[Also at: ]{\Coimbra}
\affiliation{\MPP}
\author{E.~Bossio\,\orcidlink{0000-0001-9304-1829}} 
\thanks{Corresponding authors: mcorona@roma2.infn.it; elisabetta.bossio@cea.fr}
\affiliation{\CEA}
\author{F.~Buchsteiner} 
\affiliation{\HEPHY}
\author{J.~Burkhart\,\orcidlink{0000-0002-1989-7845}} \affiliation{\HEPHY}
\author{F.~Cappella\,\orcidlink{0000-0003-0900-6794}} \affiliation{\INFNRoma}
\author{M.~Cappelli\,\orcidlink{0009-0002-6148-5964}} \affiliation{\Sapienza} 
\affiliation{\INFNRoma}
\author{N.~Casali\,\orcidlink{0000-0003-3669-8247}} \affiliation{\INFNRoma}
\author{R.~Cerulli\,\orcidlink{0000-0003-2051-3471}} \affiliation{\INFNTorVergata}
\author{A.~Cruciani\,\orcidlink{0000-0003-2247-8067}} \affiliation{\INFNRoma}
\author{G.~Del~Castello\,\orcidlink{0000-0001-7182-358X}} \affiliation{\INFNRoma}
\author{M.~del~Gallo~Roccagiovine} 
\affiliation{\Sapienza} 
\affiliation{\INFNRoma}
\author{S.~Dorer\,\orcidlink{0009-0001-1670-5780}} 
\affiliation{\TUW}
\author{A.~Erhart\,\orcidlink{0000-0002-8721-177X}} 
\affiliation{\TUM}
\author{M.~Friedl\,\orcidlink{0000-0002-7420-2559}} 
\affiliation{\HEPHY}
\author{S.~Fichtinger} 
\affiliation{\HEPHY}
\author{V.M.~Ghete\,\orcidlink{0000-0002-9595-6560}} \affiliation{\HEPHY}
\author{M.~Giammei\,\orcidlink{0009-0006-9104-2055}} \affiliation{\TorVergata} 
\affiliation{\INFNTorVergata}
\author{C.~Goupy\,\orcidlink{0000-0003-4954-5311}} 
\altaffiliation[Now at: ]{\MPIK} 
\affiliation{\CEA}
\author{J.~Hakenm\"{u}ller\,\orcidlink{0000-0003-0470-3320}}\affiliation{\HEPHY}
\author{D.~Hauff} 
\affiliation{\MPP} 
\affiliation{\TUM}
\author{F.~Jeanneau\,\orcidlink{0000-0002-6360-6136}} 
\affiliation{\CEA}
\author{E.~Jericha\,\orcidlink{0000-0002-8663-0526}} 
\affiliation{\TUW}
\author{M.~Kaznacheeva\,\orcidlink{0000-0002-2712-1326}} \affiliation{\TUM}
\author{H.~Kluck\,\orcidlink{0000-0003-3061-3732}} 
\affiliation{\HEPHY}
\author{A.~Langenk\"{a}mper} 
\affiliation{\MPP}
\author{T.~Lasserre\,\orcidlink{0000-0002-4975-2321}} 
\altaffiliation[Now at: ]{\MPIK}
\affiliation{\CEA} 
\affiliation{\TUM} 
\author{D.~Lhuillier\,\orcidlink{0000-0003-2324-0149}} \affiliation{\CEA}
\author{M.~Mancuso\,\orcidlink{0000-0001-9805-475X}} 
\affiliation{\MPP}
\author{R.~Martin} 
\affiliation{\CEA} 
\affiliation{\TUW}
\author{B.~Mauri} 
\affiliation{\MPP}
\author{A.~Mazzolari} 
\affiliation{\INFNFerrara} 
\affiliation{\Ferrara}
\author{L.~McCallin} 
\affiliation{\CEA}
\author{H.~Neyrial} 
\affiliation{\CEA}
\author{C.~Nones} 
\affiliation{\CEA}
\author{L.~Oberauer} 
\affiliation{\TUM}
\author{L.~Peters\,\orcidlink{0000-0002-1649-8582}}
\altaffiliation[Now at: ]{\MPIK}
\affiliation{\TUM} 
\affiliation{\CEA} 
\author{F.~Petricca\,\orcidlink{0000-0002-6355-2545}} 
\affiliation{\MPP}
\author{W.~Potzel} 
\affiliation{\TUM}
\author{F.~Pr\"{o}bst} 
\affiliation{\MPP}
\author{F.~Pucci} 
\altaffiliation[Now at: ]{\INFNLnGS} 
\affiliation{\MPP}
\author{F.~Reindl\,\orcidlink{0000-0003-0151-2174}} \affiliation{\HEPHY} 
\affiliation{\TUW}
\author{M.~Romagnoni} 
\affiliation{\INFNFerrara} 
\affiliation{\Ferrara}
\author{J.~Rothe\,\orcidlink{0000-0001-5748-7428}} 
\altaffiliation[Now at: ]{\Kiutra} 
\affiliation{\TUM}
\author{N.~Schermer\,\orcidlink{0009-0004-4213-5154}} 
\affiliation{\TUM}
\author{J.~Schieck\,\orcidlink{0000-0002-1058-8093}} \affiliation{\HEPHY} 
\affiliation{\TUW}
\author{S.~Sch\"{o}nert\,\orcidlink{0000-0001-5276-2881}} \affiliation{\TUM}
\author{C.~Schwertner} 
\affiliation{\HEPHY} 
\affiliation{\TUW}
\author{L.~Scola} 
\affiliation{\CEA}
\author{G.~Soum-Sidikov\,\orcidlink{0000-0003-1900-1794}} \affiliation{\CEA}
\author{L.~Stodolsky} 
\affiliation{\MPP}
\author{A.~Schr\"{o}der\,\orcidlink{0009-0005-1598-1635}}
\affiliation{\TUM}
\author{R.~Strauss\,\orcidlink{0000-0002-5589-9952}} 
\affiliation{\TUM}
\author{R.~Thalmeier\,\orcidlink{0009-0003-4480-0990}} \affiliation{\HEPHY}
\author{C.~Tomei} 
\affiliation{\INFNRoma}
\author{L.~Valla\,\orcidlink{0009-0003-7140-9196}} \affiliation{\HEPHY}
\author{M.~Vignati\,\orcidlink{0000-0002-8945-1128}} \affiliation{\Sapienza} 
\affiliation{\INFNRoma}
\author{M.~Vivier\,\orcidlink{0000-0003-2199-0958}} 
\affiliation{\CEA}
\author{A.~Wallach\,\orcidlink{0009-0009-1703-9634}}
\affiliation{\TUM}
\author{P.~Wasser\,\orcidlink{0009-0004-7650-7307}}
\affiliation{\TUM}
\author{A.~Wex\,\orcidlink{0009-0003-5371-2466}} 
\affiliation{\TUM}
\author{L.~Wienke\,\orcidlink{0009-0006-5548-2109}}
\affiliation{\TUM}

\collaboration{NUCLEUS Collaboration} \noaffiliation

\title{Prospect of the NUCLEUS Experiment at Chooz for Coherent Elastic Neutrino-Nucleus Scattering and New Physics Searches}

\begin{abstract}

The NUCLEUS experiment aims to measure coherent elastic neutrino–nucleus scattering (\cenns) at unprecedentedly low nuclear recoil energies using gram-scale cryogenic calorimeters operated at the Chooz nuclear power plant in France. Access to recoil energies at the $\mathcal{O}(10~\mathrm{eV})$ scale enables \cenns studies at extremely low momentum transfer and provides enhanced sensitivity to new physics.
In this work, we present sensitivity projections for the upcoming NUCLEUS technical and physics runs, incorporating a data-driven treatment of the low-energy excess (LEE) observed during commissioning. We develop a likelihood framework that exploits reactor-power variation to disentangle signal and background in a low signal-to-background regime and to assess the impact of the dominant systematic uncertainties.
For the \textit{Technical Run} with a 7\,g \cawo target, we find competitive sensitivity to several scenarios beyond the Standard Model, which do not require a \cenns observation. For the \textit{Physics Run}, assuming complete suppression of the LEE, we project a $4.7\,\sigma$ observation of \cenns with a statistical precision of about 20\% in 1 year, enabling a determination of the weak mixing angle at the lowest momentum transfer probed to date with CE$\nu$NS and leading \cenns-based constraints on the neutrino charge radius and new mediator models.

\end{abstract}

\maketitle  
%\newpage
%\linenumbers

\section{Introduction}
\label{sec:intro}

Neutrinos interact extremely weakly with matter, making them notoriously difficult to detect. Yet, coherent elastic neutrino-nucleus scattering (\cenns), occurring at neutrino energies below $\sim50$\,MeV, offers a rare opportunity to observe neutrino interactions through a purely weak neutral-current process.
Predicted by Freedman in 1974~\cite{PhysRevD.9.1389}, and further studied in Refs.~\cite{Kopeliovich:1974mv,Drukier:1984vhf}, \cenns is a Standard Model (SM) process involving the coherent scattering of a neutrino off an entire nucleus via $Z^0$ exchange. Its relatively large cross-section compared to other low-energy neutrino interactions makes it a powerful tool for exploring a broad range of physics scenarios, from nuclear structure and electroweak parameters to physics beyond the SM (BSM).

The COHERENT Collaboration provided the first experimental observation of this process using a cesium-iodine (CsI) detector and neutrinos from the Spallation Neutron Source (SNS)~\cite{COHERENT:2017ipa,COHERENT:2020ybo}, followed by measurements on argon nuclei ~\cite{COHERENT:2020ybo,COHERENT:2020iec} and more recently on germanium~\cite{COHERENT:2025vuz,Adhikari:2026uvz}. 
Complementary searches have also emerged in the dark-matter community, where experiments such as PandaX~\cite{PandaX:2024muv}, XENONnT~\cite{XENON:2024ijk}, and LUX-ZEPLIN~\cite{LZ:2025igz} have begun probing \cenns from solar neutrinos.
Within this framework, reactor-based \cenns measurements play a crucial role: nuclear power plants provide intense, continuous, and well-localized sources of low-energy $\bar\nu_e$, enabling measurements deep in the fully coherent regime. Recently, CONUS+ reported a $3.7\,\sigma$ \cenns detection using high-purity germanium detectors near a commercial reactor~\cite{Ackermann:2025obx}.
Several other \cenns experiments are operating at reactor facilities~\cite{nGeN:2025hsd,TEXONO:2024vfk,CONNIE:2021ggh,Ricochet:2025bpk,RED-100:2024izi,Mondal:2025odw}, with the objective of precision measurements.

The NUCLEUS Collaboration aims to measure the \cenns cross section in the fully coherent regime with high precision by deploying gram-scale cryogenic calorimeters in close proximity to the Chooz nuclear power plant in France~\cite{NUCLEUS:2019igx}. NUCLEUS will employ \cawo target crystals operated at 10\,mK and instrumented with tungsten transition-edge sensors (W-TES), targeting nuclear-recoil energy thresholds down to $\mathcal{O}(10\,\mathrm{eV})$ thanks to the excellent intrinsic energy resolution of cryogenic calorimetry~\cite{NUCLEUS:2017gvo}.
This detector concept builds directly on the technology developed by the CRESST experiment for low-mass dark-matter searches, also employing \cawo targets and TES-based cryogenic calorimetry~\cite{CRESST:2019jnq,CRESST:2024cpr}. A precise measurement of reactor \cenns with NUCLEUS will also provide important input for future dark-matter searches with similar detectors, as coherent neutrino scattering will ultimately constitute an irreducible background once sensitivities reach the neutrino fog~\cite{OHare:2021utq}.

The experiment will be hosted in Chooz at the so-called very near site (VNS), a $\sim 24\,\mathrm{m}^2$ experimental room located between the two reactor cores at distances of 72\,m and 102\,m, respectively. Each core operates at a nominal thermal power of $4.25\,\mathrm{GW_{th}}$ and is shut down for refueling for approximately one month per year. For the reactor antineutrino flux prediction, we adopt the model of Ref.~\cite{Perisse:2023efm} and the average fission fractions reported therein, obtaining an integrated flux at the VNS of $(2.15\pm0.08)\times10^{12}\,\mathrm{cm^{-2} s^{-1}}$ when both reactors are operated at full power.
Being located in the basement of a five-story office building, the modest overburden of the VNS corresponds to about 3\,m water-equivalent. As a consequence, the experiment operates in an environment with significant exposure to cosmic-ray–induced backgrounds, requiring a dedicated and highly optimized shielding strategy. The background mitigation concept combines passive shielding with multiple active veto systems, whose performance and interplay have been extensively studied using detailed Monte Carlo simulations~\cite{Abele:2025yca}. In particular, muon-induced backgrounds are suppressed by a muon-veto system~\cite{NUCLEUS:2022rcz} that includes a cryogenic muon veto~\cite{Erhart:2023vam}, while external $\gamma$ rays and secondary particles are further mitigated by a cryogenic outer veto (COV) based on high-purity Ge detectors providing close to $4\pi$ coverage of the cryogenic target~\cite{goupyPhd,wexPhd}. Together, these systems are designed to compensate for the limited overburden at the VNS and to suppress particle-induced backgrounds in the \cenns region of interest to a level that preserves sensitivity to the expected signal. 

A simplified version of the NUCLEUS experiment was commissioned in 2024 at the Technical University of Munich (TUM)~\cite{NUCLEUS:2025ymr}, demonstrating simultaneous operation of all components with stable performance over about two months.

Building on this milestone, the experiment will be relocated to Chooz in 2026 and operated in two successive phases.
An upcoming \textit{Technical Run} will deploy four \cawo detectors, each instrumented with two TESs, with a total target mass of about 7\,g. The primary goals of this phase are to validate detector performance under VNS conditions and to characterize the particle background in situ. Based on the commissioning results, the sensitivity during the \textit{Technical Run} is expected to remain limited by the low-energy excess (LEE), an excess of sub-keV events not explained by known backgrounds or detector noise~\cite{Fuss:2022fxe, Angloher:2022pas,GRAPES-3:2024yym}. 
One leading hypothesis is that a significant fraction of the LEE is stress-induced, arising from relaxation processes in the detector crystal or surrounding mechanical structures. The double-TES technology, pioneered by CRESST and TESSERACT~\cite{GRAPES-3:2024yym,Anthony-Petersen:2024vdh} and implemented in NUCLEUS during the commissioning run~\cite{NUCLEUS:2025ymr}, has demonstrated the capability to discriminate a distinct LEE component that predominantly couples to the sensor rather than to the absorber phonon system. This supports a detector-intrinsic origin for at least part of the LEE. Nevertheless, additional components remain, and a comprehensive study of their origin and mitigation is presented in Ref.~\cite{Abele:2026sqm}.
The \textit{Technical Run} will be followed by a \textit{Physics Run} with the same \cawo target mass, in which the detectors will be integrated into an instrumented holder (inner veto) currently under development within the Collaboration~\cite{NUCLEUS:2017htt,NUCLEUS:2019kxv}. 
This represents a key upgrade with respect to the \textit{Technical Run}, as the inner veto is expected to provide additional event rejection capabilities, in particular mitigating LEE contributions arising from holder-related or external stress on the crystals~\cite{NUCLEUS:2017htt, NUCLEUS:2019kxv}. In parallel, complementary strategies to further reduce the LEE are under investigation, including approaches exploiting its time dependence and the detector cooling conditions~\cite{Abele:2026sqm}. 

The goal of this paper is to assess the physics potential of NUCLEUS at the Chooz VNS during the upcoming \textit{Technical Run} and subsequent \textit{Physics Run}, and to quantify its reach for the SM \cenns signal and selected BSM scenarios enabled by the unprecedentedly low recoil-energy threshold. 
The phenomenological LEE model adopted for the sensitivity studies of the \textit{Technical Run} is introduced in Sec.~\ref{sec:tec-run}.
Importantly, the two-reactor configuration at Chooz naturally induces reactor-power variation over time through scheduled outages and periods of reduced power. As discussed in Sec.~\ref{sec:reactormodulation}, this time dependence can be exploited to enhance sensitivity in the low signal-to-background regime encountered during this phase.
For the \textit{Physics Run}, we consider an optimistic scenario in which the LEE is suppressed to a negligible level, such that the dominant background is given by the particle-induced component estimated from simulations~\cite{Abele:2025yca}. Under this assumption, a measurement of the SM \cenns signal becomes feasible, with a significantly improved sensitivity to BSM scenarios. For both runs, a total measurement period of one year is assumed.
If the LEE cannot be fully suppressed, but is nevertheless reduced compared to the currently observed level, an intermediate physics reach between the \textit{Technical Run} and \textit{Physics Run} scenarios can be expected. A quantitative assessment of such intermediate scenarios is beyond the scope of this work, as it would require a more detailed understanding of the LEE properties and of the effectiveness of the ongoing mitigation strategies.

The paper is structured as follows. In Sec.~\ref{sec:Theo} we describe the theoretical framework considered in this work. In Sec.~\ref{sec:Sens} we present the sensitivity strategy, including a likelihood framework that incorporates the anticipated reactor power variation. In Sec.~\ref{sec:Res} we report the projected sensitivities, and we conclude in Sec.~\ref{sec:conclusions}. The appendices provide details on the validation of the statistical framework (App.~\ref{app:A}) and on the treatment of systematic uncertainties (App.~\ref{app:syst}).

\section{\cenns Theory Framework}
\label{sec:Theo}
\subsection{Standard Model \cenns Cross Section}\label{sec:theory}

The coherent elastic scattering of a neutrino $\nu_\ell$ ($\ell = e,\mu,\tau$) off a nucleus $\mathcal{N}$ is described, within the SM, by the differential cross section~\cite{PhysRevD.9.1389,Kopeliovich:1974mv,Drukier:1984vhf,Sierra:2023pnf}
\begin{equation}
    \dfrac{d\sigma_{\nu_{\ell}\text{-}\mathcal{N}}}{d T_\mathrm{nr}} = 
    \dfrac{G_{\text{F}}^2 M}{\pi} 
    \left( 1 -\frac{T_{\rm nr}}{E_\nu} - \dfrac{M T_\mathrm{nr}}{2 E_\nu^2} \right)
    \left( Q^{V}_{\ell, \mathrm{SM}} \right)^2 \, ,
    \label{eq:cexsec}
\end{equation}
where $G_F$ is the Fermi constant, $M$ is the nuclear mass, $E_\nu$ is the neutrino energy, and $T_{\rm nr}$ denotes the nuclear recoil energy.
The effective weak nuclear charge, which parametrizes the strength of the neutrino-nucleus coupling, is given by
\begin{equation}
    Q^{V}_{\ell, \mathrm{SM}} = \left[ g_{V}^{p}(\nu_\ell) Z F_Z(|\vec{q}|^2) + g_{V}^{n} N F_N(|\vec{q}|^2) \right] \, ,
    \label{eq:weakcharge}
\end{equation}
where $Z$ and $N$ are the proton and neutron numbers of the nucleus, $g_{V}^{n}$ and $g_{V}^{p}$ represent the weak neutral-current couplings of neutrinos with protons and neutrons, and $F_Z(q^2)$ and $F_N(q^2)$ are the corresponding nuclear form factors, describing the loss of coherence at high momentum transfer $|\vec{q}| \simeq \sqrt{2 M T_{\rm nr}}$~\cite{AtzoriCorona:2023ktl}.
To account for the isotopic composition of the target nuclei, a weighted sum over the CE$\nu$NS cross sections of the individual isotopes is performed, specifically
\begin{equation}
\left\langle \frac{d\sigma_{\nu_\ell-\mathcal{N}}}{dT_{\rm nr}} \right\rangle
=
\sum_i f_i \,
\frac{d\sigma_{\nu_\ell-\mathcal{N}_i}}{dT_{\rm nr}} \, ,
\end{equation}
where $f_i$ is the isotopic abundance of the $i-$th isotope of the $\mathcal{N}$ nuclear species.
The isotopic abundances of Ca, W, O and their respective masses are taken from Ref.~\cite{BerglundWieser+2011+397+410} and~\cite{PubChemAtomicMass} respectively.
At the sub-keV recoil energies relevant for NUCLEUS, the momentum transfer is sufficiently small that nuclear coherence is essentially preserved and the form factors are very close to unity. Nevertheless, we model both proton and neutron distributions using the Helm parametrization~\cite{Helm:1956zz}, with nuclear radii following the prescription of Lewin and Smith~\cite{Lewin:1995rx}.

In the SM, the coefficients $g_{V}^{n}$ and $g_{V}^{p}$, including radiative corrections under the so-called minimal subtraction scheme ($\rm \overline{MS}$)~\cite{Fanchiotti:1992tu}, are given by~\cite{AtzoriCorona:2024rtv,AtzoriCorona:2023ktl,Erler:2013xha,PhysRevD.110.030001}
\begin{align}\label{eq:couplings}
g_{V}^{p}(\nu_{e}) &= 0.0379\, , &
g_{V}^{n} &= -0.5117\, ,
\end{align}
which should be compared with the tree-level (TL) predictions of $g_{V,\rm TL}^{p}=1/2-2\sin^2\vartheta_W\simeq0.0225$ and $g_{V,\rm TL}^n=-0.5$ respectively.
The strong suppression of $g_V^p$ relative to $g_V^n$ originates from the weak mixing angle, whose low-momentum-transfer value is $\sin^2\vartheta_W(|\vec{q}|\to0)=0.23873(5)$~\cite{ParticleDataGroup:2024cfk}. As a consequence, \cenns predominantly probes the neutron content of the nucleus, making the nuclear weak charge highly sensitive to variations of $\sin^2\vartheta_W$ at low momentum transfer, a regime where only few experimental measurements exist. 
The extraction of the weak mixing angle from \cenns data is typically affected by a degeneracy with the average neutron radius of the target nucleus~\cite{AtzoriCorona:2023ktl}, which must be inferred from nuclear-structure models in the absence of precise experimental measurements. At the very low nuclear recoil energies probed by NUCLEUS, however, the momentum transfer is sufficiently small that nuclear-structure effects play only a marginal role, enabling a cleaner sensitivity to electroweak parameters.

In Eq.~\ref{eq:couplings}, the values for the neutrino-proton coupling are derived by taking into account their flavor dependence. In fact, radiative corrections induce a flavor-dependent contribution to the neutrino-proton coupling through the neutrino charge radius $\langle r^2_{\nu_e}\rangle$ (NCR), which represents the only non-vanishing electromagnetic property of neutrinos within the SM. For the electronic flavor, this is equal to $\langle r^2_{\nu_e}\rangle\simeq-0.83\times10^{-32}\;\rm cm^2$~\cite{Giunti:2014ixa}. The NCR modifies the effective proton coupling as~\cite{Giunti:2014ixa,AtzoriCorona:2022qrf,Cadeddu:2019eta}
\begin{align}\nonumber
    g_V^p(\nu_\ell)\rightarrow & \tilde{g}_V^p-\dfrac{2}{3}M_W^2 \langle r^2_{\nu_\ell}\rangle\sin^2\vartheta_W\\&= \tilde{g}_V^p-\dfrac{\sqrt{2}\pi\alpha_{\rm EM}}{3G_F}\langle r^2_{\nu_\ell}\rangle \, ,    \label{NCRradiative}
\end{align}
where $M_W$ is the mass of the $W^\pm$ boson, $\alpha_{\rm EM}$ is the electromagnetic fine structure constant. At the same time, $\tilde g_V^p \simeq 0.0184$ is the neutrino-proton coupling without the contribution of the SM NCR, but including the other radiative corrections as described in Ref.~\cite{AtzoriCorona:2023ktl}. In this work, we also account for the energy dependence of the NCR-related radiative corrections~\cite{AtzoriCorona:2024rtv}. In this study, we consider only the diagonal neutrino charge radius $\langle r^2_{\nu_e}\rangle$, neglecting possible BSM effects that could induce off-diagonal, or so-called transition NCRs in which the neutrino changes flavor during the interaction~\cite{Giunti:2014ixa}.
Any deviation of the NCR from its SM prediction would therefore constitute a clear signal of BSM physics.
Although the allowed parameter space is tightly constrained around the SM values~\cite{AtzoriCorona:2025xwr}, a precise determination of the NCR is still lacking. Future high-precision \cenns measurements thus offer a promising avenue to probe this fundamental neutrino property.\\
\par
\begin{figure}[tbp]
    \centering
    \includegraphics[width=\linewidth]{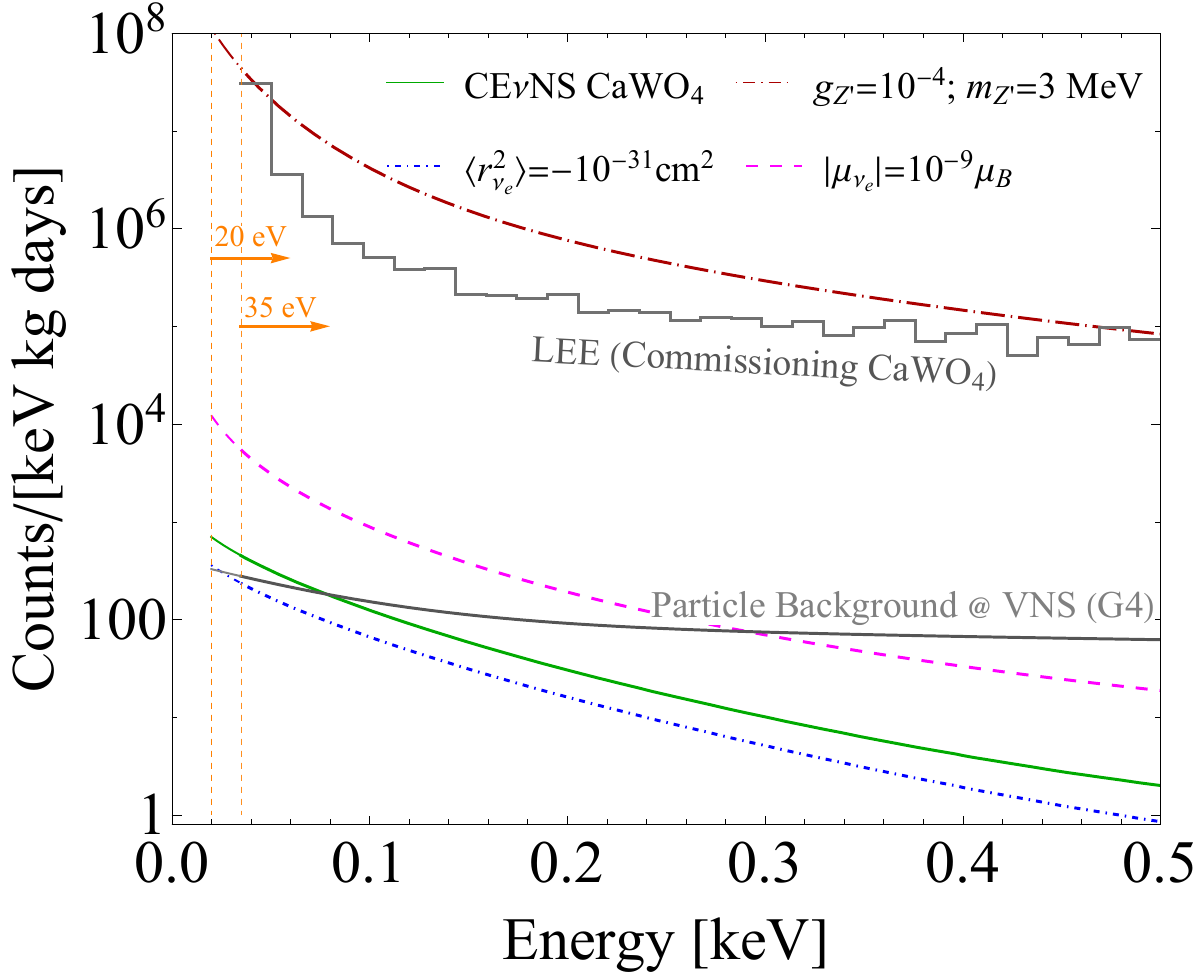}
    \caption{    Expected event rate in the NUCLEUS \cawo target at Chooz as a function of the observed energy for different physics scenarios. The green line shows the SM \cenns prediction, while the dashed and dot-dashed curves correspond to a representative value of neutrino magnetic moment (magenta), modified neutrino charge radius (blue), and the presence of a new light universal mediator (red). The theoretical predictions are qualitatively compared with the LEE background measured during the NUCLEUS commissioning run at TUM~\cite{NUCLEUS:2025ymr} (dark gray line), and the particle background contribution estimated for the NUCLEUS detectors at VNS from Geant4 (G4) simulations~\cite{Abele:2025yca} (light gray line). Vertical lines mark recoil-energy thresholds of 20\,eV and 35\,eV.}
    \label{fig:DataVsModel}
\end{figure}

 Figure~\ref{fig:DataVsModel} illustrates the expected \cenns recoil spectrum in \cawo for the SM (solid green) and for an enhanced value of the NCR (dot-dashed blue). The NCR enters as a correction to the neutrino-proton coupling and can therefore interfere either constructively or destructively with the SM weak charge. For a single nuclear target probed at very low momentum transfer, such an effect would largely manifest as an overall rescaling of the \cenns rate when the form factors approach unity. In contrast, for a multi-component target such as \cawo, the different nuclear species contribute with distinct weights and recoil-energy distributions, leading to a non-trivial distortion of the observed spectrum rather than a simple normalization shift.

\subsection{Beyond the Standard Model Scenarios}
The ultra-low nuclear recoil energy threshold targeted by NUCLEUS provides enhanced sensitivity to new-physics scenarios that modify the \cenns spectrum at low energies. In this work, we consider two well-motivated classes of BSM scenarios: non-standard neutrino interactions (NSI) and the neutrino magnetic moment.

Non-standard neutrino interactions with quarks can be parameterized at low energies through effective four-fermion operators. 
Assuming flavor-conserving neutral-current interactions, the corresponding NSI Lagrangian term can be written as~\cite{Barranco:2005yy,Giunti:2019xpr,Coloma:2023ixt}
\begin{equation}
\mathcal{L}_{\text{NSI}}^{\text{NC}}
=
- 2 \sqrt{2} G_{\text{F}}\sum_{\ell=e,\mu}
\left( \overline{\nu_{\ell L}} \gamma^{\rho} \nu_{\ell L} \right)
\sum_{f=u,d}
\epsilon_{\ell\ell}^{fV}
\left( \overline{f} \gamma_{\rho} f \right)
\, ,
\label{lagrangian}
\end{equation}
where $\nu_{\ell \text{L}}$ and $f$ represent the neutrino and the fermion fields, respectively.
The parameters
$\epsilon_{\ell\ell}^{fV}$, where $f=u,d$ stands for the flavor of the quark and $\ell=e,\mu$ is the neutrino flavor, describe the strength of non-standard interactions relative to standard neutral-current weak interactions. This NSI Lagrangian modifies the weak charge in Eq.~(\ref{eq:weakcharge}), which becomes~\cite{Barranco:2005yy}
\begin{eqnarray}
Q_{\ell,\mathrm{NSI}}^{V}
&=&
\left( g_{V}^{p}(\nu_{\ell}) + 2 \epsilon_{\ell\ell}^{uV} + \epsilon_{\ell\ell}^{dV} \right)
Z
F_{Z}(|\vet{q}|^2)
+ \nonumber \\ 
&+&\left( g_{V}^{n} + \epsilon_{\ell\ell}^{uV} + 2 \epsilon_{\ell\ell}^{dV} \right)
N
F_{N}(|\vet{q}|^2) \, .
\label{eq:Qalpha2}
\end{eqnarray}
In this work, we investigate the sensitivity to NSI parameters in the flavor-preserving framework, where neutrino flavor remains unchanged during scattering, and the parameters $\epsilon_{ee}^{uV}$ and $\epsilon_{ee}^{dV}$ are allowed to vary simultaneously. This approach enables a direct comparison with existing studies and sets the stage for the physics reach of the experiment under alternative NSIs frameworks, e.g.~\cite{Giunti:2019xpr,Amaral:2023tbs,RES-NOVA:2026fii}.

A particularly well-studied realization of NSIs arises when they are mediated by a light vector boson $Z'$, associated with an additional $U(1)'$ gauge symmetry. In this case, the NSI parameters acquire a momentum-dependent form~\cite{Cadeddu:2020nbr,AtzoriCorona:2022moj}
\begin{equation}
    \epsilon_{\ell \ell}^{fV}=\dfrac{g_{Z'}^2\,Q'_\ell Q'_f}{\sqrt{2}G_F\, (|\vec{q}|^2+m_{Z'}^2)} \, ,
    \label{eq:LM}
\end{equation}
where $m_{Z'}$ and $g_{Z'}$ denote the mediator mass and coupling strength, respectively, and $Q'$ are the fermion charges under the new gauge symmetry. 
As a reference scenario, in this work, we consider a universal model with $Q'_\ell = Q'_f = 1$, which, while not gauge invariant without additional field content~\cite{Allanach:2018vjg}, provides a convenient reference for comparison with existing experimental constraints. Other gauge structures, such as $B-L$ or $B-3L_e$~\cite{DeRomeri:2022twg,AtzoriCorona:2022moj,Demirci:2023tui}, or scenarios with a scalar mediator~\cite{Lindner:2016wff}, can be straightforwardly accommodated once experimental data become available. 
Owing to the $1/(|\vec{q}|^2 + m_{Z'}^2)$ enhancement of the mediator propagator at low momentum transfer, light vector mediator scenarios are particularly sensitive to the ultra-low recoil-energy regime probed by NUCLEUS. This is illustrated in Fig.~\ref{fig:DataVsModel}, where the red dot-dashed curve shows the theoretical prediction for a representative point in the $\left(g_{Z'},\, m_{Z'}\right)$ parameter space, which features a strong enhancement at low recoil energies.
This allows NUCLEUS to set competitive constraints on light mediator models even in the absence of a direct \cenns observation.

A fundamental BSM neutrino property is provided by the neutrino magnetic moment ($\mu_{\nu}$), which is expected to arise because neutrinos are massive~\cite{Giunti:2014ixa,Giunti:2015gga,Giunti:2024gec}. In the minimally extended SM, the expected value of the neutrino magnetic moment for Dirac neutrinos is of the order of $\mu_\nu\sim10^{-19}\mu_B$~\cite{Giunti:2024gec}, which lies far below current experimental sensitivities. Therefore, any positive experimental observation would constitute an unambiguous signal of new physics, demanding further theoretical investigation to clarify its origin. The magnetic-moment contribution does not interfere with the SM \cenns amplitude and adds incoherently to Eq.~(\ref{eq:cexsec}), namely
\begin{equation}
\dfrac{d\sigma_{\nu_{\ell}\text{-}\mathcal{N}}^{\text{MM}}}{d T_\mathrm{nr}}
=
\dfrac{ \pi \alpha_{\rm EM}^2 }{ m_{e}^2 }
\left( \dfrac{1}{T_\mathrm{nr}} - \dfrac{1}{E_\nu} \right)
Z^2 F_{Z}^2(|\vec{q}|^2)
\left| \dfrac{\mu_{\nu_{\ell}}}{\mu_{\text{B}}} \right|^2
\, ,
\label{cs-mag}
\end{equation}
where $\mu_{\nu_\ell}$ is the effective magnetic moment of the neutrino flavor $\nu_\ell$ and $\mu_B$ is the Bohr magneton~\cite{Giunti:2014ixa}.
Owing to its characteristic $1/T_{\rm nr}$ enhancement, this contribution is particularly relevant at very low recoil energies, making NUCLEUS sensitive to competitive constraints despite its small target mass. An example of the
corresponding theoretical prediction, assuming a neutrino magnetic moment of
$\mu_{\nu_e} = 10^{-9}\,\mu_B$, is shown in Figure~\ref{fig:DataVsModel}.

In this work, we restrict the study of neutrino magnetic moment to the \cenns channel. This additional interaction channel is particularly relevant in scenarios predicting an enhanced low-energy signal, and has been successfully exploited in several experiments employing Ge~\cite{CONUS:2022qbb,Beda:2009kx,AtzoriCorona:2025ygn,Coloma:2022avw}, CsI~\cite{Coloma:2022avw,TEXONO:2002pra,AtzoriCorona:2022qrf} and Xe~\cite{XENON:2022ltv,LZ:2023poo} targets. In fact, the neutrino-electron scattering ($\nu$ES) signal in the $\mathcal{O}(\mathrm{keV})$ recoil range and can be modeled with good theoretical control by scaling the free electron-neutrino cross section by the number of ionized electrons for a given energy deposit~\cite{Kouzakov:2017hbc}. 
In contrast, for the $\mathcal{O}(10~\mathrm{eV})$ energy scale relevant for NUCLEUS, a reliable calculation of neutrino-electron scattering in \cawo is hindered by poorly known atomic many-body effects. For this reason, we do not consider the $\nu$ES channel further in this analysis.

\section{Sensitivity Strategy}
\label{sec:Sens}

In this section, we describe the strategy adopted to derive sensitivity projections for the different phases of the NUCLEUS experiment at the Chooz site. The analysis framework accounts for the distinct detector configurations and background conditions expected during the upcoming \textit{Technical Run} and the subsequent \textit{Physics Run}.

\subsection{Signal Prediction and Background Model}

The signal and background models used in the sensitivity studies depend on the specific detector configuration considered. In this work, we distinguish between two main running phases: the initial \textit{Technical Run} at Chooz and the following \textit{Physics Run}.

\subsubsection{\cenns Signal Prediction}

The expected \cenns recoil spectrum is obtained by convolving the differential cross section with the reactor antineutrino flux, integrating over all possible neutrino energies, and folding in the detector energy response. The resulting differential event rate is given by
\begin{align}\label{eq:RateExp}\nonumber
\frac{dR}{dT_{\rm nr}}= & N_T \sum_\mathcal{N}
\int_{0}^{T^{\prime\text{max}}_{\text{nr}}}
\hspace{-0.3cm}
dT'_{\text{nr}}\;\mathcal{R}(T_{\text{nr}},T'_{\text{nr}}) 
\int_{E_{\nu}^{\text{min}}(T'_{\text{nr}})}^{E_{\nu}^{\text{max}}}
\hspace{-0.3cm}
d E_\nu \times \\
&
\frac{d \Phi}{d E_\nu}(E_\nu)
\left\langle\frac{d \sigma_{\bar \nu_e-\mathcal{N}}}{d T'_{\mathrm{nr}}}(E_\nu, T'_{\mathrm{nr}})\right\rangle\, , \end{align}
where the sum runs over the nuclear species $\mathcal{N}$ in \cawo, and $N_T$ denotes the number of target nuclei in the detector. The maximum recoil energy is $T^{\prime\,\mathrm{max}}_{\rm nr} \simeq 2\left(E_{\nu}^{\rm max}\right)^2/M$, with $E_{\nu}^{\rm max}=12$~MeV~\cite{Perisse:2023efm}, while the minimum neutrino energy required to produce a recoil $T'_{\rm nr}$ is $E_\nu^{\rm min}(T'_{\rm nr}) \simeq \sqrt{M T'_{\rm nr}/2}$.

The detector response function $\mathcal{R}$ is modeled as a Gaussian, with an energy-dependent width $\sigma(T_{\rm nr})$, parameterized following Ref.~\cite{Abele:2025rrl}
\begin{equation}\label{eq:energyres}
   \sigma(T_{\rm nr})=\sqrt{\left(\frac{\eta}{e_{\rm ath}}\right)^{-1}T_{\rm nr} + (\beta\, T_{\rm nr})^2 + \sigma_0^2}\, .
\end{equation}
Here $\eta/e_{\rm ath}=0.0045\pm0.0016\,\mathrm{meV}^{-1}$ denotes the ratio between the athermal phonon collection efficiency and the mean energy of an athermal phonon, while $\beta=0.013\pm0.0011$ encodes systematic effects such as non-uniformities and position-dependence effects~\cite{Abele:2025rrl}. We assume a baseline resolution of $\sigma_0=4$\,eV, consistent with the energy-threshold targets considered in this work. Owing to the excellent intrinsic energy resolution of NUCLEUS detectors, the impact of the energy response on the total event rate is at the sub-percent level and constitutes a negligible source of systematic uncertainty.
Since the trigger threshold is typically set at five times the detector's baseline resolution, the finite energy resolution also determines the effective trigger efficiency near threshold: for an optimal-filter trigger, the efficiency follows the Gaussian smearing and is therefore 50\% at the trigger threshold by construction~\cite{DiDomizio:2010ph}. This behavior was experimentally validated in the commissioning run analysis~\cite{NUCLEUS:2025ymr}. No additional energy-dependent trigger efficiency is applied beyond this resolution folding; only a flat overall efficiency factor is included in the overall analysis efficiency quoted below (Sec.~\ref{sec:Res}).

\subsubsection{Technical Run}\label{sec:tec-run}

During the \textit{Technical Run} at Chooz, the detector module will consist of four \cawo double-TES detectors, for a total target mass of 7\,g. In this configuration, the \cenns region of interest is expected to be dominated by the LEE, while the particle-induced background estimated from simulations is negligible in comparison (see Figure~\ref{fig:DataVsModel}).

The rate and spectral shape of the LEE are modeled using data from the NUCLEUS commissioning run~\cite{NUCLEUS:2025ymr}. Since an energy threshold of 35\,eV was achieved during commissioning, the same analysis threshold is adopted for the technical-run sensitivity projections. The analysis range extends up to 500\,eV. Within this range, the LEE spectral shape is parameterized by a double-exponential function,
\begin{equation}
      S_{\rm LEE}(T_{\rm nr}) = B \times \left[ f\,e^{-T_{\rm nr}/\varepsilon_1} + (1-f)\,e^{-T_{\rm nr}/\varepsilon_2} \right] \,,
\end{equation}
where $S_{\rm LEE}(T_{\rm nr})$ describes the normalized energy distribution of LEE events in the analysis window. The normalization constant $B$ is chosen such that $\int S_{\rm LEE}(T_{\rm nr})\, dT_{\rm nr} = 1$ over the 35--500\,eV range. The parameters $f \simeq 0.99$, $\varepsilon_1 \simeq 28$\,eV, and $\varepsilon_2 \simeq 320$\,eV are obtained from fits to the \cawo commissioning data over the wider 35--7000\,eV energy range~\cite{NUCLEUS:2025ymr}, where the fit model includes an additional constant component that becomes relevant only well above 500\,eV and is therefore neglected in the present analysis.

In addition, a time-dependent decay of the LEE rate was observed during the commissioning run and is modeled by a power-law function,
\begin{equation}
R_{\rm LEE}(t) = R_{0, \rm LEE} \, (t-t_0)^{-k} \, ,
\end{equation}
where $R_{0,\rm LEE}$ is the initial LEE rate and $k$ is the exponent governing the temporal decay. The parameter $t_0$ denotes a common reference time defined as the moment when the detectors reach 4\,K, which is used to parameterize the onset of the LEE decay. During the commissioning run, this occurred after approximately 9\,days, and a similar timescale is expected at Chooz. Further details on the definition of $t_0$ and the LEE time evolution are provided in Ref.~\cite{Abele:2026sqm}. When the time difference $t - t_0$ is expressed in days, the exponent was found to be $k \simeq 0.59$ during the commissioning run, and was observed to consistently describe multiple runs with a different NUCLEUS detector~\cite{Abele:2026sqm}; the same functional form is therefore adopted in this work.
The initial LEE rate observed in the commissioning \cawo detector, operated as a single-TES device, between 35 and 500\,eV is 3650\,events/day. In this work, this rate is directly adopted for the \textit{Technical Run} detectors. Although these will be operated with a double-TES configuration, which is expected to provide additional discrimination against the LEE, we conservatively do not assume any suppression from this technique. The same initial rate is therefore assigned to each of the four detectors. Furthermore, despite the smaller target mass of the commissioning detector (0.76\,g) compared to the 1.75\,g of each \textit{Technical Run} detector, we assume that the LEE rate does not scale with detector mass. This assumption is supported by observations from other cryogenic experiments operating at comparable energy thresholds~\cite{Angloher:2022pas, Baxter:2025odk, Abele:2026sqm}. \footnote{We note that the phenomenological assumptions adopted for the LEE model in this work represent one possible, conservative scenario motivated by the currently available data. The scaling of the LEE with detector mass, material, and detector design will need to be validated experimentally with dedicated measurements using the \textit{Technical Run} detectors.}  
In addition, the same initial rate is adopted for the Chooz site despite the different overburden with respect to the shallow underground commissioning setup. This assumption is motivated by dedicated studies within NUCLEUS, which did not observe a correlation between the LEE rate and variations in the ambient particle background or shielding conditions, thereby disfavoring a dominant cosmogenic origin of the LEE~\cite{Abele:2026sqm}.

\subsubsection{Physics Run (LEE-free)}\label{sec:phys-run}

\begin{table}[tb]
\centering
\begin{tabular}{c|c|c}
 & \multicolumn{2}{c}{Rate (events/day)} \\
$T_{\rm nr}$ range (eV) & \cenns\ Signal & Background \\
\hline
 20-200 & 0.214 & 0.209 \\
200-500 & 0.019 & 0.167  \\
\end{tabular}
\caption{Expected signal and background rates for the NUCLEUS \textit{Physics Run} with a 7\,g \cawo target, assuming complete suppression of the LEE. The signal rates correspond to operation at maximum reactor power. The quoted background rate includes all background components considered in Ref.~\cite{Abele:2025yca}, and is derived from the corresponding background simulations.} \label{tab:rate-physics}
\end{table}

The \textit{Physics Run} is expected to start immediately after the \textit{Technical Run} and will employ the same total \cawo target mass of about 7\,g. In this phase, the detectors will be integrated into an instrumented holder (inner veto)~\cite{NUCLEUS:2017htt, NUCLEUS:2019kxv}, representing a key upgrade aimed at enhancing the rejection of the LEE background, in particular for contributions arising from holder-related or external stress on the crystals. Complementary mitigation strategies, such as exploiting the time dependence of the LEE and optimising detector cooling conditions, are also under investigation. For the sensitivity projections presented in this work, we consider an optimistic scenario in which the LEE is suppressed to a negligible level, such that the dominant background during the \textit{Physics Run} is given by the particle-induced component estimated from simulations~\cite{Abele:2025yca}. Because the detectors are located at distances of order $70$-$100$\,m from the reactor cores, reactor-correlated particle backgrounds are expected to be negligible, and this component is therefore assumed to be constant in time.

An improved recoil-energy threshold of 20\,eV is considered achievable and is adopted for the physics-run sensitivity projections. 
This lower threshold does not rely on a modified detector design, but on already demonstrated energy resolutions~\cite{NUCLEUS:2017gvo} and on improved analysis methods developed within the Collaboration~\cite{NUCLEUS:2026utx}.
The analysis range extends up to 500\,eV, and the expected signal and background rates~\cite{Abele:2025yca} in this range are summarized in Table~\ref{tab:rate-physics}.

\subsection{Reactor-Power Variation Analysis}\label{sec:reactormodulation}

At the Chooz nuclear power plant, periods in which both reactor cores are simultaneously OFF are unlikely, meaning background subtraction based solely on dedicated reactor-OFF data is not possible. 
However, scheduled maintenance shutdowns of individual reactors and periods of reduced reactor power naturally induce a variation of the total thermal power, and hence of the antineutrino flux impinging on the detector.

The total antineutrino flux at the VNS can be expressed in terms of a single source characterized by an effective thermal power and distance. For the two Chooz cores, this is given by
\begin{equation}\label{eq:eff-power-dist}
\frac{1}{L_{\text{eff}}^2} = \frac{1}{L_1^2}+\frac{1}{L_2^2}\, ,
\qquad
P_{\text{eff}} = \frac{P_1 L_2^2 + P_2 L_1^2}{L_1^2+L_2^2} \, ,
\end{equation}
where $L_1 = 72$\,m and $L_2 = 102$\,m are the distances to reactor cores B1 and B2, respectively, and $P_1$ and $P_2$ denote their thermal powers. This yields an effective distance of $L_{\mathrm{eff}} \simeq 58.8$\,m. Each reactor operates at a nominal thermal power of 4.25\,GW$_\mathrm{th}$, such that the effective power varies between zero, when both cores are OFF, and 4.25\,GW$_\mathrm{th}$ when both cores are ON. Intermediate effective powers amount to approximately one-third of the maximum power when B1 is OFF and two-thirds when B2 is OFF.

The time dependence of the effective reactor power induces a corresponding variation of the antineutrino flux, which can be exploited to identify signal and background in the time domain. This strategy was successfully employed by the Double Chooz experiment to obtain a background–model-independent measurement of the mixing angle $\theta_{13}$ with a precision comparable to that achieved using full spectral information~\cite{DoubleChooz:2020vtr}. In that case, the method benefited from a very high signal rate and a large signal-to-background ratio (approximately 1000 signal events per day at full power compared to fewer than 20 background events per day, corresponding to S/B $\sim 50$).
In contrast, the situation for NUCLEUS is markedly different. When the LEE is taken into account, the signal-to-background ratio in the \cenns recoil-energy region of interest is $\mathrm{S/B} \ll 1$, and even under the ideal background conditions of the \textit{Physics Run}, the total event rate at full reactor power remains below one count per day\footnote{The much lower event rate in NUCLEUS is primarily a consequence of the gram-scale target mass, in contrast to the $\sim 8.3$\,t liquid scintillator target mass of each Double Chooz detector~\cite{DoubleChooz:2022ukr}, even though the \cenns cross section at reactor energies is orders of magnitude larger than the one of the inverse beta decay.}.
To retain sensitivity in this low-rate, background-dominated regime, we develop a likelihood-based analysis framework that exploits both the time dependence induced by reactor-power variation and the recoil-energy information of individual events.

The unbinned log-likelihood function ($\mathcal{L}$) used in this work is defined as
\begin{align}\label{eq:loglike_comb}
-\log \mathcal{L} &= \left[ \alpha \int_{T_1}^{T_2} P(t)\; dt + \int_{T_1}^{T_2} \beta(t)\; dt \right] \nonumber \\
&- \sum_i \log \left[ \alpha \, P(t_i) \, f_s(E_i) + \beta(t_i) \, f_b(E_i) \right] \, , 
\end{align}
where each term has a well-defined physical interpretation:
\begin{itemize}
    \item $\alpha$ is the rate of the \cenns signal, expressed in counts per day per MW$_\mathrm{th}$;
    \item $P(t)$ is the time-dependent reactor power, proportional to the instantaneous antineutrino flux at the detector, and $t_i$ denotes the timestamp of the $i$-th event;
    \item $f_s(E)$ is the \cenns recoil-energy probability density function (PDF) within the analysis window, including detector response effects;
    \item $\beta(t)$ is the rate of the background component, in counts per day. In the \textit{Technical Run}, $\beta(t)$ follows the temporal decay of the LEE as introduced in Sec.~\ref{sec:tec-run}, while in the \textit{Physics Run} it is assumed to be constant;
    \item $f_b(E)$ is the background recoil-energy PDF. It corresponds to the LEE spectral shape for the \textit{Technical Run} and to the particle-induced background spectrum estimated from simulations for the \textit{Physics Run}.
\end{itemize}

The first term in square brackets in Eq.~(\ref{eq:loglike_comb}) represents the Poisson expectation value for the total number of events in the observation time window $[T_1, T_2]$, given by the sum of the signal contribution, proportional to the time-integrated reactor power, and the time-integrated background rate.
The second term describes the likelihood of the observed dataset, where each event $(t_i, E_i)$ is assigned a probability composed of a time-modulated signal component and a background component whose rate may vary in time but whose spectral shape is assumed to be stationary.
\begin{figure}[tbp]
    \centering
    \includegraphics[width=\linewidth]{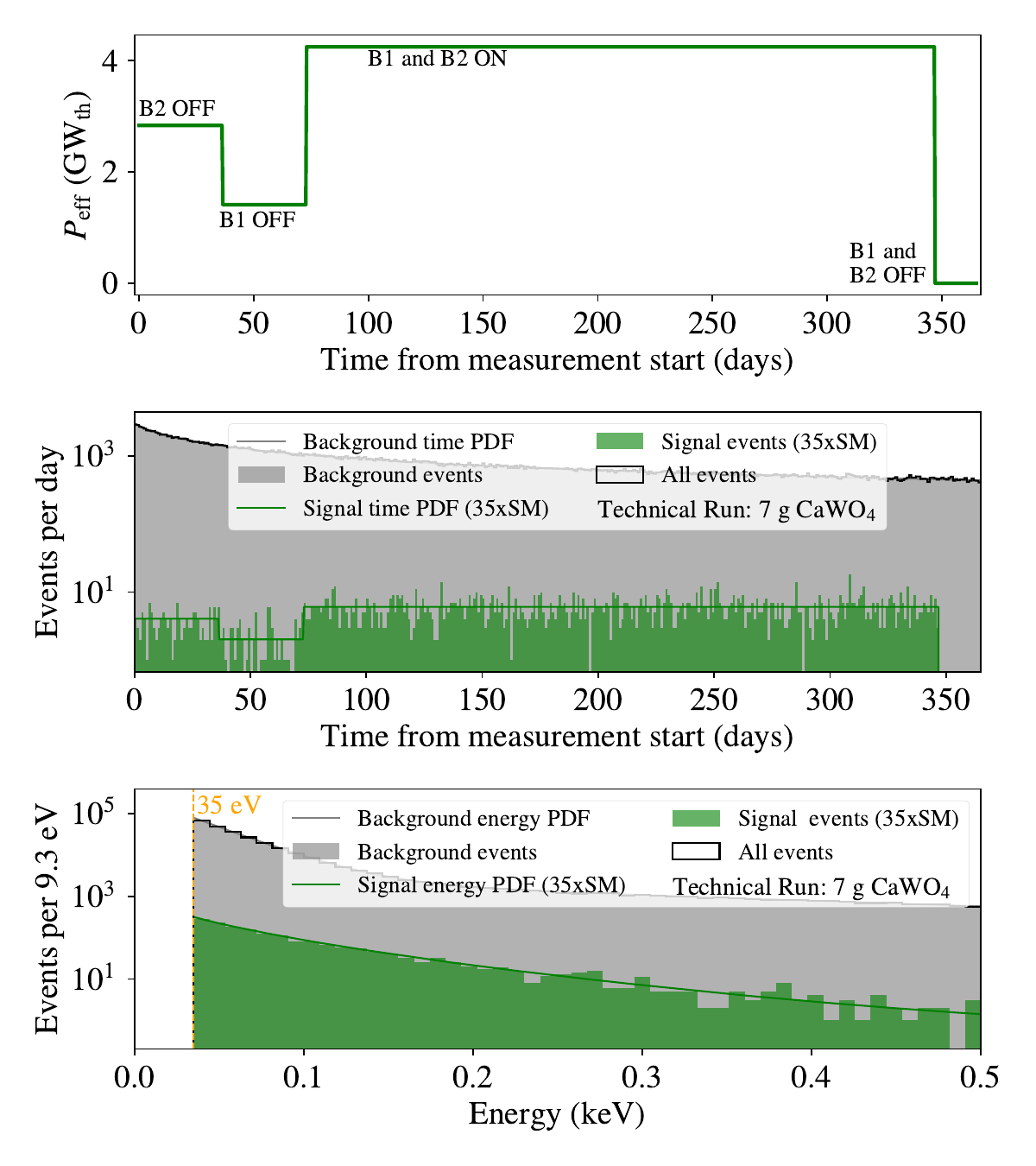}
    \caption{\textbf{Top:} Assumed time evolution of the effective reactor thermal power (Eq.~\ref{eq:eff-power-dist}) at the Chooz VNS over one year, reflecting realistic operational conditions. The two reactor cores are treated as a single effective source at an effective distance of $L_{\mathrm{eff}} \simeq 58.8$\,m.
    \textbf{Middle:} Representative Monte Carlo pseudo-experiment for the \textit{Technical Run} configuration, generated assuming a \cenns signal normalized to 35 times the SM prediction for illustrative purposes. Time distribution of events over one year, showing the signal and background components separately, together with their expected PDFs. The signal follows the reactor-power variation, while the background reflects the time decay of the LEE. \textbf{Bottom:} Recoil-energy spectrum of the same pseudo-experiment, displaying the signal and background contributions and their corresponding energy PDFs.}
    \label{fig:ex-power-profile}
\end{figure}
Figure~\ref{fig:ex-power-profile} illustrates the reactor-power configuration adopted for the sensitivity study, constructed assuming a realistic one-year data-taking period. During this time, the reactor operational schedule includes two single-core refueling and maintenance shutdowns, each lasting approximately one month, during which one of the two reactors (B1 or B2) is OFF while the other operates at nominal thermal power. In addition, a two-week period in which both reactors are simultaneously OFF is assumed. For the remainder of the year, both reactors are taken to operate at nominal power. This results in an overall effective reactor duty cycle of approximately 85\%. The assumed schedule is representative of typical Chooz operating conditions and is consistent with historical reactor operation patterns inferred from publicly available electricity production records averaged over the past decade~\cite{rte-data}.
This structured variation of the antineutrino flux provides time intervals with enhanced signal-to-background contrast, despite the presence of the dominant LEE component. 
In the present study, stable detector operation after the initial cool-down is assumed, such that the LEE time evolution can be described by a single power-law behavior over the full data-taking period. Possible interruptions requiring a warm-up and subsequent cool-down cycle could be incorporated into the phenomenological LEE model and included in the likelihood analysis if needed.

Monte Carlo pseudo-experiments are generated by randomly sampling Poisson-distributed toy datasets from the signal and background PDFs defined above, accounting for the time dependence of the reactor power and, where applicable, of the background rate. Figure~\ref{fig:ex-power-profile} shows an example pseudo-experiment for the \textit{Technical Run} configuration, illustrating the resulting time and energy distributions of signal and background events.
For each pseudo-experiment, parameter inference is performed using a profile likelihood ratio test statistic~\cite{Cowan:2010js},
\begin{equation}\label{eq:profile-likelihood}
q(\theta) = -2 \log \frac{\mathcal{L}(\theta, \hat{\hat{\boldsymbol{\nu}}})}{\mathcal{L}(\hat{\theta}, \hat{\boldsymbol{\nu}})} \, ,
\end{equation}
where $\theta$ denotes the parameter of interest (e.g.\ the \cenns signal normalization or a BSM coupling), $\boldsymbol{\nu}$ represents the set of nuisance parameters, $\hat{\hat{\boldsymbol{\nu}}}$ are the values of the nuisance parameters that maximize the likelihood for fixed $\theta$, and $(\hat{\theta}, \hat{\boldsymbol{\nu}})$ correspond to the global maximum of the likelihood.
All relevant nuisance parameters are profiled in the fit. These include the normalization of the background component, as well as the parameters describing the background energy spectrum and, for the \textit{Technical Run}, the temporal evolution of the LEE rate. No external constraints are imposed on these parameters.
The distributions of the test statistic obtained from ensembles of pseudo-experiments are used to determine median expected discovery significances for the SM \cenns signal or, in the absence of a signal, to derive projected upper limits on the parameters of interest at the desired confidence level. Additional details on the statistical procedure and its validation are provided in App.~\ref{app:A}.

\section{Results}\label{sec:Res}

In this section, we present the results of the sensitivity studies described above. Unless otherwise stated, all projections assume one year of data acquisition with an overall 80\% efficiency, which accounts for detector live time and analysis efficiency. Illustrative examples on the extraction of median expected upper limits, confidence intervals, and discovery significances are presented in App.~\ref{app:A}.

\subsection{Sensitivity to Standard Model \cenns}\label{sec:SensSM}

\begin{figure}[tb]
        \centering
        \includegraphics[width=0.45\textwidth]{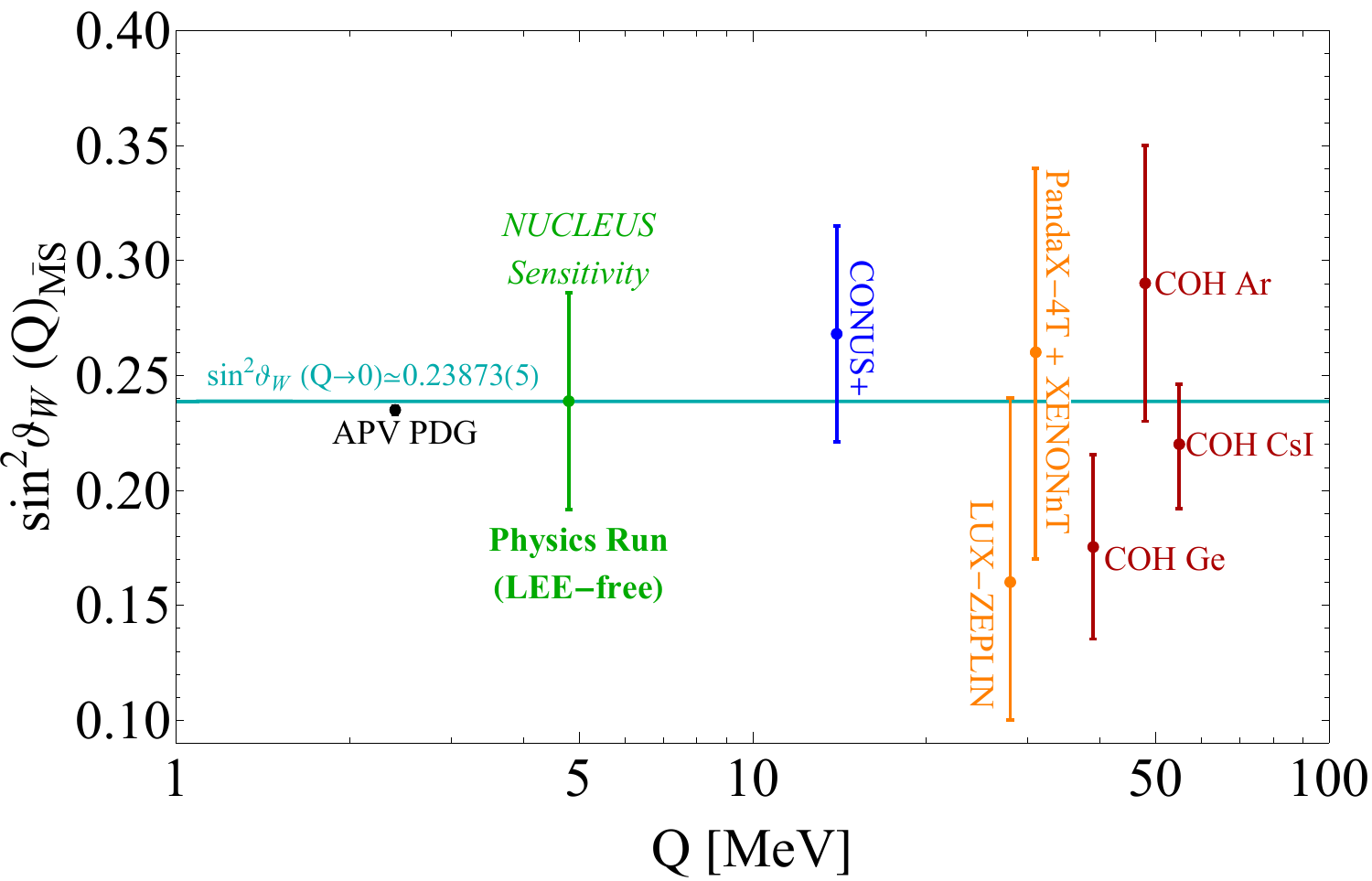}
        \caption{Projected sensitivity of the NUCLEUS experiment to the weak mixing angle $\sin^2\theta_W(Q)_{\mathrm{\overline{MS}}}$ as a function of the momentum transfer $|\vec{q}|\doteq\rm Q$. The solid line represents the latest theoretical determination at zero momentum transfer~\cite{ParticleDataGroup:2024cfk}. The green point indicate the expected precision for the \textit{Physics Run}, assuming complete suppression of the LEE. This is compared to the precision from atomic parity violation (APV) on cesium~\cite{ParticleDataGroup:2024cfk} and existing measurements from \cenns experiments, specifically: CONUS+~\cite{Alpizar-Venegas:2025wor}, COHERENT CsI~\cite{COHERENT:2021xmm}, Ar~\cite{Cadeddu:2020lky}, Ge~\cite{AtzoriCorona:2025xgj}, a combined analysis of XENONnT~\cite{XENON:2024ijk} and PandaX~\cite{PandaX:2024muv} data~\cite{DeRomeri:2024iaw} and the determination from LUX-ZEPLIN~\cite{LZ:2025igz}.}
        \label{fig:runningweak}
    \end{figure}

Owing to the presence of the LEE, which exceeds the expected \cenns signal by several orders of magnitude, a detection of the SM \cenns signal is not expected during the upcoming \textit{Technical Run}. Using the reactor-power variation analysis introduced in Sec.~\ref{sec:reactormodulation}, together with the signal and background models described in Sec.~\ref{sec:tec-run}, we obtain a median 90\%~confidence level (CL) sensitivity corresponding to a signal normalization approximately 35 times larger than the SM prediction. To put this number into context, we can compare it with results obtained from other reactor \cenns experiments. At 90\% CL, CONNIE~\cite{CONNIE:2019xid} and RED-100~\cite{RED-100:2024izi} reported observed upper limits around 70 and 60 times the SM expectation, respectively, while MINER~\cite{Mondal:2025odw} currently stands at roughly 1500 times the predicted SM value. In contrast, the CONUS+ experiment reported a $3.7\,\sigma$ observation of reactor \cenns with a total uncertainty of about 27\% on the extracted neutrino signal~\cite{Ackermann:2025obx}.
For comparison, a likelihood analysis based solely on spectral-shape information yields a substantially weaker sensitivity, at the level of about 200 times the SM rate, due to the strong degeneracy between the \cenns spectrum and the LEE shape. An analysis based only on the time variation of the total event rate performs better, reaching a sensitivity of roughly 50 times the SM expectation at the 90\%~CL. These results highlight the substantial gain achieved by combining temporal and spectral information in the likelihood, as illustrated in Fig.~\ref{fig:profileL-tec-run} of App.~\ref{app:A}.
Because the sensitivity of the \textit{Technical Run} is limited by the dominant LEE background, it improves only slowly with exposure, and extending the data-taking period beyond one year would not lead to a substantial gain in sensitivity within realistic timescales.

In contrast, during the \textit{Physics Run} -- where the LEE contribution is assumed to be suppressed to negligible levels -- a clear observation of the SM \cenns signal becomes feasible. For the reference configuration considered here, a median discovery significance of $4.7\,\sigma$ is expected, corresponding to approximately 58 expected CE$\nu$NS events and a statistical uncertainty of around 20\% on the measured cross section.
This precision enables a competitive determination of the weak mixing angle at low momentum transfer, with $\sin^2\vartheta_W$ treated as the only free electroweak parameter in the fit. 
We obtain an expected 1$\sigma$ (68.27\%~CL) interval of
\begin{equation}
    0.187 <\sin^2\vartheta_W(|\vec{q}|\sim5\,\text{MeV})< 0.286 \,,
\end{equation}
where the quoted momentum transfer corresponds to the characteristic scale probed by NUCLEUS. This would represent the lowest momentum-transfer determination of the weak mixing angle to date from a CE$\nu$NS experiment. A comparison with existing measurements is shown in Figure~\ref{fig:runningweak}. 
The impact of the dominant systematic uncertainties on these results has also been evaluated. Their contribution is found to be at the level of a few percent, and therefore subdominant with respect to the statistical uncertainty. Details of this study are provided in App.~\ref{app:syst}.
\begin{figure}[tb]
        \centering
        \includegraphics[width=0.45\textwidth]{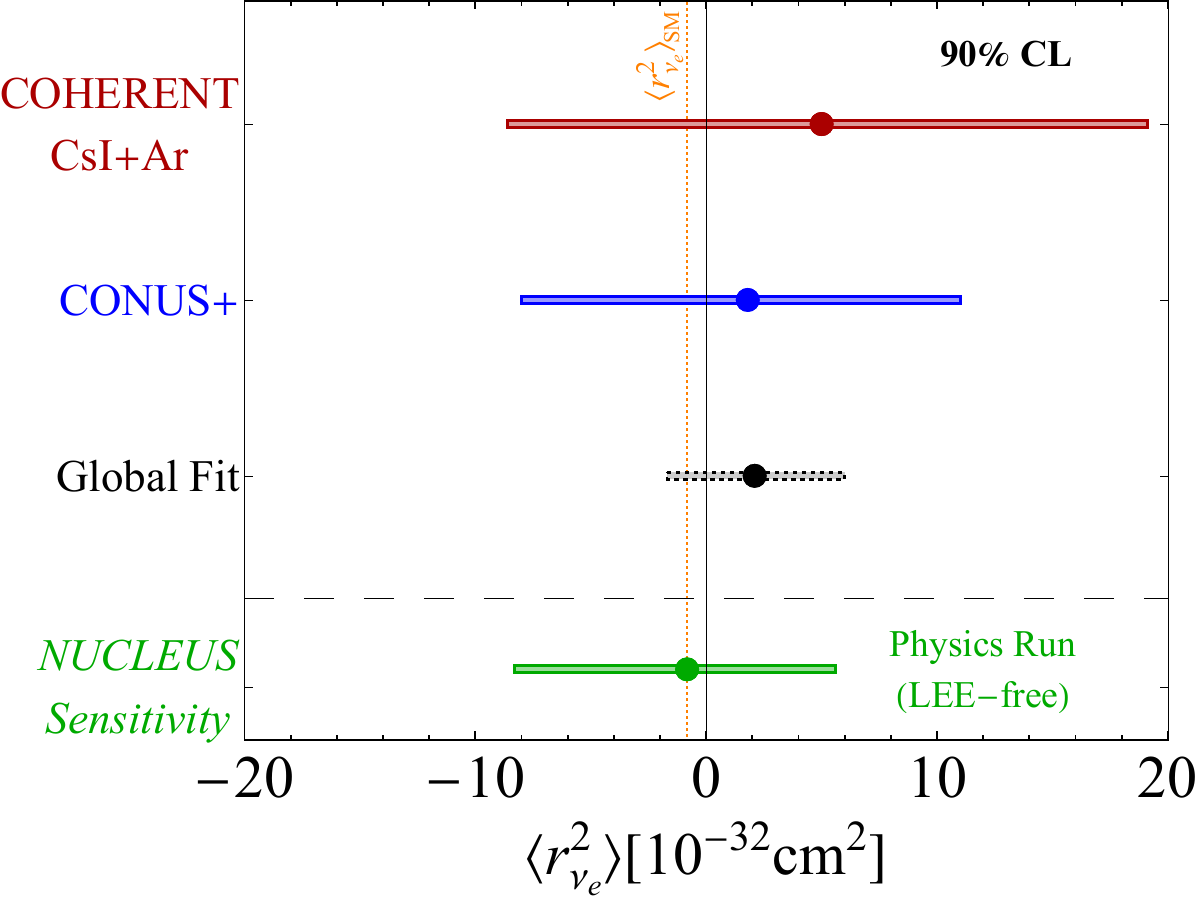}
        \caption{Projected sensitivity of the NUCLEUS experiment to the diagonal electronic neutrino charge radius $\langle r^2_{\nu_e}\rangle$. The green point indicate the expected precision for the \textit{Physics Run}, assuming complete suppression of the LEE, compared to the precision from CONUS+~\cite{AtzoriCorona:2025ygn} and COHERENT CsI+Ar~\cite{AtzoriCorona:2024rtv}, and a global fit of existing neutrino data~\cite{AtzoriCorona:2025xwr}.}
        \label{fig:limits-cr}
    \end{figure}

In Figure~\ref{fig:limits-cr} we present the projected sensitivities to the neutrino charge radius. For the \textit{Physics Run}, NUCLEUS is expected to constrain the NCR within\footnote{In this study, we do not extend the fit to the degenerate solution appearing at large and negative values of  $\langle r^2_{\nu_e}\rangle$, as this region has been ruled out by a recent global analysis of neutrino data~\cite{AtzoriCorona:2025xwr}.}
\begin{equation}
    (-8.3 < \langle r_{\nu_e}^2\rangle < 5.6 )\times 10^{-32}\,\text{cm}^2
\end{equation}
at 90\%~CL. This would represent one of the most stringent constraints to date, when compared to the existing \cenns measurement shown in Figure~\ref{fig:limits-cr}. We note that while at the SNS it is possible to simultaneously constrain both the electronic and muonic flavors, an intrinsic degeneracy between the two parameters weakens the sensitivity on the electronic flavor. Furthermore, since the neutrino flux at COHERENT is dominated by $\nu_\mu$, reactor neutrinos offer a highly complementary probe, exceptionally sensitive to the electronic flavor. This complementarity helps to break degeneracies when combining data from different neutrino sources.

\subsection{Sensitivity To Physics Beyond the Standard Model}\label{sec:BSMResults}

\begin{figure*}[th!]
    \centering
    \includegraphics[width=0.5\linewidth]{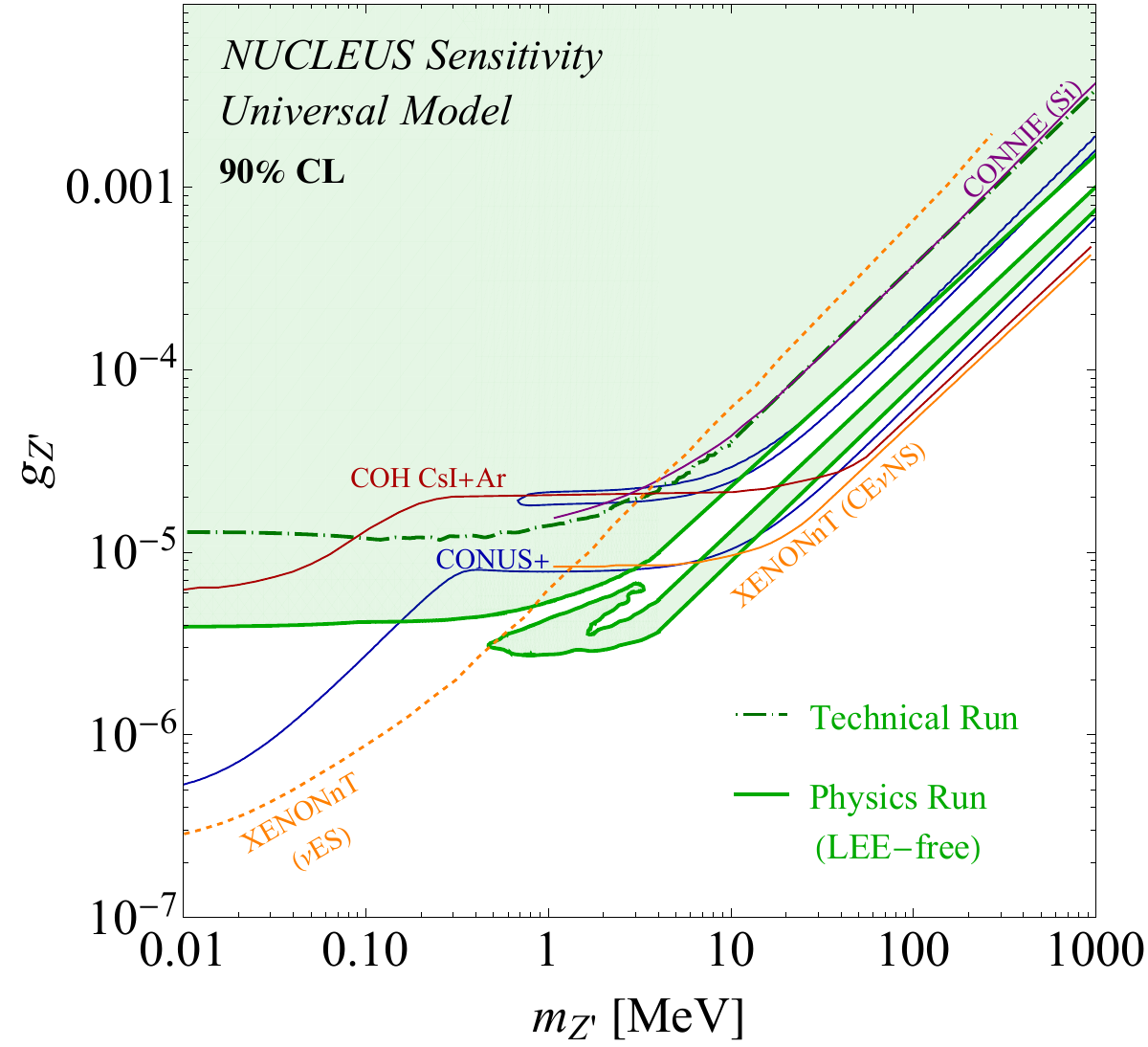}
\includegraphics[width=0.48\linewidth]{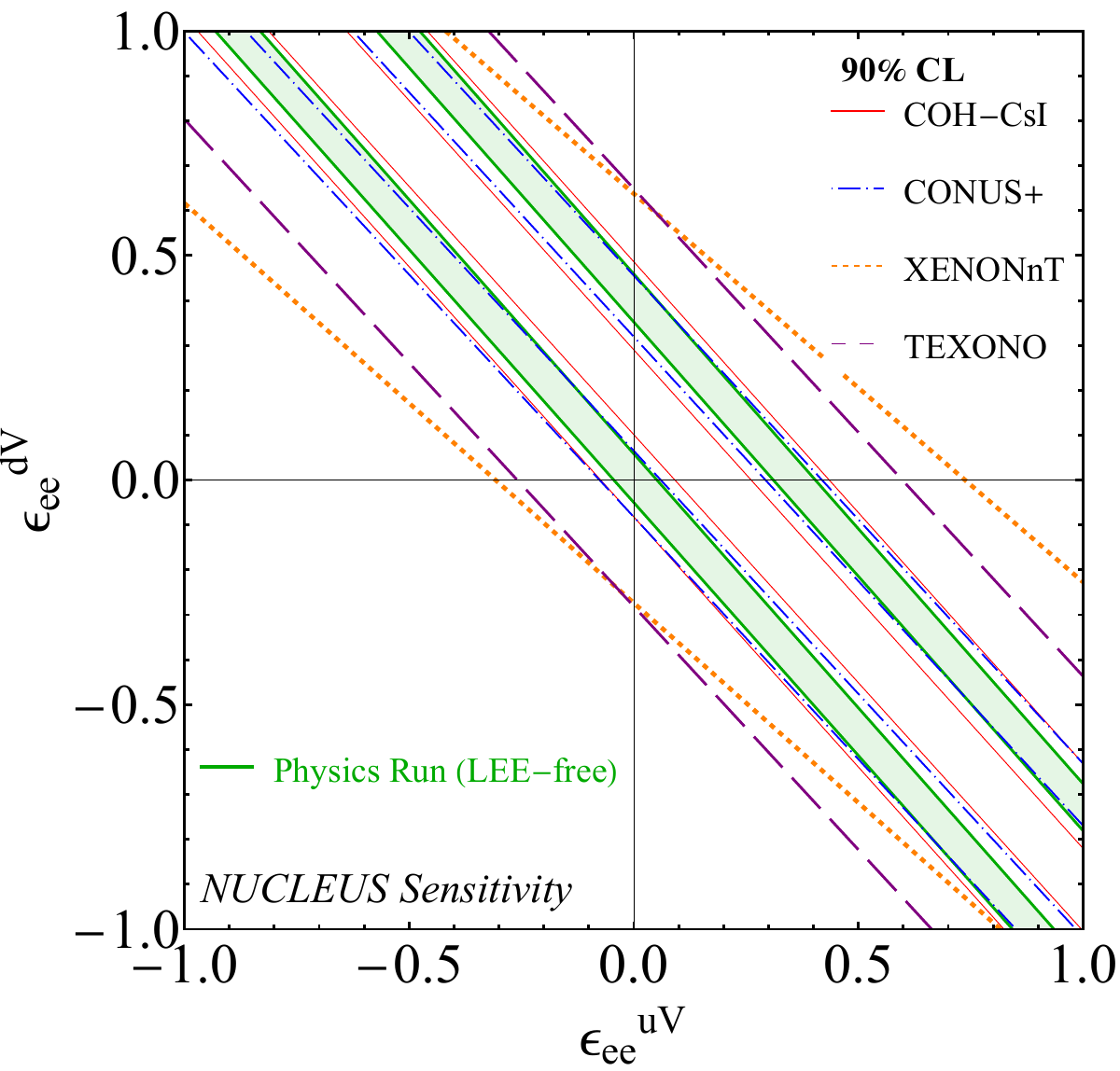}
    \caption{Projected sensitivity of the NUCLEUS experiment to a universal light mediator model (\textbf{left}) and to a flavor preserving NSI scenario (\textbf{right}). The green regions indicate the parameter space that can be probed by NUCLEUS at 90\% CL, with curves corresponding to the \textit{Technical Run} (only for the light mediator case) and \textit{Physics Run} (filled region), assuming complete suppression of the LEE for the latter. (\textbf{left}) The sensitivity to the light mediator model is compared to existing constraints at the 90\% CL from CONUS+~\cite{DeRomeri:2025csu,AtzoriCorona:2025ygn}, COHERENT CsI+Ar~\cite{DeRomeri:2022twg}, CONNIE~\cite{CONNIE:2019xid}, XENONnT electron recoil ($\nu$ES)~\cite{A:2022acy} and CE$\nu$NS data ~\cite{Blanco-Mas:2024ale}. (\textbf{right}) We compare the sensitivity to NSI with the results from COHERENT CsI~\cite{COHERENT:2021xmm}, TEXONO~\cite{TEXONO:2025sub}, CONUS+~\cite{DeRomeri:2025csu} and XENONnT~\cite{AristizabalSierra:2024nwf} at the 90\% CL.}
    \label{fig:VNSNuLimit}
\end{figure*}

Figure~\ref{fig:VNSNuLimit} shows the projected sensitivity of NUCLEUS to a universal light vector mediator, parameterized according to Eq.~\ref{eq:LM}, for both the \textit{Technical Run} and the \textit{Physics Run}, compared with existing constraints from other \cenns experiments. Despite the presence of the dominant LEE background, competitive sensitivities are already achieved during the \textit{Technical Run}, owing to the strong enhancement of light-mediator effects at low momentum transfer and the exploitation of reactor-power variation.
The \textit{Physics Run} significantly extends the reach of NUCLEUS, allowing previously unexplored regions of the parameter space to be probed, particularly for mediator masses between approximately 0.1 and 10\,MeV when compared to existing \cenns experiments. The sensitivity exhibits a characteristic degeneracy band, corresponding to regions where destructive interference occurs between the SM nuclear weak charge and the contribution from the new vector mediator~\cite{AtzoriCorona:2022moj}.
\par The right panel of Figure~\ref{fig:VNSNuLimit} shows the projected NUCLEUS sensitivity in the
$(\epsilon_{ee}^{uV},\,\epsilon_{ee}^{dV})$ plane achievable during the \textit{Physics Run}, compared with 90\% CL constraints from a selection of existing \cenns experiments.
The NUCLEUS sensitivity bands display a pronounced anti-correlation between the up- and down-quark vector NSI couplings, arising from the phenomenology of the model and from the different proton and neutron contributions of the target nuclei. 
Again, the precision foreseen is comparable, or even better, than that from leading \cenns experiments, highlighting the
capability of NUCLEUS to test subleading non-standard interaction effects
and to help resolve degeneracies present in existing measurements.
\par
\begin{figure}[th!]
    \centering
    \includegraphics[width=1.\linewidth]{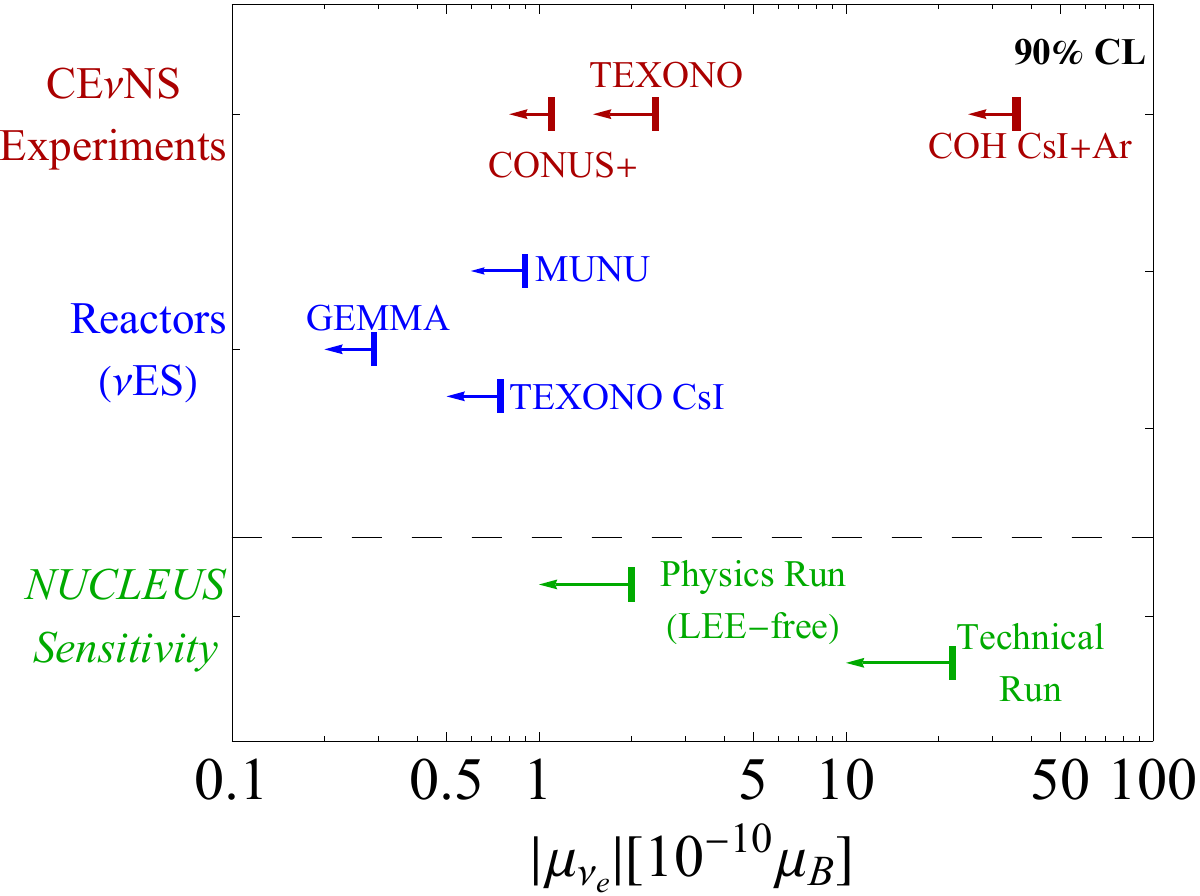}
    \caption{Projected sensitivity of the NUCLEUS experiment to the effective electronic neutrino magnetic moment $\mu_{\nu_e}$ at the 90\% CL. The green arrows indicate the expected precision for the \textit{Technical Run} and \textit{Physics Run}, assuming complete suppression of the LEE for the latter. The results of the sensitivity are compared to existing constraints on $\mu_{\nu_e}$ from CE$\nu$NS experiments (COHERENT CsI+Ar~\cite{DeRomeri:2022twg}, TEXONO~\cite{AtzoriCorona:2025ygn} and CONUS+~\cite{Chattaraj:2025fvx}) and data that exploit $\nu$ES at the $\sim$MeV range at reactor plants (MUNU~\cite{MUNU:2005xnz}, GEMMA~\cite{Beda:2012zz} and TEXONO CsI~\cite{TEXONO:2006xds}).}
    \label{fig:limits-mm}
\end{figure}

 In addition to this, already during the \textit{Technical Run}, NUCLEUS is expected to place a constraint on the effective electron-flavor neutrino magnetic moment that is more stringent than the current limit obtained from the combined COHERENT CsI+Ar dataset~\cite{AtzoriCorona:2022qrf,DeRomeri:2022twg}, namely $\mu_{\nu_e} < 22.3\times 10^{-10}\mu_B$ at 90\% CL as shown in Fig.~\ref{fig:limits-mm}. This sensitivity is driven by the characteristic $1/T_{\rm nr}$ enhancement of the magnetic-moment contribution at low recoil energies, which compensates for the small target mass of the experiment. 
For the \textit{Physics Run}, an additional improvement of approximately one order of magnitude is expected, yielding a projected sensitivity of
$\mu_{\nu_e} < 2.0\times 10^{-10}\mu_B$ at 90\% CL. This places NUCLEUS among the most sensitive \cenns-based probes of the effective electron neutrino magnetic moment, even though CONUS+ constraints remain stronger when $\nu$ES is included on top of the CE$\nu$NS signal. 
We note, however, that direct constraints on $\mu_{\nu_e}$ exploiting reactor neutrino–electron scattering data at $\sim$ MeV recoil energies currently provide constraints that are approximately one order of magnitude stronger~\cite{MUNU:2005xnz,Beda:2012zz,TEXONO:2006xds,CONUS:2022qbb}.

In addition, the effective neutrino magnetic moment can be tightly constrained using solar-neutrino data. However, due to the phenomenon of neutrino oscillations, the extracted constraints for solar neutrino magnetic moments correspond to different combinations of the magnetic contributions relative to the mass eigenstate~\cite{Ternes:2025lqh}. For this reason, the resulting limits from solar neutrinos are not directly comparable to those derived in this work and are therefore not shown in Fig.~\ref{fig:limits-mm}.
For a qualitative comparison, the strongest constraints on the effective solar magnetic moment ($\mu_{\nu_s}$) have been obtained by XENONnT, $\mu_{\nu_s} \lesssim 6.4\times10^{-12}\mu_B$~\cite{XENON:2023cxc}, LUX-ZEPLIN, $\mu_{\nu_s} \lesssim 8.3\times10^{-12}\mu_B$~\cite{LZ:2025zpw}, and by BOREXINO, $\mu_{\nu_s} \lesssim 2.8\times10^{-11}\mu_B$~\cite{Borexino:2017fbd}, all of which rely on the $\nu$ES channel.

\subsection{Impact of Increased Target Mass}

\begin{figure}[th!]
    \centering
    \includegraphics[width=0.98\linewidth]{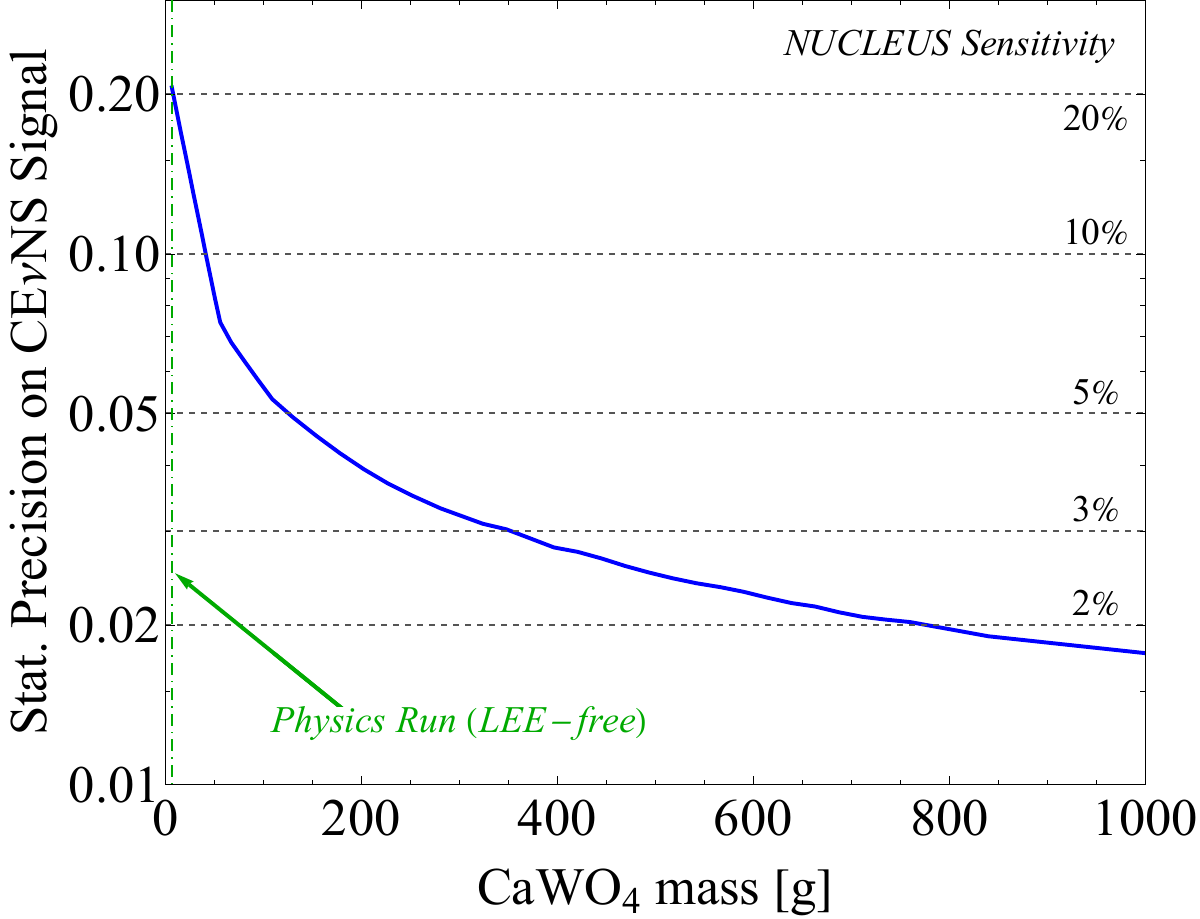}
    \caption{Median statistical precision on the SM CE$\nu$NS signal as a function of the total \cawo target mass, assuming the configuration of the \textit{Physics Run} with negligible LEE background, 20 eV threshold and 1 year of exposure. The blue curve shows the median expected 1 $\sigma$ statistical uncertainty obtained from ensembles of pseudo‑experiments. The green dashed line indicates the nominal 7 g target mass for the \textit{Physics Run} assuming complete suppression of the LEE. Horizontal dashed lines mark representative precision levels (20\%, 10\%, 5\%, 3\%, 2\%).}
    \label{fig:exposure}
\end{figure}

The results presented above highlight the distinctive strength of the NUCLEUS detector concept: thanks to its ultra-low nuclear recoil energy threshold, competitive sensitivity to \cenns and to several BSM scenarios can be achieved with a target mass at the gram scale. In particular, a \cenns observation with a statistical uncertainty of about 20\% is possible with only 7\,g of \cawo, a mass up to several orders of magnitude smaller than that employed by other \cenns experiments operating at higher recoil-energy thresholds. For comparison, experiments such as COHERENT and CONUS+ achieved comparable statistical precision using target masses at the kilogram scale or above, reflecting the strong dependence of \cenns sensitivity on the accessible recoil-energy threshold.
However, increasing the \cawo target mass offers substantial margin to improve the precision of the CE$\nu$NS cross‑section measurement. Such improvements would directly enhance the experiment’s sensitivity to the physics parameters considered in this work, making higher‑precision measurements particularly valuable. To quantify this potential, we examine how the statistical sensitivity scales with target mass under the assumptions of the \textit{Physics Run}, considering statistical uncertainties only. This projection is intended as a simplified estimate of the achievable statistical improvement and does not account for possible changes in detector design, background composition, or mitigation strategies that could accompany a scaling to significantly larger target masses.
The results are shown in Fig.~\ref{fig:exposure}, which reports the 1~$\sigma$ median statistical precision obtained through the sensitivity procedure described in Sec.~\ref{sec:reactormodulation}.
We find that improving the statistical precision of the SM \cenns measurement by a factor of two, to reach similar precision as current COHERENT measurements~\cite{COHERENT:2021xmm}, requires an increase in target mass by approximately a factor six (about 40 g of \cawo). 
Pushing the precision further will eventually bring the experiment into a regime where systematic uncertainties dominate over the statistical ones. In the configuration considered here, this transition occurs for a \cawo target mass of approximately 340~g (see Fig.~\ref{fig:exposure}), at which point the statistical precision reaches about 3\%, becoming comparable to reactor‑related systematic uncertainties (see App.~\ref{app:syst}).

\section{Conclusions}
\label{sec:conclusions}

In this work, we have presented a comprehensive sensitivity study for the NUCLEUS experiment at the Chooz VNS, focusing on its physics reach during the upcoming technical and physics runs. The study exploits two key features of the experiment: an unprecedentedly low nuclear recoil energy threshold enabled by cryogenic calorimetry, and the use of reactor-power variation to disentangle signal and background in a challenging low signal-to-background environment.

During the \textit{Technical Run}, the sensitivity is limited by the presence of the LEE, which dominates the recoil-energy spectrum at sub-keV energies. Nevertheless, by combining recoil-energy information with the time dependence induced by reactor-power variation, NUCLEUS is expected to reach competitive sensitivity to several BSM scenarios, including light vector mediators and the neutrino magnetic moment, even in the absence of a SM \cenns observation.

In view of the subsequent \textit{Physics Run}, the collaboration is currently pursuing a dedicated program to suppress the LEE background. This includes the integration of the detectors into an instrumented inner veto, as well as studies of its time evolution and dependence on detector cooling conditions. Assuming that these mitigation strategies will reduce the LEE contribution to negligible levels compared to the particle-induced background estimated from simulations, NUCLEUS is expected to achieve a clear observation of SM \cenns with a statistical uncertainty of about 20\% using only 7\,g of \cawo. This measurement would enable a competitive determination of the weak mixing angle at the lowest momentum transfer probed to date by \cenns experiments and to put leading constraints on the neutrino charge radius. At the same time, the ultra-low recoil-energy threshold provides strong sensitivity to new physics effects that modify the \cenns spectrum at low energies, allowing NUCLEUS to explore previously unexplored regions of light-mediator parameter space.
Overall, this study demonstrates that the NUCLEUS experiment occupies a unique position in the landscape of \cenns measurements. By accessing the lowest nuclear recoil energies achieved to date, NUCLEUS can compensate for its small target mass and provide competitive sensitivity to both SM and new-physics scenarios. These results highlight the strong complementarity between ultra-low-threshold cryogenic detectors and larger-mass experiments operating at higher recoil-energy thresholds, and establish NUCLEUS as a powerful probe of \cenns and low-energy neutrino physics at nuclear reactors.

\begin{acknowledgments}
This work has been financed by the CEA, the INFN, the ÖAW and partially supported by the TU Munich and the MPI für Physik. NUCLEUS members acknowledge additional funding by the DFG through the SFB1258 and the Excellence Cluster ORIGINS, by the European Commission through the ERC-StG2018-804228 ``NU-CLEUS'', by the P2IO LabEx (ANR-10-LABX-0038) in the framework ``Investissements d'Avenir'' (ANR-11-IDEX-0003-01) managed by the Agence Nationale de la Recherche (ANR), France, by the Austrian Science Fund (FWF) through the ``P 34778-N, ELOISE'', and by Max-Planck-Institut  für Kernphysik (MPIK), Germany. This research was funded in whole or in part by the Austrian Science Fund (FWF) I6955 DOI 10.55776/I6955.
\end{acknowledgments}

%\newpage

%\appendix
\appendix

\section{Validation of the Statistical Framework}\label{app:A}
In this Appendix, we validate the likelihood-based inference framework described in Sec.~\ref{sec:reactormodulation} and illustrate the construction of upper limits, confidence intervals, and discovery significances used throughout this work.

\subsection{Fit Parameter Reconstruction}

\begin{figure*}[tbp]
    \centering
    \includegraphics[width=0.9\linewidth]{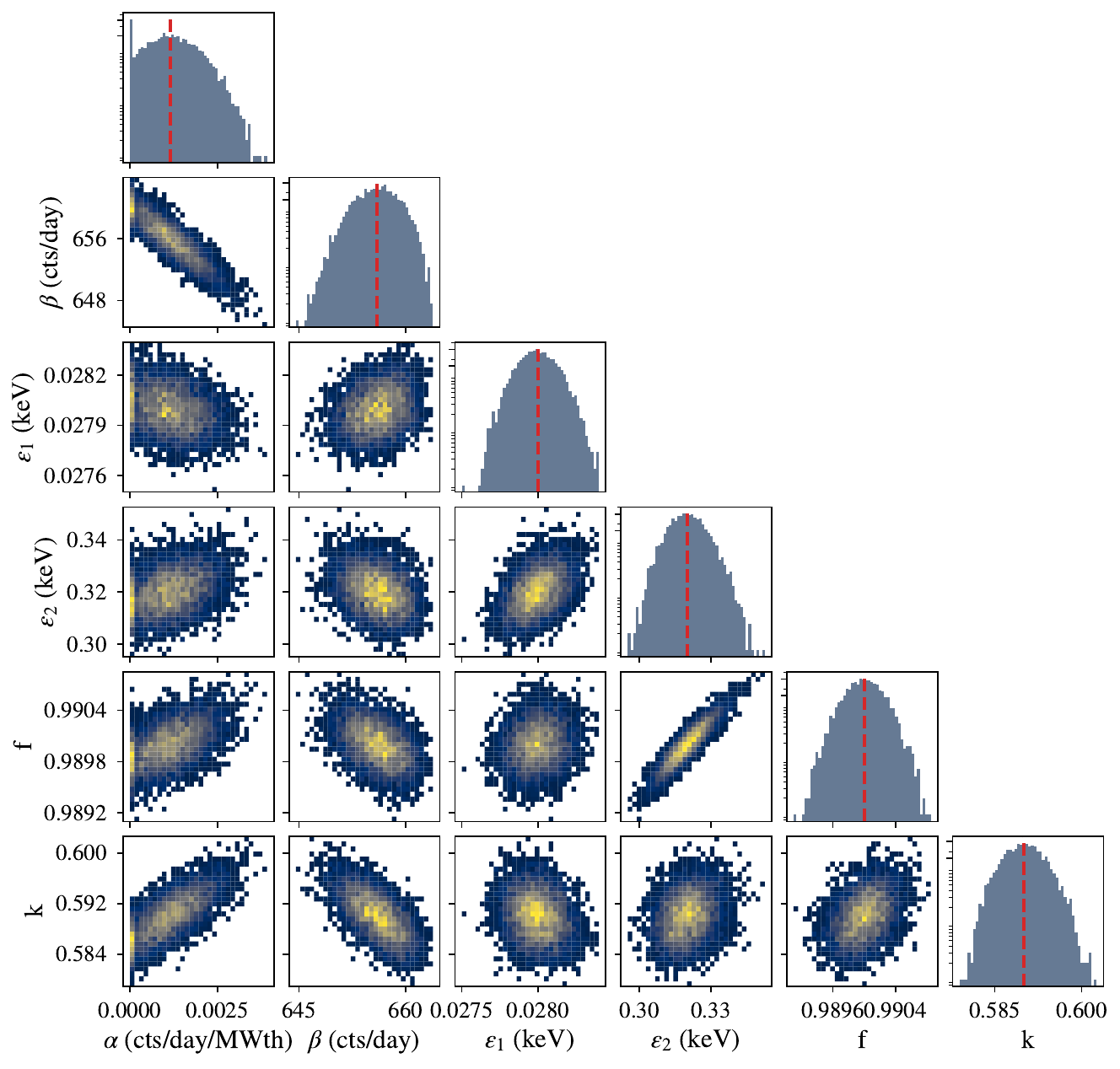}
    \caption{Distributions of reconstructed best-fit parameters and their pairwise correlations for the \textit{Technical Run} configuration, obtained from ensembles of $\sim 5000$ pseudo-experiments generated with fixed injected values. The red dashed lines indicate the injected (true) parameter values.}
    \label{fig:triangleplot}
\end{figure*}

To validate the fit procedure based on the likelihood defined in Eq.~\ref{eq:loglike_comb}, we use ensembles of pseudo-experiments generated from the nominal signal and background model. For each pseudo-experiment, the full likelihood is maximized and the reconstructed best-fit parameters and their uncertainties are recorded. This procedure allows us to assess unbiasedness, parameter correlations and the overall behavior of the likelihood in the relevant regions of parameter space. 

As a representative example, Fig.~\ref{fig:triangleplot} shows the distributions of the reconstructed best-fit parameters and their pairwise correlations for the \textit{Technical Run} configuration, assuming an injected signal strength $\alpha_{\rm true}=35\times \alpha_{\rm SM}$, where $\alpha_{\mathrm{SM}}$ denotes the normalization of the SM CE$\nu$NS signal, and the LEE background model defined in Sec.~\ref{sec:tec-run}. 
The distributions peak at the injected (true) values (red dashed lines), demonstrating unbiased parameter reconstruction. The one-dimensional projections are approximately symmetric and show no secondary modes or visible distortions, indicating that the likelihood is well behaved in the explored parameter region. 

The off-diagonal panels reveal the correlation structure among the fitted parameters. A pronounced anti-correlation is observed between the signal normalization $\alpha$ and the overall background normalization $\beta$, reflecting their partial spectral degeneracy in the low-energy region where both contribute. Strong correlations are also visible among the parameters governing the LEE model. In particular, the spectral shape parameters $\varepsilon_1$, $\varepsilon_2$, and $f$ exhibit correlations, consistent with their coupled role in defining the LEE spectral behavior. The parameter $k$, which controls the time decay of the LEE component, shows correlations with both the signal and background normalization. These arise from the interplay between spectral and time-dependent components of the likelihood, as variations in the time evolution of the background can partially mimic changes in normalization. 

The observed correlation pattern is well understood and fully accounted for in the analysis through likelihood profiling.

\subsection{Test Statistic Distribution}
Confidence intervals in this work are derived from the profile-likelihood ratio test statistic defined in Eq.~\ref{eq:profile-likelihood}. Under regularity conditions and for sufficient large event counts, Wilks' theorem predicts that the test statistic evaluated at the true parameter value follows a $\chi^2$ distribution with one degree of freedom~\cite{Wilks:1938dza,Cowan:2010js}. In this section we examine the behavior of the test statistic in both the \textit{Technical Run} and \textit{Physics Run} configurations using ensembles of pseudo-experiments.

\begin{figure*}[tbp]
    \centering
    \includegraphics[width=0.49\linewidth]{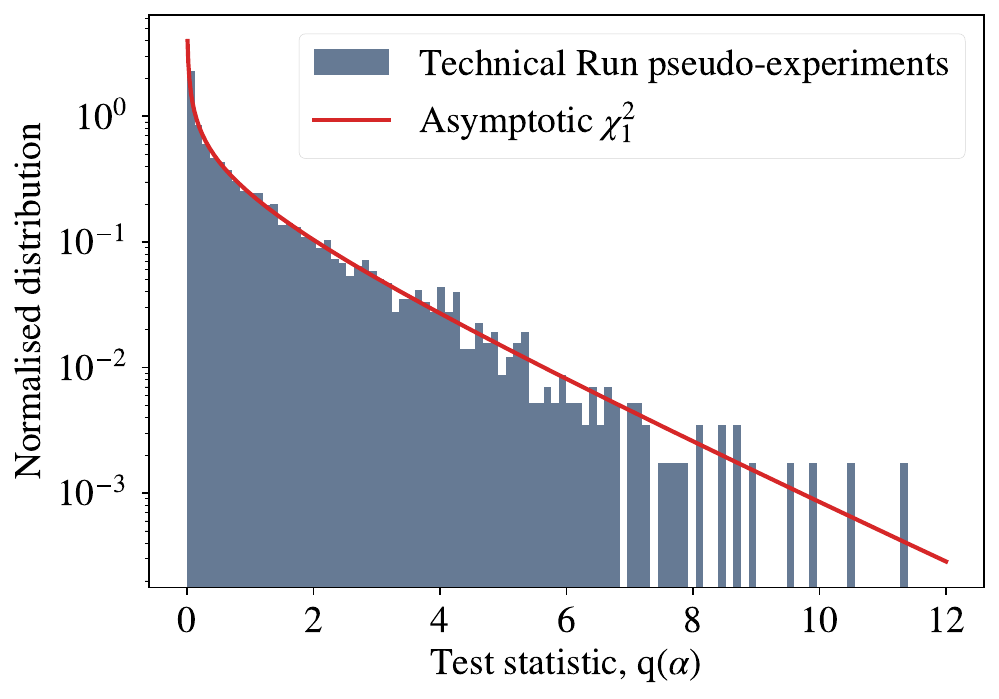}
\includegraphics[width=0.49\linewidth]{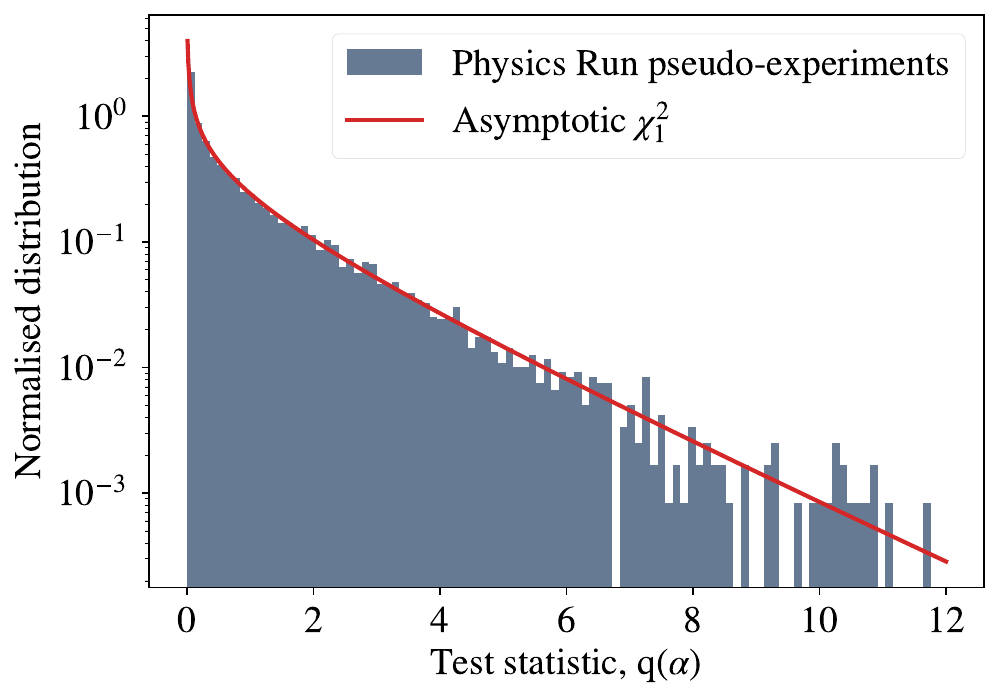}
    \caption{\textbf{Left:} distribution of the profile-likelihood ratio test statistic for the \textit{Technical Run} configuration, obtained from $\sim 5000$ pseudo-experiments generated with an injected signal strength $\alpha_{\rm true} = 35\times \alpha_{\rm SM}$. 
    \textbf{Right:} same distribution for the \textit{Physics Run} configuration, obtained from $\sim 10^4$ pseudo‑experiments generated with an injected signal strength $\alpha_{\rm true} = \alpha_{\rm SM}$.    
    The histogram shows the normalized empirical distribution, while the red curve represents the asymptotic $\chi^2$ distribution with one degree of freedom ($\chi^2_1$).}
    \label{fig:TS-dist-tec-run}
\end{figure*}

The distribution of the test statistic $q(\alpha_{\rm true})$, with $\alpha_{\rm true}=35\times \alpha_{\rm SM}$, for the \textit{Technical Run} configuration is shown in the left panel of Fig.~\ref{fig:TS-dist-tec-run}. The distribution is obtained from the same ensemble of $\sim 5000$ pseudo-experiments used for the parameter-reconstruction study described above, with the injected signal strength of $\alpha_{\rm true}=35\times \alpha_{\rm SM}$. For each pseudo-experiment, the profile-likelihood ratio is evaluated at the injected parameter value, and the resulting values of $q(\alpha_{\rm true})$ are collected.

The empirical 90\% quantile of the simulated distribution is found to be $2.67 \pm 0.04$, compared to the asymptotic expectation of 2.71 for a $\chi^2$ distribution with one degree of freedom. The observed agreement is satisfactory within statistical precision in the region relevant for limit construction and indicates that the use of asymptotic thresholds provides an adequate approximation for the \textit{Technical Run} sensitivity projections presented in this work.

\begin{figure}[tbp]
    \centering
    \includegraphics[width=0.95\linewidth]{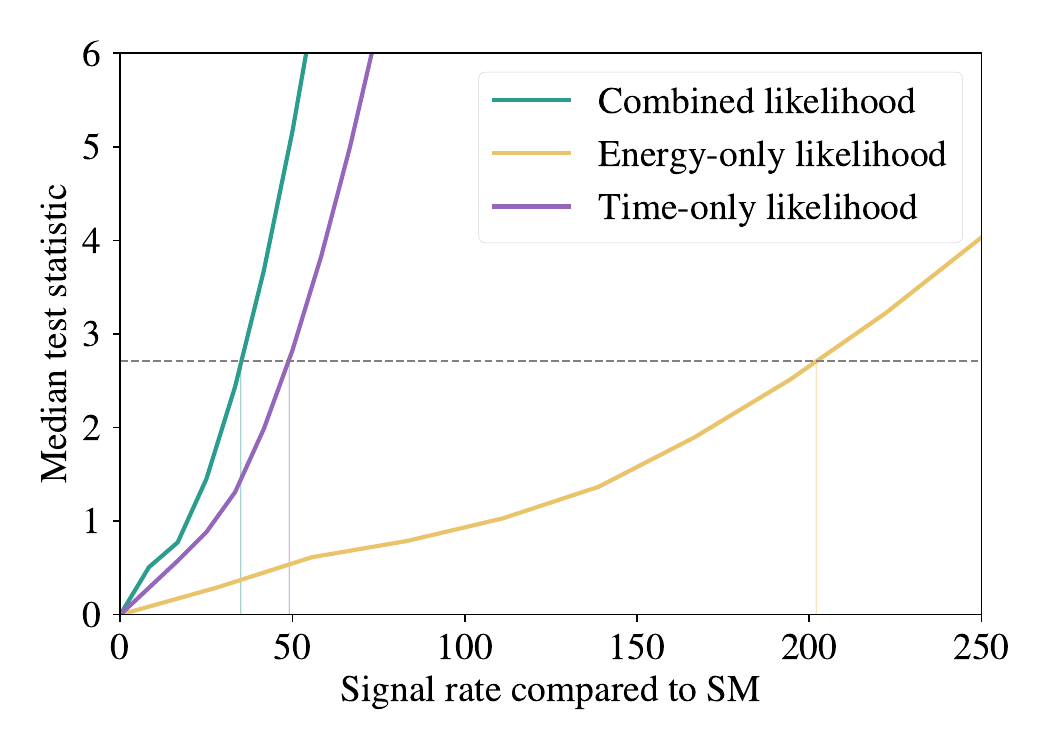}
    \caption{Median profile-likelihood ratio test statistic as a function of the injected \cenns signal strength (expressed relative to the SM prediction) for the \textit{Technical Run} configuration. For each signal hypothesis, the median value of $q(\alpha)$ is obtained from ensembles of 500 pseudo-experiments fitted with the energy-only, time-only, and combined likelihoods. 90\% CL upper limits are determined from the intersection of the curves with the asymptotic threshold $q=2.71$ (gray dashed line).}
    \label{fig:profileL-tec-run}
\end{figure}

Figure~\ref{fig:profileL-tec-run} illustrates the procedure used to extract upper limits on the \cenns signal strength in the \textit{Technical Run} configuration.
For each signal hypothesis, ensembles of 500 pseudo-experiments are generated and analyzed with the corresponding likelihood model. The profile likelihood ratio $q(\alpha)$ is constructed for every pseudo-experiment, and the median value is reported. This corresponds to the standard definition of expected sensitivity, i.e. behavior of the test statistic in 50\% of repeated experiments under the assumed signal hypothesis~\cite{Cowan:2010js}. 

Three analysis strategies are compared: an energy-only likelihood, a time-only likelihood, and the combined time-energy likelihood adopted in the main analysis. For all injected signal strengths, the combined likelihood yields the largest median test statistic, reflecting the complementarity of spectral and reactor-power information. The energy-only analysis exploits differences in recoil spectra between signal and background, while the time-only analysis relies on the distinct time dependence of signal and background, with only the former correlated with reactor power. Their combination reduces parameter degeneracies and produces a steeper increase of the test statistic with signal strength. 

The intersection with $q(\alpha)\simeq2.71$ defines the expected median 90\% CL upper limit. The more stringent upper limit obtained in the combined analysis directly quantifies the gain achieved by exploiting both observables in the \textit{Technical Run} scenario.

We now consider the \textit{Physics Run} configuration. Ensemble of $\sim 10^4$ pseudo-experiments are generated assuming an injected signal strength $\alpha_{\rm true}=\alpha_{\rm SM}$ and the background model defined in Sec.~\ref{sec:phys-run}. 
The distribution of the test statistic $q(\alpha_{\rm true})$ is shown in the right panel of Fig.~\ref{fig:TS-dist-tec-run}, together with the $\chi^2$ distribution with one degree of freedom. The empirical 68.27\% (1$\sigma$) quantile of the simulated distribution is found to be $1.009\pm 0.019$, in good agreement with the asymptotic expectation of unity. 
This level of agreement indicates that the asymptotic thresholds provide an adequate approximation for the sensitivity projections of the \textit{Physics Run} configuration presented in this work.

\begin{figure}[tbp]
    \centering
    \includegraphics[width=0.95\linewidth]{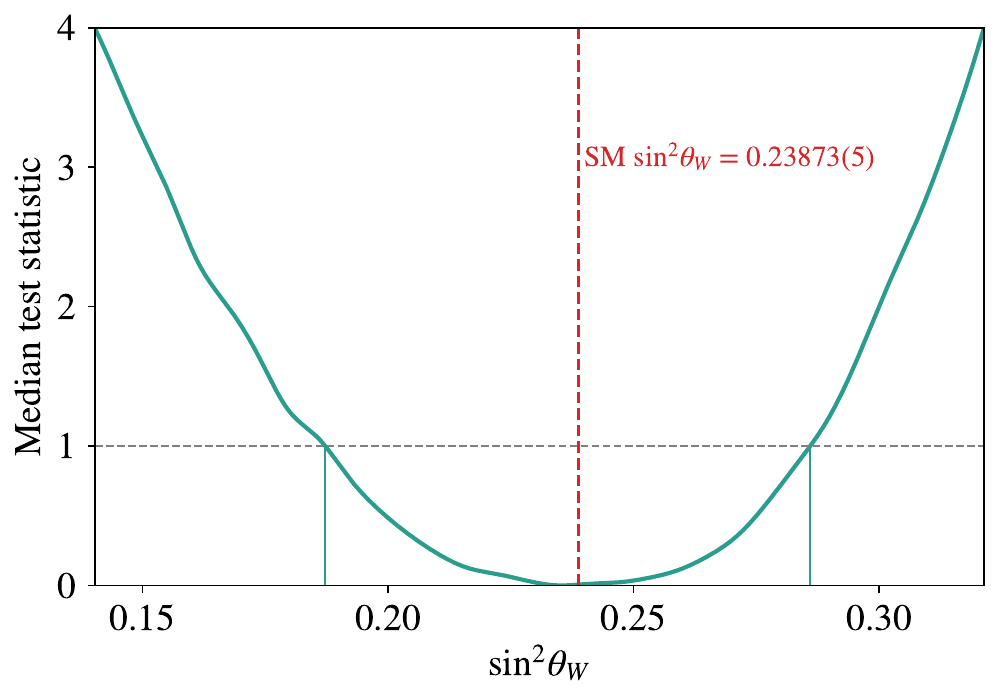}
    \caption{Median profile-likelihood ratio test statistic as a function of the weak mixing angle $\sin^2\theta_W$ for the \textit{Physics Run} configuration. The median is obtained from $\sim 10^4$ pseudo-experiments generated under the SM hypothesis. The expected 1$\sigma$ confidence interval is determined from the intersection with $q=1$ (gray dashed line).}
    \label{fig:profileL-phys-run}
\end{figure}

To illustrate how confidence intervals on a measured parameter are extracted in this work, Fig.~\ref{fig:profileL-phys-run} shows the median profile-likelihood ratio test statistic as a function of the weak mixing angle $\sin^2\theta_W$ for the \textit{Physics Run} configuration. The curve is obtained from an ensemble of $10^4$ pseudo-experiments generated under the SM hypothesis. For each pseudo-experiment, the profile likelihood ratio $q(\sin^2\theta_W)$ is evaluated on a fixed grid in $\sin^2\theta_W$, and the median value at each grid point is reported. 
The minimum of the median curve coincides with the injected SM value within statistical precision, and the likelihood is approximately quadratic in its vicinity. The intersections with $q(\sin^2\theta_W) = 1$ define the expected (median) 1$\sigma$ confidence interval. 

In addition to interval estimation, we evaluate the expected discovery significance for the SM \cenns signal in the \textit{Physics Run} configuration. For this purpose, we determine the distribution of the test statistic $q_0=q(\alpha=0)$ using ensembles of pseudo-experiments generated under the SM hypothesis. The significance is then defined as $Z=\sqrt{q_0}$, corresponding to the asymptotic relation for a one-sided test in the regime where the profile-likelihood ratio follows the expected $\chi^2$ behavior~\cite{Cowan:2010js}.

\begin{figure}[tpb]
    \centering
    \includegraphics[width=0.95\linewidth]{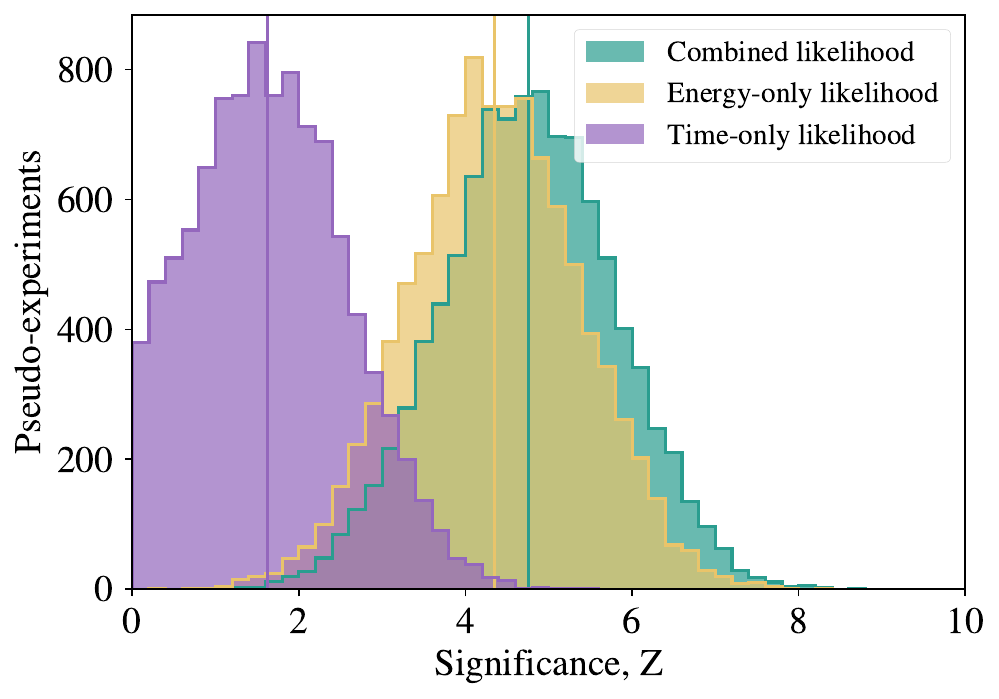}
    \caption{Distribution of the discovery significance $Z=\sqrt{q_0}$ for the SM \cenns signal in the \textit{Physics Run} configuration. Results are obtained from ensembles of $\sim 10^4$ pseudo-experiments generated under the SM hypothesis and analyzed with the energy-only, time-only, and combined likelihoods. The median of each distribution defines the expected discovery significance.}
    \label{fig:sign-phys-run}
\end{figure}

Figure~\ref{fig:sign-phys-run} shows the resulting distribution of the significance $Z$, obtained from ensembles of $\sim 10^4$ pseudo-experiments generated under the SM hypothesis. The median of this distribution defines the expected discovery significance, corresponding to the sensitivity achieved in 50\% of repeated experiments under the SM hypothesis.
Again, the three analysis strategies are compared: an energy-only likelihood, a time-only likelihood, and the combined time-energy likelihood adopted in the main analysis.. In contrast to the \textit{Technical Run} configuration, the \textit{Physics Run} sensitivity is primarily driven by spectral information, reflecting the stronger separation between signal and background once the overall background level is reduced. The time-only analysis provides a comparatively weaker constraint; however, its inclusion in the combined likelihood increases the median significance by further constraining background components not correlated with reactor power.

\section{Systematic Uncertainties}\label{app:syst}

In this Appendix, we evaluate the impact of the dominant sources of systematic uncertainty on the projected sensitivities, focusing in particular on the nuclear recoil energy scale and the reactor antineutrino flux.
We first consider the uncertainty associated with the absolute energy calibration at low recoil energies, which is a known challenge for experiments operating at ultra-low thresholds. The NUCLEUS Collaboration is addressing this issue through dedicated calibration campaigns, including measurements with controlled nuclear recoils using the CRAB facility~\cite{CRAB:2022rcm,Abele:2025vrj} and low-energy X-ray calibration sources~\cite{Abele:2025rrl}. These approaches are pursued in addition to the LED calibration based on Poisson statistics~\cite{NUCLEUS:2025ymr}, with the goal of cross-checking the energy scale and constraining the associated systematic uncertainties.
In particular, CRAB provides a direct calibration of nuclear recoils in \cawo, thereby avoiding model-dependent assumptions and offering a robust handle on the nuclear-recoil energy scale. 
Current CRAB measurements are performed offsite~\cite{Abele:2025vrj}, and studies are ongoing to assess the feasibility of an onsite calibration at the VNS, which would allow a direct in-situ determination of the nuclear recoil response and is expected to significantly reduce the associated systematic uncertainty. During the commissioning run, a discrepancy at the level of 25\% was observed between the LED calibration based on Poisson statistics and the position of the copper X-ray line at 8.04\,keV~\cite{NUCLEUS:2025ymr}. Although this uncertainty is expected to be reduced in future runs, we conservatively assume a 25\% uncertainty on the absolute energy scale in the present study.

Unlike ionization- or scintillation-based detectors, cryogenic calorimeters measure almost the entire nuclear recoil energy as heat. As a result, they do not rely on quenching-factor models, which constitute a major source of systematic uncertainty in other \cenns experiments -- particularly at the very low recoil energies where the \cenns signal is concentrated and where quenching effects are poorly constrained experimentally~\cite{Li:2025pfw}. This feature is further reinforced by calibration approaches such as CRAB, which directly probe nuclear recoils and are intrinsically insensitive to quenching effects. Consequently, the dominant detector-related systematic uncertainty for NUCLEUS arises from the absolute calibration of the energy scale rather than from energy-dependent quenching effects.

The impact of the energy-scale uncertainty is modeled by introducing a linear relation between the observed and true recoil energies
\begin{equation}
T_{\rm nr,obs} = T_{\rm nr,true}(1+\delta)\, .
\end{equation}
For the evaluation of this systematic effect, the transformation is applied to the \cenns signal prediction in Eq.~(\ref{eq:RateExp}) and to the background model only when generating pseudo-experiments. The fits to these pseudo-data sets are performed with the nominal (unshifted) energy model. Since the differential rate and the analysis threshold depend explicitly on the recoil energy, the shifted model differs from the nominal prediction both in spectral shape and in the total number of events above threshold. 
Pseudo-experiments are generated under two scenarios, corresponding to energy-scale shifts of $\delta= + 25\%$ and $\delta= - 25\%$, and then fitted using the nominal energy model ($\delta= 0$). The reconstructed signal normalization is compared to that obtained from pseudo-experiments both generated and fitted with the nominal model, yielding an asymmetric systematic bias of $_{-4.0}^{+0.5}$\% on the reconstructed \cenns signal, with a similar variation observed in the median expected discovery significance.
This bias therefore quantifies the impact of the energy-scale uncertainty on the extracted SM normalization, including the change in the predicted event rate induced by the shift.

A similar procedure is used to estimate the systematic uncertainty associated with the reactor antineutrino flux. Propagating the $1\,\sigma$ flux uncertainty reported in Ref.~\cite{Perisse:2023efm} to the \cenns signal model results in a relative uncertainty of $\pm2$\% on the extracted signal normalization. Both systematic effects are therefore subdominant compared to the expected $\sim20$\% statistical uncertainty of the \cenns measurement in the \textit{Physics Run} and have a negligible impact on both the SM observables presented in Sec.~\ref{sec:SensSM} and on the projected sensitivities to the BSM scenarios discussed in Sec.~\ref{sec:BSMResults}.

In addition, the uncertainty on the reactor thermal power determination has been considered. The Double Chooz collaboration reports an uncertainty of 0.5\% on the instantaneous thermal power at full reactor power when using operator-provided EDF data~\cite{DoubleChooz:2014kuw}. In the present study, we rely on publicly available reactor power information, for which the estimated uncertainty is approximately 5\%. This larger value reflects the use of public data in the current sensitivity study rather than an intrinsic limitation: access to operator-provided reactor information would allow a determination at the $\sim0.5\%$ level. However, this refinement is not required for the projected sensitivity of the \textit{Physics Run}, as the overall signal normalization uncertainty remains subleading compared to the expected $\sim20\%$ statistical uncertainty. For future higher-exposure phases and mass scaling, the use of operator-provided reactor information and a detailed treatment of the fission inventory evolution will become necessary and will be addressed at that stage.

Finally, we do not assign an additional theoretical uncertainty to the \cenns cross-section prediction. As discussed in Sec.~\ref{sec:theory}, nuclear form-factor effects are negligible in the recoil-energy range relevant for NUCLEUS, and the uncertainty on the SM value of the weak mixing angle induces only a sub-percent variation in the normalization. These effects are therefore negligible compared to the statistical and experimental systematic uncertainties considered above.

\bibliography{ref}

\begin{thebibliography}{105}%
\makeatletter
\providecommand \@ifxundefined [1]{%
 \@ifx{#1\undefined}
}%
\providecommand \@ifnum [1]{%
 \ifnum #1\expandafter \@firstoftwo
 \else \expandafter \@secondoftwo
 \fi
}%
\providecommand \@ifx [1]{%
 \ifx #1\expandafter \@firstoftwo
 \else \expandafter \@secondoftwo
 \fi
}%
\providecommand \natexlab [1]{#1}%
\providecommand \enquote  [1]{``#1''}%
\providecommand \bibnamefont  [1]{#1}%
\providecommand \bibfnamefont [1]{#1}%
\providecommand \citenamefont [1]{#1}%
\providecommand \href@noop [0]{\@secondoftwo}%
\providecommand \href [0]{\begingroup \@sanitize@url \@href}%
\providecommand \@href[1]{\@@startlink{#1}\@@href}%
\providecommand \@@href[1]{\endgroup#1\@@endlink}%
\providecommand \@sanitize@url [0]{\catcode `\\12\catcode `\$12\catcode
  `\&12\catcode `\#12\catcode `\^12\catcode `\_12\catcode `\%12\relax}%
\providecommand \@@startlink[1]{}%
\providecommand \@@endlink[0]{}%
\providecommand \url  [0]{\begingroup\@sanitize@url \@url }%
\providecommand \@url [1]{\endgroup\@href {#1}{\urlprefix }}%
\providecommand \urlprefix  [0]{URL }%
\providecommand \Eprint [0]{\href }%
\providecommand \doibase [0]{https://doi.org/}%
\providecommand \selectlanguage [0]{\@gobble}%
\providecommand \bibinfo  [0]{\@secondoftwo}%
\providecommand \bibfield  [0]{\@secondoftwo}%
\providecommand \translation [1]{[#1]}%
\providecommand \BibitemOpen [0]{}%
\providecommand \bibitemStop [0]{}%
\providecommand \bibitemNoStop [0]{.\EOS\space}%
\providecommand \EOS [0]{\spacefactor3000\relax}%
\providecommand \BibitemShut  [1]{\csname bibitem#1\endcsname}%
\let\auto@bib@innerbib\@empty
%</preamble>
\bibitem [{\citenamefont {Freedman}(1974)}]{PhysRevD.9.1389}%
  \BibitemOpen
  \bibfield  {author} {\bibinfo {author} {\bibfnamefont {D.~Z.}\ \bibnamefont
  {Freedman}},\ }\bibfield  {title} {\bibinfo {title} {Coherent effects of a
  weak neutral current},\ }\href {https://doi.org/10.1103/PhysRevD.9.1389}
  {\bibfield  {journal} {\bibinfo  {journal} {Phys. Rev. D}\ }\textbf {\bibinfo
  {volume} {9}},\ \bibinfo {pages} {1389} (\bibinfo {year} {1974})}\BibitemShut
  {NoStop}%
\bibitem [{\citenamefont {Kopeliovich}\ and\ \citenamefont
  {Frankfurt}(1974)}]{Kopeliovich:1974mv}%
  \BibitemOpen
  \bibfield  {author} {\bibinfo {author} {\bibfnamefont {V.~B.}\ \bibnamefont
  {Kopeliovich}}\ and\ \bibinfo {author} {\bibfnamefont {L.~L.}\ \bibnamefont
  {Frankfurt}},\ }\bibfield  {title} {\bibinfo {title} {{Isotopic and chiral
  structure of neutral current}},\ }\href@noop {} {\bibfield  {journal}
  {\bibinfo  {journal} {JETP Lett.}\ }\textbf {\bibinfo {volume} {19}},\
  \bibinfo {pages} {145} (\bibinfo {year} {1974})}\BibitemShut {NoStop}%
\bibitem [{\citenamefont {Drukier}\ and\ \citenamefont
  {Stodolsky}(1984)}]{Drukier:1984vhf}%
  \BibitemOpen
  \bibfield  {author} {\bibinfo {author} {\bibfnamefont {A.}~\bibnamefont
  {Drukier}}\ and\ \bibinfo {author} {\bibfnamefont {L.}~\bibnamefont
  {Stodolsky}},\ }\bibfield  {title} {\bibinfo {title} {{Principles and
  Applications of a Neutral Current Detector for Neutrino Physics and
  Astronomy}},\ }\href {https://doi.org/10.1103/PhysRevD.30.2295} {\bibfield
  {journal} {\bibinfo  {journal} {Phys. Rev. D}\ }\textbf {\bibinfo {volume}
  {30}},\ \bibinfo {pages} {2295} (\bibinfo {year} {1984})}\BibitemShut
  {NoStop}%
\bibitem [{\citenamefont {Akimov}\ \emph {et~al.}(2017)\citenamefont {Akimov}
  \emph {et~al.}}]{COHERENT:2017ipa}%
  \BibitemOpen
  \bibfield  {author} {\bibinfo {author} {\bibfnamefont {D.}~\bibnamefont
  {Akimov}} \emph {et~al.} (\bibinfo {collaboration} {COHERENT}),\ }\bibfield
  {title} {\bibinfo {title} {{Observation of Coherent Elastic Neutrino-Nucleus
  Scattering}},\ }\href {https://doi.org/10.1126/science.aao0990} {\bibfield
  {journal} {\bibinfo  {journal} {Science}\ }\textbf {\bibinfo {volume}
  {357}},\ \bibinfo {pages} {1123} (\bibinfo {year} {2017})},\ \Eprint
  {https://arxiv.org/abs/1708.01294} {arXiv:1708.01294 [nucl-ex]} \BibitemShut
  {NoStop}%
\bibitem [{\citenamefont {Akimov}\ \emph {et~al.}(2020)\citenamefont {Akimov}
  \emph {et~al.}}]{COHERENT:2020ybo}%
  \BibitemOpen
  \bibfield  {author} {\bibinfo {author} {\bibfnamefont {D.}~\bibnamefont
  {Akimov}} \emph {et~al.} (\bibinfo {collaboration} {COHERENT}),\ }\href
  {https://doi.org/10.5281/zenodo.3903810} {\bibinfo {title} {{COHERENT
  Collaboration data release from the first detection of coherent elastic
  neutrino-nucleus scattering on argon}}} (\bibinfo {year} {2020}),\ \Eprint
  {https://arxiv.org/abs/2006.12659} {arXiv:2006.12659 [nucl-ex]} \BibitemShut
  {NoStop}%
\bibitem [{\citenamefont {Akimov}\ \emph {et~al.}(2021)\citenamefont {Akimov}
  \emph {et~al.}}]{COHERENT:2020iec}%
  \BibitemOpen
  \bibfield  {author} {\bibinfo {author} {\bibfnamefont {D.}~\bibnamefont
  {Akimov}} \emph {et~al.} (\bibinfo {collaboration} {COHERENT}),\ }\bibfield
  {title} {\bibinfo {title} {{First Measurement of Coherent Elastic
  Neutrino-Nucleus Scattering on Argon}},\ }\href
  {https://doi.org/10.1103/PhysRevLett.126.012002} {\bibfield  {journal}
  {\bibinfo  {journal} {Phys. Rev. Lett.}\ }\textbf {\bibinfo {volume} {126}},\
  \bibinfo {pages} {012002} (\bibinfo {year} {2021})},\ \Eprint
  {https://arxiv.org/abs/2003.10630} {arXiv:2003.10630 [nucl-ex]} \BibitemShut
  {NoStop}%
\bibitem [{\citenamefont {Adamski}\ \emph {et~al.}(2025)\citenamefont {Adamski}
  \emph {et~al.}}]{COHERENT:2025vuz}%
  \BibitemOpen
  \bibfield  {author} {\bibinfo {author} {\bibfnamefont {S.}~\bibnamefont
  {Adamski}} \emph {et~al.} (\bibinfo {collaboration} {COHERENT}),\ }\bibfield
  {title} {\bibinfo {title} {{Evidence of Coherent Elastic Neutrino-Nucleus
  Scattering with COHERENT{\textquoteright}s Germanium Array}},\ }\href
  {https://doi.org/10.1103/PhysRevLett.134.231801} {\bibfield  {journal}
  {\bibinfo  {journal} {Phys. Rev. Lett.}\ }\textbf {\bibinfo {volume} {134}},\
  \bibinfo {pages} {231801} (\bibinfo {year} {2025})}\BibitemShut {NoStop}%
\bibitem [{\citenamefont {Adhikari}\ \emph {et~al.}(2026)\citenamefont
  {Adhikari} \emph {et~al.}}]{Adhikari:2026uvz}%
  \BibitemOpen
  \bibfield  {author} {\bibinfo {author} {\bibfnamefont {M.}~\bibnamefont
  {Adhikari}} \emph {et~al.},\ }\href@noop {} {\bibinfo {title} {{Measurement
  of coherent elastic neutrino nucleus scattering on germanium by COHERENT}}}
  (\bibinfo {year} {2026}),\ \Eprint {https://arxiv.org/abs/2603.17951}
  {arXiv:2603.17951 [hep-ex]} \BibitemShut {NoStop}%
\bibitem [{\citenamefont {Bo}\ \emph {et~al.}(2024)\citenamefont {Bo} \emph
  {et~al.}}]{PandaX:2024muv}%
  \BibitemOpen
  \bibfield  {author} {\bibinfo {author} {\bibfnamefont {Z.}~\bibnamefont {Bo}}
  \emph {et~al.} (\bibinfo {collaboration} {PandaX}),\ }\bibfield  {title}
  {\bibinfo {title} {{First Indication of Solar B8 Neutrinos through Coherent
  Elastic Neutrino-Nucleus Scattering in PandaX-4T}},\ }\href
  {https://doi.org/10.1103/PhysRevLett.133.191001} {\bibfield  {journal}
  {\bibinfo  {journal} {Phys. Rev. Lett.}\ }\textbf {\bibinfo {volume} {133}},\
  \bibinfo {pages} {191001} (\bibinfo {year} {2024})},\ \Eprint
  {https://arxiv.org/abs/2407.10892} {arXiv:2407.10892 [hep-ex]} \BibitemShut
  {NoStop}%
\bibitem [{\citenamefont {Aprile}\ \emph {et~al.}(2024)\citenamefont {Aprile}
  \emph {et~al.}}]{XENON:2024ijk}%
  \BibitemOpen
  \bibfield  {author} {\bibinfo {author} {\bibfnamefont {E.}~\bibnamefont
  {Aprile}} \emph {et~al.} (\bibinfo {collaboration} {XENON}),\ }\bibfield
  {title} {\bibinfo {title} {{First Indication of Solar B8 Neutrinos via
  Coherent Elastic Neutrino-Nucleus Scattering with XENONnT}},\ }\href
  {https://doi.org/10.1103/PhysRevLett.133.191002} {\bibfield  {journal}
  {\bibinfo  {journal} {Phys. Rev. Lett.}\ }\textbf {\bibinfo {volume} {133}},\
  \bibinfo {pages} {191002} (\bibinfo {year} {2024})},\ \Eprint
  {https://arxiv.org/abs/2408.02877} {arXiv:2408.02877 [nucl-ex]} \BibitemShut
  {NoStop}%
\bibitem [{\citenamefont {Akerib}\ \emph
  {et~al.}(2025{\natexlab{a}})\citenamefont {Akerib} \emph
  {et~al.}}]{LZ:2025igz}%
  \BibitemOpen
  \bibfield  {author} {\bibinfo {author} {\bibfnamefont {D.~S.}\ \bibnamefont
  {Akerib}} \emph {et~al.} (\bibinfo {collaboration} {LZ}),\ }\href
  {https://doi.org/10.17182/hepdata.167350.v1} {\bibinfo {title} {{Searches for
  Light Dark Matter and Evidence of Coherent Elastic Neutrino-Nucleus
  Scattering of Solar Neutrinos with the LUX-ZEPLIN (LZ) Experiment}}}
  (\bibinfo {year} {2025}{\natexlab{a}}),\ \Eprint
  {https://arxiv.org/abs/2512.08065} {arXiv:2512.08065 [hep-ex]} \BibitemShut
  {NoStop}%
\bibitem [{\citenamefont {Ackermann}\ \emph {et~al.}(2025)\citenamefont
  {Ackermann} \emph {et~al.}}]{Ackermann:2025obx}%
  \BibitemOpen
  \bibfield  {author} {\bibinfo {author} {\bibfnamefont {N.}~\bibnamefont
  {Ackermann}} \emph {et~al.},\ }\bibfield  {title} {\bibinfo {title} {{Direct
  observation of coherent elastic antineutrino{\textendash}nucleus
  scattering}},\ }\href {https://doi.org/10.1038/s41586-025-09322-2} {\bibfield
   {journal} {\bibinfo  {journal} {Nature}\ }\textbf {\bibinfo {volume}
  {643}},\ \bibinfo {pages} {1229} (\bibinfo {year} {2025})},\ \Eprint
  {https://arxiv.org/abs/2501.05206} {arXiv:2501.05206 [hep-ex]} \BibitemShut
  {NoStop}%
\bibitem [{\citenamefont {Belov}\ \emph {et~al.}(2025)\citenamefont {Belov}
  \emph {et~al.}}]{nGeN:2025hsd}%
  \BibitemOpen
  \bibfield  {author} {\bibinfo {author} {\bibfnamefont {V.}~\bibnamefont
  {Belov}} \emph {et~al.} (\bibinfo {collaboration}
  {({\ensuremath{\nu}}GeN),}),\ }\bibfield  {title} {\bibinfo {title} {{New
  constraints on coherent elastic neutrino{\textendash}nucleus scattering by
  the {\ensuremath{\nu}}GeN experiment*}},\ }\href
  {https://doi.org/10.1088/1674-1137/adb9c8} {\bibfield  {journal} {\bibinfo
  {journal} {Chin. Phys. C}\ }\textbf {\bibinfo {volume} {49}},\ \bibinfo
  {pages} {053004} (\bibinfo {year} {2025})},\ \Eprint
  {https://arxiv.org/abs/2502.18502} {arXiv:2502.18502 [hep-ex]} \BibitemShut
  {NoStop}%
\bibitem [{\citenamefont {Karmakar}\ \emph {et~al.}(2025)\citenamefont
  {Karmakar} \emph {et~al.}}]{TEXONO:2024vfk}%
  \BibitemOpen
  \bibfield  {author} {\bibinfo {author} {\bibfnamefont {S.}~\bibnamefont
  {Karmakar}} \emph {et~al.} (\bibinfo {collaboration} {TEXONO}),\ }\bibfield
  {title} {\bibinfo {title} {{New Limits on the Coherent Neutrino-Nucleus
  Elastic Scattering Cross Section at the Kuo-Sheng Reactor-Neutrino
  Laboratory}},\ }\href {https://doi.org/10.1103/PhysRevLett.134.121802}
  {\bibfield  {journal} {\bibinfo  {journal} {Phys. Rev. Lett.}\ }\textbf
  {\bibinfo {volume} {134}},\ \bibinfo {pages} {121802} (\bibinfo {year}
  {2025})},\ \Eprint {https://arxiv.org/abs/2411.18812} {arXiv:2411.18812
  [nucl-ex]} \BibitemShut {NoStop}%
\bibitem [{\citenamefont {Aguilar-Arevalo}\ \emph {et~al.}(2022)\citenamefont
  {Aguilar-Arevalo} \emph {et~al.}}]{CONNIE:2021ggh}%
  \BibitemOpen
  \bibfield  {author} {\bibinfo {author} {\bibfnamefont {A.}~\bibnamefont
  {Aguilar-Arevalo}} \emph {et~al.} (\bibinfo {collaboration} {CONNIE}),\
  }\bibfield  {title} {\bibinfo {title} {{Search for coherent elastic
  neutrino-nucleus scattering at a nuclear reactor with CONNIE 2019 data}},\
  }\href {https://doi.org/10.1007/JHEP05(2022)017} {\bibfield  {journal}
  {\bibinfo  {journal} {JHEP}\ }\textbf {\bibinfo {volume} {05}},\ \bibinfo
  {pages} {017}},\ \Eprint {https://arxiv.org/abs/2110.13033} {arXiv:2110.13033
  [hep-ex]} \BibitemShut {NoStop}%
\bibitem [{\citenamefont {Armatol}\ \emph {et~al.}(2025)\citenamefont {Armatol}
  \emph {et~al.}}]{Ricochet:2025bpk}%
  \BibitemOpen
  \bibfield  {author} {\bibinfo {author} {\bibfnamefont {A.}~\bibnamefont
  {Armatol}} \emph {et~al.} (\bibinfo {collaboration} {Ricochet}),\ }\bibfield
  {title} {\bibinfo {title} {{Characterization of mini-CryoCube detectors from
  the RICOCHET experiment commissioning at the Institut Laue-Langevin}},\
  }\href {https://doi.org/10.1103/7xy6-jq3c} {\bibfield  {journal} {\bibinfo
  {journal} {Phys. Rev. D}\ }\textbf {\bibinfo {volume} {112}},\ \bibinfo
  {pages} {112019} (\bibinfo {year} {2025})},\ \Eprint
  {https://arxiv.org/abs/2507.22751} {arXiv:2507.22751 [hep-ex]} \BibitemShut
  {NoStop}%
\bibitem [{\citenamefont {Akimov}\ \emph {et~al.}(2025)\citenamefont {Akimov}
  \emph {et~al.}}]{RED-100:2024izi}%
  \BibitemOpen
  \bibfield  {author} {\bibinfo {author} {\bibfnamefont {D.~Y.}\ \bibnamefont
  {Akimov}} \emph {et~al.} (\bibinfo {collaboration} {RED-100}),\ }\bibfield
  {title} {\bibinfo {title} {{First constraints on the coherent elastic
  scattering of reactor antineutrinos off xenon nuclei}},\ }\href
  {https://doi.org/10.1103/PhysRevD.111.072012} {\bibfield  {journal} {\bibinfo
   {journal} {Phys. Rev. D}\ }\textbf {\bibinfo {volume} {111}},\ \bibinfo
  {pages} {072012} (\bibinfo {year} {2025})},\ \Eprint
  {https://arxiv.org/abs/2411.18641} {arXiv:2411.18641 [hep-ex]} \BibitemShut
  {NoStop}%
\bibitem [{\citenamefont {Mondal}\ \emph {et~al.}(2025)\citenamefont {Mondal}
  \emph {et~al.}}]{Mondal:2025odw}%
  \BibitemOpen
  \bibfield  {author} {\bibinfo {author} {\bibfnamefont {D.}~\bibnamefont
  {Mondal}} \emph {et~al.},\ }\href@noop {} {\bibinfo {title} {{CE$\nu$NS
  Search with Cryogenic Sapphire Detectors at MINER: Results from the TRIGA
  reactor data and Future Sensitivity at HFIR}}} (\bibinfo {year} {2025}),\
  \Eprint {https://arxiv.org/abs/2510.15999} {arXiv:2510.15999 [nucl-ex]}
  \BibitemShut {NoStop}%
\bibitem [{\citenamefont {Angloher}\ \emph {et~al.}(2019)\citenamefont
  {Angloher} \emph {et~al.}}]{NUCLEUS:2019igx}%
  \BibitemOpen
  \bibfield  {author} {\bibinfo {author} {\bibfnamefont {G.}~\bibnamefont
  {Angloher}} \emph {et~al.} (\bibinfo {collaboration} {NUCLEUS}),\ }\bibfield
  {title} {\bibinfo {title} {{Exploring $\hbox {CE}\nu \hbox {NS}$ with NUCLEUS
  at the Chooz nuclear power plant}},\ }\href
  {https://doi.org/10.1140/epjc/s10052-019-7454-4} {\bibfield  {journal}
  {\bibinfo  {journal} {Eur. Phys. J. C}\ }\textbf {\bibinfo {volume} {79}},\
  \bibinfo {pages} {1018} (\bibinfo {year} {2019})},\ \Eprint
  {https://arxiv.org/abs/1905.10258} {arXiv:1905.10258 [physics.ins-det]}
  \BibitemShut {NoStop}%
\bibitem [{\citenamefont {Strauss}\ \emph
  {et~al.}(2017{\natexlab{a}})\citenamefont {Strauss} \emph
  {et~al.}}]{NUCLEUS:2017gvo}%
  \BibitemOpen
  \bibfield  {author} {\bibinfo {author} {\bibfnamefont {R.}~\bibnamefont
  {Strauss}} \emph {et~al.} (\bibinfo {collaboration} {NUCLEUS}),\ }\bibfield
  {title} {\bibinfo {title} {{Gram-scale cryogenic calorimeters for rare-event
  searches}},\ }\href {https://doi.org/10.1103/PhysRevD.96.022009} {\bibfield
  {journal} {\bibinfo  {journal} {Phys. Rev. D}\ }\textbf {\bibinfo {volume}
  {96}},\ \bibinfo {pages} {022009} (\bibinfo {year} {2017}{\natexlab{a}})},\
  \Eprint {https://arxiv.org/abs/1704.04317} {arXiv:1704.04317
  [physics.ins-det]} \BibitemShut {NoStop}%
\bibitem [{\citenamefont {Abdelhameed}\ \emph {et~al.}(2019)\citenamefont
  {Abdelhameed} \emph {et~al.}}]{CRESST:2019jnq}%
  \BibitemOpen
  \bibfield  {author} {\bibinfo {author} {\bibfnamefont {A.~H.}\ \bibnamefont
  {Abdelhameed}} \emph {et~al.} (\bibinfo {collaboration} {CRESST}),\
  }\bibfield  {title} {\bibinfo {title} {{First results from the CRESST-III
  low-mass dark matter program}},\ }\href
  {https://doi.org/10.1103/PhysRevD.100.102002} {\bibfield  {journal} {\bibinfo
   {journal} {Phys. Rev. D}\ }\textbf {\bibinfo {volume} {100}},\ \bibinfo
  {pages} {102002} (\bibinfo {year} {2019})},\ \Eprint
  {https://arxiv.org/abs/1904.00498} {arXiv:1904.00498 [astro-ph.CO]}
  \BibitemShut {NoStop}%
\bibitem [{\citenamefont {Angloher}\ \emph
  {et~al.}(2024{\natexlab{a}})\citenamefont {Angloher} \emph
  {et~al.}}]{CRESST:2024cpr}%
  \BibitemOpen
  \bibfield  {author} {\bibinfo {author} {\bibfnamefont {G.}~\bibnamefont
  {Angloher}} \emph {et~al.} (\bibinfo {collaboration} {CRESST}),\ }\bibfield
  {title} {\bibinfo {title} {{First observation of single photons in a CRESST
  detector and new dark matter exclusion limits}},\ }\href
  {https://doi.org/10.1103/PhysRevD.110.083038} {\bibfield  {journal} {\bibinfo
   {journal} {Phys. Rev. D}\ }\textbf {\bibinfo {volume} {110}},\ \bibinfo
  {pages} {083038} (\bibinfo {year} {2024}{\natexlab{a}})},\ \Eprint
  {https://arxiv.org/abs/2405.06527} {arXiv:2405.06527 [astro-ph.CO]}
  \BibitemShut {NoStop}%
\bibitem [{\citenamefont {O'Hare}(2021)}]{OHare:2021utq}%
  \BibitemOpen
  \bibfield  {author} {\bibinfo {author} {\bibfnamefont {C.~A.~J.}\
  \bibnamefont {O'Hare}},\ }\bibfield  {title} {\bibinfo {title} {{New
  Definition of the Neutrino Floor for Direct Dark Matter Searches}},\ }\href
  {https://doi.org/10.1103/PhysRevLett.127.251802} {\bibfield  {journal}
  {\bibinfo  {journal} {Phys. Rev. Lett.}\ }\textbf {\bibinfo {volume} {127}},\
  \bibinfo {pages} {251802} (\bibinfo {year} {2021})},\ \Eprint
  {https://arxiv.org/abs/2109.03116} {arXiv:2109.03116 [hep-ph]} \BibitemShut
  {NoStop}%
\bibitem [{\citenamefont {Periss{\'e}}\ \emph {et~al.}(2023)\citenamefont
  {Periss{\'e}}, \citenamefont {Onillon}, \citenamefont {Mougeot},
  \citenamefont {Vivier}, \citenamefont {Lasserre}, \citenamefont {Letourneau},
  \citenamefont {Lhuillier},\ and\ \citenamefont {Mention}}]{Perisse:2023efm}%
  \BibitemOpen
  \bibfield  {author} {\bibinfo {author} {\bibfnamefont {L.}~\bibnamefont
  {Periss{\'e}}}, \bibinfo {author} {\bibfnamefont {A.}~\bibnamefont
  {Onillon}}, \bibinfo {author} {\bibfnamefont {X.}~\bibnamefont {Mougeot}},
  \bibinfo {author} {\bibfnamefont {M.}~\bibnamefont {Vivier}}, \bibinfo
  {author} {\bibfnamefont {T.}~\bibnamefont {Lasserre}}, \bibinfo {author}
  {\bibfnamefont {A.}~\bibnamefont {Letourneau}}, \bibinfo {author}
  {\bibfnamefont {D.}~\bibnamefont {Lhuillier}},\ and\ \bibinfo {author}
  {\bibfnamefont {G.}~\bibnamefont {Mention}},\ }\bibfield  {title} {\bibinfo
  {title} {{Comprehensive revision of the summation method for the prediction
  of reactor {\ensuremath{\nu}}{\textasciimacron}e fluxes and spectra}},\
  }\href {https://doi.org/10.1103/PhysRevC.108.055501} {\bibfield  {journal}
  {\bibinfo  {journal} {Phys. Rev. C}\ }\textbf {\bibinfo {volume} {108}},\
  \bibinfo {pages} {055501} (\bibinfo {year} {2023})},\ \Eprint
  {https://arxiv.org/abs/2304.14992} {arXiv:2304.14992 [nucl-ex]} \BibitemShut
  {NoStop}%
\bibitem [{\citenamefont {Abele}\ \emph
  {et~al.}(2026{\natexlab{a}})\citenamefont {Abele} \emph
  {et~al.}}]{Abele:2025yca}%
  \BibitemOpen
  \bibfield  {author} {\bibinfo {author} {\bibfnamefont {H.}~\bibnamefont
  {Abele}} \emph {et~al.},\ }\bibfield  {title} {\bibinfo {title} {{Particle
  background characterization and prediction for the NUCLEUS reactor CE$\nu $NS
  experiment}},\ }\href {https://doi.org/10.1140/epjc/s10052-025-15168-9}
  {\bibfield  {journal} {\bibinfo  {journal} {Eur. Phys. J. C}\ }\textbf
  {\bibinfo {volume} {86}},\ \bibinfo {pages} {29} (\bibinfo {year}
  {2026}{\natexlab{a}})},\ \Eprint {https://arxiv.org/abs/2509.03559}
  {arXiv:2509.03559 [physics.ins-det]} \BibitemShut {NoStop}%
\bibitem [{\citenamefont {Wagner}\ \emph {et~al.}(2022)\citenamefont {Wagner}
  \emph {et~al.}}]{NUCLEUS:2022rcz}%
  \BibitemOpen
  \bibfield  {author} {\bibinfo {author} {\bibfnamefont {V.}~\bibnamefont
  {Wagner}} \emph {et~al.} (\bibinfo {collaboration} {NUCLEUS}),\ }\bibfield
  {title} {\bibinfo {title} {{Development of a compact muon veto for the
  Nucleus experiment}},\ }\href
  {https://doi.org/10.1088/1748-0221/17/05/T05020} {\bibfield  {journal}
  {\bibinfo  {journal} {JINST}\ }\textbf {\bibinfo {volume} {17}}\bibfield
  {number} {\bibinfo  {number} { (05)},\ \bibinfo {pages} {T05020}},\ }\Eprint
  {https://arxiv.org/abs/2202.03991} {arXiv:2202.03991 [physics.ins-det]}
  \BibitemShut {NoStop}%
\bibitem [{\citenamefont {Erhart}\ \emph {et~al.}(2024)\citenamefont {Erhart}
  \emph {et~al.}}]{Erhart:2023vam}%
  \BibitemOpen
  \bibfield  {author} {\bibinfo {author} {\bibfnamefont {A.}~\bibnamefont
  {Erhart}} \emph {et~al.},\ }\bibfield  {title} {\bibinfo {title} {{A plastic
  scintillation muon veto for sub-Kelvin temperatures}},\ }\href
  {https://doi.org/10.1140/epjc/s10052-023-12375-0} {\bibfield  {journal}
  {\bibinfo  {journal} {Eur. Phys. J. C}\ }\textbf {\bibinfo {volume} {84}},\
  \bibinfo {pages} {70} (\bibinfo {year} {2024})},\ \Eprint
  {https://arxiv.org/abs/2310.08457} {arXiv:2310.08457 [physics.ins-det]}
  \BibitemShut {NoStop}%
\bibitem [{\citenamefont {Goupy}(2024)}]{goupyPhd}%
  \BibitemOpen
  \bibfield  {author} {\bibinfo {author} {\bibfnamefont {C.}~\bibnamefont
  {Goupy}},\ }\href {https://theses.fr/s310276?domaine=theses} {\bibinfo
  {title} {{Background mitigation strategy for the detection of coherent
  elastic scattering of reactor antineutrinos on nuclei with the NUCLEUS
  experiment}}} (\bibinfo {year} {2024})\BibitemShut {NoStop}%
\bibitem [{\citenamefont {Wex}(2025)}]{wexPhd}%
  \BibitemOpen
  \bibfield  {author} {\bibinfo {author} {\bibfnamefont {A.}~\bibnamefont
  {Wex}},\ }\href {https://mediatum.ub.tum.de/?id=1775381} {\bibinfo {title}
  {{Background Suppression and Cryogenic Vibration Decoupling for the Coherent
  Elastic Neutrino Scattering Experiment NUCLEUS}}} (\bibinfo {year}
  {2025})\BibitemShut {NoStop}%
\bibitem [{\citenamefont {Abele}\ \emph
  {et~al.}(2025{\natexlab{a}})\citenamefont {Abele} \emph
  {et~al.}}]{NUCLEUS:2025ymr}%
  \BibitemOpen
  \bibfield  {author} {\bibinfo {author} {\bibfnamefont {H.}~\bibnamefont
  {Abele}} \emph {et~al.} (\bibinfo {collaboration} {NUCLEUS}),\ }\bibfield
  {title} {\bibinfo {title} {{Commissioning of the NUCLEUS Experiment at the
  Technical University of Munich}},\ }\href {https://doi.org/10.1103/c95p-8kh2}
  {\bibfield  {journal} {\bibinfo  {journal} {Phys. Rev. D}\ }\textbf {\bibinfo
  {volume} {112}},\ \bibinfo {pages} {072013} (\bibinfo {year}
  {2025}{\natexlab{a}})},\ \Eprint {https://arxiv.org/abs/2508.02488}
  {arXiv:2508.02488 [hep-ex]} \BibitemShut {NoStop}%
\bibitem [{\citenamefont {Adari}\ \emph {et~al.}(2022)\citenamefont {Adari}
  \emph {et~al.}}]{Fuss:2022fxe}%
  \BibitemOpen
  \bibfield  {author} {\bibinfo {author} {\bibfnamefont {P.}~\bibnamefont
  {Adari}} \emph {et~al.},\ }\bibfield  {title} {\bibinfo {title} {{EXCESS
  workshop: Descriptions of rising low-energy spectra}},\ }\href
  {https://doi.org/10.21468/SciPostPhysProc.9.001} {\bibfield  {journal}
  {\bibinfo  {journal} {SciPost Phys. Proc.}\ }\textbf {\bibinfo {volume}
  {9}},\ \bibinfo {pages} {001} (\bibinfo {year} {2022})},\ \Eprint
  {https://arxiv.org/abs/2202.05097} {arXiv:2202.05097 [astro-ph.IM]}
  \BibitemShut {NoStop}%
\bibitem [{\citenamefont {Angloher}\ \emph {et~al.}(2023)\citenamefont
  {Angloher} \emph {et~al.}}]{Angloher:2022pas}%
  \BibitemOpen
  \bibfield  {author} {\bibinfo {author} {\bibfnamefont {G.}~\bibnamefont
  {Angloher}} \emph {et~al.},\ }\bibfield  {title} {\bibinfo {title} {{Latest
  observations on the low energy excess in CRESST-III}},\ }\href
  {https://doi.org/10.21468/SciPostPhysProc.12.013} {\bibfield  {journal}
  {\bibinfo  {journal} {SciPost Phys. Proc.}\ }\textbf {\bibinfo {volume}
  {12}},\ \bibinfo {pages} {013} (\bibinfo {year} {2023})},\ \Eprint
  {https://arxiv.org/abs/2207.09375} {arXiv:2207.09375 [astro-ph.CO]}
  \BibitemShut {NoStop}%
\bibitem [{\citenamefont {Angloher}\ \emph
  {et~al.}(2024{\natexlab{b}})\citenamefont {Angloher} \emph
  {et~al.}}]{GRAPES-3:2024yym}%
  \BibitemOpen
  \bibfield  {author} {\bibinfo {author} {\bibfnamefont {G.}~\bibnamefont
  {Angloher}} \emph {et~al.} (\bibinfo {collaboration} {CRESST}),\ }\bibfield
  {title} {\bibinfo {title} {{DoubleTES detectors to investigate the CRESST low
  energy background: results from above-ground prototypes}},\ }\href
  {https://doi.org/10.1140/epjc/s10052-024-13282-8} {\bibfield  {journal}
  {\bibinfo  {journal} {Eur. Phys. J. C}\ }\textbf {\bibinfo {volume} {84}},\
  \bibinfo {pages} {1001} (\bibinfo {year} {2024}{\natexlab{b}})},\ \bibinfo
  {note} {[Erratum: Eur.Phys.J.C 84, 1227 (2024)]},\ \Eprint
  {https://arxiv.org/abs/2404.02607} {arXiv:2404.02607 [physics.ins-det]}
  \BibitemShut {NoStop}%
\bibitem [{\citenamefont {Anthony-Petersen}\ \emph {et~al.}(2025)\citenamefont
  {Anthony-Petersen} \emph {et~al.}}]{Anthony-Petersen:2024vdh}%
  \BibitemOpen
  \bibfield  {author} {\bibinfo {author} {\bibfnamefont {R.}~\bibnamefont
  {Anthony-Petersen}} \emph {et~al.},\ }\bibfield  {title} {\bibinfo {title}
  {{Low energy backgrounds and excess noise in a two-channel low-threshold
  calorimeter}},\ }\href {https://doi.org/10.1063/5.0247343} {\bibfield
  {journal} {\bibinfo  {journal} {Appl. Phys. Lett.}\ }\textbf {\bibinfo
  {volume} {126}},\ \bibinfo {pages} {102601} (\bibinfo {year} {2025})},\
  \Eprint {https://arxiv.org/abs/2410.16510} {arXiv:2410.16510
  [physics.ins-det]} \BibitemShut {NoStop}%
\bibitem [{\citenamefont {Abele}\ \emph
  {et~al.}(2026{\natexlab{b}})\citenamefont {Abele} \emph
  {et~al.}}]{Abele:2026sqm}%
  \BibitemOpen
  \bibfield  {author} {\bibinfo {author} {\bibfnamefont {H.}~\bibnamefont
  {Abele}} \emph {et~al.},\ }\href@noop {} {\bibinfo {title} {{Characterization
  of the Low Energy Excess using a NUCLEUS $Al_2O_3$ detector}}} (\bibinfo
  {year} {2026}{\natexlab{b}}),\ \Eprint {https://arxiv.org/abs/2603.07687}
  {arXiv:2603.07687 [physics.ins-det]} \BibitemShut {NoStop}%
\bibitem [{\citenamefont {Strauss}\ \emph
  {et~al.}(2017{\natexlab{b}})\citenamefont {Strauss} \emph
  {et~al.}}]{NUCLEUS:2017htt}%
  \BibitemOpen
  \bibfield  {author} {\bibinfo {author} {\bibfnamefont {R.}~\bibnamefont
  {Strauss}} \emph {et~al.} (\bibinfo {collaboration} {NUCLEUS}),\ }\bibfield
  {title} {\bibinfo {title} {{The $\nu$-cleus experiment: A gram-scale
  fiducial-volume cryogenic detector for the first detection of coherent
  neutrino-nucleus scattering}},\ }\href
  {https://doi.org/10.1140/epjc/s10052-017-5068-2} {\bibfield  {journal}
  {\bibinfo  {journal} {Eur. Phys. J. C}\ }\textbf {\bibinfo {volume} {77}},\
  \bibinfo {pages} {506} (\bibinfo {year} {2017}{\natexlab{b}})},\ \Eprint
  {https://arxiv.org/abs/1704.04320} {arXiv:1704.04320 [physics.ins-det]}
  \BibitemShut {NoStop}%
\bibitem [{\citenamefont {Rothe}\ \emph {et~al.}(2019)\citenamefont {Rothe}
  \emph {et~al.}}]{NUCLEUS:2019kxv}%
  \BibitemOpen
  \bibfield  {author} {\bibinfo {author} {\bibfnamefont {J.}~\bibnamefont
  {Rothe}} \emph {et~al.} (\bibinfo {collaboration} {NUCLEUS}),\ }\bibfield
  {title} {\bibinfo {title} {{NUCLEUS: Exploring Coherent Neutrino-Nucleus
  Scattering with Cryogenic Detectors}},\ }\href
  {https://doi.org/10.1007/s10909-019-02283-7} {\bibfield  {journal} {\bibinfo
  {journal} {J. Low Temp. Phys.}\ }\textbf {\bibinfo {volume} {199}},\ \bibinfo
  {pages} {433} (\bibinfo {year} {2019})}\BibitemShut {NoStop}%
\bibitem [{\citenamefont {Sierra}(2023)}]{Sierra:2023pnf}%
  \BibitemOpen
  \bibfield  {author} {\bibinfo {author} {\bibfnamefont {D.~A.}\ \bibnamefont
  {Sierra}},\ }\bibfield  {title} {\bibinfo {title} {{Extraction of neutron
  density distributions from high-statistics coherent elastic neutrino-nucleus
  scattering data}},\ }\href {https://doi.org/10.1016/j.physletb.2023.138140}
  {\bibfield  {journal} {\bibinfo  {journal} {Phys. Lett. B}\ }\textbf
  {\bibinfo {volume} {845}},\ \bibinfo {pages} {138140} (\bibinfo {year}
  {2023})},\ \Eprint {https://arxiv.org/abs/2301.13249} {arXiv:2301.13249
  [hep-ph]} \BibitemShut {NoStop}%
\bibitem [{\citenamefont {Atzori~Corona}\ \emph {et~al.}(2023)\citenamefont
  {Atzori~Corona}, \citenamefont {Cadeddu}, \citenamefont {Cargioli},
  \citenamefont {Dordei}, \citenamefont {Giunti},\ and\ \citenamefont
  {Masia}}]{AtzoriCorona:2023ktl}%
  \BibitemOpen
  \bibfield  {author} {\bibinfo {author} {\bibfnamefont {M.}~\bibnamefont
  {Atzori~Corona}}, \bibinfo {author} {\bibfnamefont {M.}~\bibnamefont
  {Cadeddu}}, \bibinfo {author} {\bibfnamefont {N.}~\bibnamefont {Cargioli}},
  \bibinfo {author} {\bibfnamefont {F.}~\bibnamefont {Dordei}}, \bibinfo
  {author} {\bibfnamefont {C.}~\bibnamefont {Giunti}},\ and\ \bibinfo {author}
  {\bibfnamefont {G.}~\bibnamefont {Masia}},\ }\bibfield  {title} {\bibinfo
  {title} {{Nuclear neutron radius and weak mixing angle measurements from
  latest COHERENT CsI and atomic parity violation Cs data}},\ }\href
  {https://doi.org/10.1140/epjc/s10052-023-11849-5} {\bibfield  {journal}
  {\bibinfo  {journal} {Eur. Phys. J. C}\ }\textbf {\bibinfo {volume} {83}},\
  \bibinfo {pages} {683} (\bibinfo {year} {2023})},\ \Eprint
  {https://arxiv.org/abs/2303.09360} {arXiv:2303.09360 [nucl-ex]} \BibitemShut
  {NoStop}%
\bibitem [{\citenamefont {Berglund}\ and\ \citenamefont
  {Wieser}(2011)}]{BerglundWieser+2011+397+410}%
  \BibitemOpen
  \bibfield  {author} {\bibinfo {author} {\bibfnamefont {M.}~\bibnamefont
  {Berglund}}\ and\ \bibinfo {author} {\bibfnamefont {M.~E.}\ \bibnamefont
  {Wieser}},\ }\bibfield  {title} {\bibinfo {title} {Isotopic compositions of
  the elements 2009 (iupac technical report)},\ }\href
  {https://doi.org/doi:10.1351/PAC-REP-10-06-02} {\bibfield  {journal}
  {\bibinfo  {journal} {Pure and Applied Chemistry}\ }\textbf {\bibinfo
  {volume} {83}},\ \bibinfo {pages} {397} (\bibinfo {year} {2011})}\BibitemShut
  {NoStop}%
\bibitem [{\citenamefont {{National Center for Biotechnology
  Information}}()}]{PubChemAtomicMass}%
  \BibitemOpen
  \bibfield  {author} {\bibinfo {author} {\bibnamefont {{National Center for
  Biotechnology Information}}},\ }\href@noop {} {\bibinfo {title} {Periodic
  table -- atomic mass}},\ \bibinfo {howpublished}
  {\url{https://pubchem.ncbi.nlm.nih.gov/ptable/atomic-mass/}},\ \bibinfo
  {note} {accessed: 2026-01-08}\BibitemShut {NoStop}%
\bibitem [{\citenamefont {Helm}(1956)}]{Helm:1956zz}%
  \BibitemOpen
  \bibfield  {author} {\bibinfo {author} {\bibfnamefont {R.~H.}\ \bibnamefont
  {Helm}},\ }\bibfield  {title} {\bibinfo {title} {{Inelastic and Elastic
  Scattering of 187-Mev Electrons from Selected Even-Even Nuclei}},\ }\href
  {https://doi.org/10.1103/PhysRev.104.1466} {\bibfield  {journal} {\bibinfo
  {journal} {Phys. Rev.}\ }\textbf {\bibinfo {volume} {104}},\ \bibinfo {pages}
  {1466} (\bibinfo {year} {1956})}\BibitemShut {NoStop}%
\bibitem [{\citenamefont {Lewin}\ and\ \citenamefont
  {Smith}(1996)}]{Lewin:1995rx}%
  \BibitemOpen
  \bibfield  {author} {\bibinfo {author} {\bibfnamefont {J.~D.}\ \bibnamefont
  {Lewin}}\ and\ \bibinfo {author} {\bibfnamefont {P.~F.}\ \bibnamefont
  {Smith}},\ }\bibfield  {title} {\bibinfo {title} {{Review of mathematics,
  numerical factors, and corrections for dark matter experiments based on
  elastic nuclear recoil}},\ }\href
  {https://doi.org/10.1016/S0927-6505(96)00047-3} {\bibfield  {journal}
  {\bibinfo  {journal} {Astropart. Phys.}\ }\textbf {\bibinfo {volume} {6}},\
  \bibinfo {pages} {87} (\bibinfo {year} {1996})}\BibitemShut {NoStop}%
\bibitem [{\citenamefont {Fanchiotti}\ \emph {et~al.}(1993)\citenamefont
  {Fanchiotti}, \citenamefont {Kniehl},\ and\ \citenamefont
  {Sirlin}}]{Fanchiotti:1992tu}%
  \BibitemOpen
  \bibfield  {author} {\bibinfo {author} {\bibfnamefont {S.}~\bibnamefont
  {Fanchiotti}}, \bibinfo {author} {\bibfnamefont {B.~A.}\ \bibnamefont
  {Kniehl}},\ and\ \bibinfo {author} {\bibfnamefont {A.}~\bibnamefont
  {Sirlin}},\ }\bibfield  {title} {\bibinfo {title} {{Incorporation of QCD
  effects in basic corrections of the electroweak theory}},\ }\href
  {https://doi.org/10.1103/PhysRevD.48.307} {\bibfield  {journal} {\bibinfo
  {journal} {Phys. Rev. D}\ }\textbf {\bibinfo {volume} {48}},\ \bibinfo
  {pages} {307} (\bibinfo {year} {1993})},\ \Eprint
  {https://arxiv.org/abs/hep-ph/9212285} {arXiv:hep-ph/9212285} \BibitemShut
  {NoStop}%
\bibitem [{\citenamefont {Atzori~Corona}\ \emph {et~al.}(2024)\citenamefont
  {Atzori~Corona}, \citenamefont {Cadeddu}, \citenamefont {Cargioli},
  \citenamefont {Dordei},\ and\ \citenamefont {Giunti}}]{AtzoriCorona:2024rtv}%
  \BibitemOpen
  \bibfield  {author} {\bibinfo {author} {\bibfnamefont {M.}~\bibnamefont
  {Atzori~Corona}}, \bibinfo {author} {\bibfnamefont {M.}~\bibnamefont
  {Cadeddu}}, \bibinfo {author} {\bibfnamefont {N.}~\bibnamefont {Cargioli}},
  \bibinfo {author} {\bibfnamefont {F.}~\bibnamefont {Dordei}},\ and\ \bibinfo
  {author} {\bibfnamefont {C.}~\bibnamefont {Giunti}},\ }\bibfield  {title}
  {\bibinfo {title} {{Momentum dependent flavor radiative corrections to the
  coherent elastic neutrino-nucleus scattering for the neutrino charge-radius
  determination}},\ }\href {https://doi.org/10.1007/JHEP05(2024)271} {\bibfield
   {journal} {\bibinfo  {journal} {JHEP}\ }\textbf {\bibinfo {volume} {05}},\
  \bibinfo {pages} {271}},\ \Eprint {https://arxiv.org/abs/2402.16709}
  {arXiv:2402.16709 [hep-ph]} \BibitemShut {NoStop}%
\bibitem [{\citenamefont {Erler}\ and\ \citenamefont
  {Su}(2013)}]{Erler:2013xha}%
  \BibitemOpen
  \bibfield  {author} {\bibinfo {author} {\bibfnamefont {J.}~\bibnamefont
  {Erler}}\ and\ \bibinfo {author} {\bibfnamefont {S.}~\bibnamefont {Su}},\
  }\bibfield  {title} {\bibinfo {title} {{The Weak Neutral Current}},\ }\href
  {https://doi.org/10.1016/j.ppnp.2013.03.004} {\bibfield  {journal} {\bibinfo
  {journal} {Prog. Part. Nucl. Phys.}\ }\textbf {\bibinfo {volume} {71}},\
  \bibinfo {pages} {119} (\bibinfo {year} {2013})},\ \Eprint
  {https://arxiv.org/abs/1303.5522} {arXiv:1303.5522 [hep-ph]} \BibitemShut
  {NoStop}%
\bibitem [{\citenamefont {Navas}\ \emph
  {et~al.}(2024{\natexlab{a}})\citenamefont {Navas} \emph
  {et~al.}}]{PhysRevD.110.030001}%
  \BibitemOpen
  \bibfield  {author} {\bibinfo {author} {\bibfnamefont {S.}~\bibnamefont
  {Navas}} \emph {et~al.} (\bibinfo {collaboration} {Particle Data Group
  Collaboration}),\ }\bibfield  {title} {\bibinfo {title} {Review of particle
  physics},\ }\href {https://doi.org/10.1103/PhysRevD.110.030001} {\bibfield
  {journal} {\bibinfo  {journal} {Phys. Rev. D}\ }\textbf {\bibinfo {volume}
  {110}},\ \bibinfo {pages} {030001} (\bibinfo {year}
  {2024}{\natexlab{a}})}\BibitemShut {NoStop}%
\bibitem [{\citenamefont {Navas}\ \emph
  {et~al.}(2024{\natexlab{b}})\citenamefont {Navas} \emph
  {et~al.}}]{ParticleDataGroup:2024cfk}%
  \BibitemOpen
  \bibfield  {author} {\bibinfo {author} {\bibfnamefont {S.}~\bibnamefont
  {Navas}} \emph {et~al.} (\bibinfo {collaboration} {Particle Data Group}),\
  }\bibfield  {title} {\bibinfo {title} {{Review of particle physics}},\ }\href
  {https://doi.org/10.1103/PhysRevD.110.030001} {\bibfield  {journal} {\bibinfo
   {journal} {Phys. Rev. D}\ }\textbf {\bibinfo {volume} {110}},\ \bibinfo
  {pages} {030001} (\bibinfo {year} {2024}{\natexlab{b}})}\BibitemShut
  {NoStop}%
\bibitem [{\citenamefont {Giunti}\ and\ \citenamefont
  {Studenikin}(2015)}]{Giunti:2014ixa}%
  \BibitemOpen
  \bibfield  {author} {\bibinfo {author} {\bibfnamefont {C.}~\bibnamefont
  {Giunti}}\ and\ \bibinfo {author} {\bibfnamefont {A.}~\bibnamefont
  {Studenikin}},\ }\bibfield  {title} {\bibinfo {title} {{Neutrino
  electromagnetic interactions: a window to new physics}},\ }\href
  {https://doi.org/10.1103/RevModPhys.87.531} {\bibfield  {journal} {\bibinfo
  {journal} {Rev. Mod. Phys.}\ }\textbf {\bibinfo {volume} {87}},\ \bibinfo
  {pages} {531} (\bibinfo {year} {2015})},\ \Eprint
  {https://arxiv.org/abs/1403.6344} {arXiv:1403.6344 [hep-ph]} \BibitemShut
  {NoStop}%
\bibitem [{\citenamefont {Atzori~Corona}\ \emph
  {et~al.}(2022{\natexlab{a}})\citenamefont {Atzori~Corona}, \citenamefont
  {Cadeddu}, \citenamefont {Cargioli}, \citenamefont {Dordei}, \citenamefont
  {Giunti}, \citenamefont {Li}, \citenamefont {Ternes},\ and\ \citenamefont
  {Zhang}}]{AtzoriCorona:2022qrf}%
  \BibitemOpen
  \bibfield  {author} {\bibinfo {author} {\bibfnamefont {M.}~\bibnamefont
  {Atzori~Corona}}, \bibinfo {author} {\bibfnamefont {M.}~\bibnamefont
  {Cadeddu}}, \bibinfo {author} {\bibfnamefont {N.}~\bibnamefont {Cargioli}},
  \bibinfo {author} {\bibfnamefont {F.}~\bibnamefont {Dordei}}, \bibinfo
  {author} {\bibfnamefont {C.}~\bibnamefont {Giunti}}, \bibinfo {author}
  {\bibfnamefont {Y.~F.}\ \bibnamefont {Li}}, \bibinfo {author} {\bibfnamefont
  {C.~A.}\ \bibnamefont {Ternes}},\ and\ \bibinfo {author} {\bibfnamefont
  {Y.~Y.}\ \bibnamefont {Zhang}},\ }\bibfield  {title} {\bibinfo {title}
  {{Impact of the Dresden-II and COHERENT neutrino scattering data on neutrino
  electromagnetic properties and electroweak physics}},\ }\href
  {https://doi.org/10.1007/JHEP09(2022)164} {\bibfield  {journal} {\bibinfo
  {journal} {JHEP}\ }\textbf {\bibinfo {volume} {09}},\ \bibinfo {pages}
  {164}},\ \Eprint {https://arxiv.org/abs/2205.09484} {arXiv:2205.09484
  [hep-ph]} \BibitemShut {NoStop}%
\bibitem [{\citenamefont {Cadeddu}\ \emph
  {et~al.}(2020{\natexlab{a}})\citenamefont {Cadeddu}, \citenamefont {Dordei},
  \citenamefont {Giunti}, \citenamefont {Li},\ and\ \citenamefont
  {Zhang}}]{Cadeddu:2019eta}%
  \BibitemOpen
  \bibfield  {author} {\bibinfo {author} {\bibfnamefont {M.}~\bibnamefont
  {Cadeddu}}, \bibinfo {author} {\bibfnamefont {F.}~\bibnamefont {Dordei}},
  \bibinfo {author} {\bibfnamefont {C.}~\bibnamefont {Giunti}}, \bibinfo
  {author} {\bibfnamefont {Y.~F.}\ \bibnamefont {Li}},\ and\ \bibinfo {author}
  {\bibfnamefont {Y.~Y.}\ \bibnamefont {Zhang}},\ }\bibfield  {title} {\bibinfo
  {title} {{Neutrino, electroweak, and nuclear physics from COHERENT elastic
  neutrino-nucleus scattering with refined quenching factor}},\ }\href
  {https://doi.org/10.1103/PhysRevD.101.033004} {\bibfield  {journal} {\bibinfo
   {journal} {Phys. Rev. D}\ }\textbf {\bibinfo {volume} {101}},\ \bibinfo
  {pages} {033004} (\bibinfo {year} {2020}{\natexlab{a}})},\ \Eprint
  {https://arxiv.org/abs/1908.06045} {arXiv:1908.06045 [hep-ph]} \BibitemShut
  {NoStop}%
\bibitem [{\citenamefont {Atzori~Corona}\ \emph
  {et~al.}(2025{\natexlab{a}})\citenamefont {Atzori~Corona}, \citenamefont
  {Cadeddu}, \citenamefont {Cargioli}, \citenamefont {Dordei}, \citenamefont
  {Giunti},\ and\ \citenamefont {Ternes}}]{AtzoriCorona:2025xwr}%
  \BibitemOpen
  \bibfield  {author} {\bibinfo {author} {\bibfnamefont {M.}~\bibnamefont
  {Atzori~Corona}}, \bibinfo {author} {\bibfnamefont {M.}~\bibnamefont
  {Cadeddu}}, \bibinfo {author} {\bibfnamefont {N.}~\bibnamefont {Cargioli}},
  \bibinfo {author} {\bibfnamefont {F.}~\bibnamefont {Dordei}}, \bibinfo
  {author} {\bibfnamefont {C.}~\bibnamefont {Giunti}},\ and\ \bibinfo {author}
  {\bibfnamefont {C.~A.}\ \bibnamefont {Ternes}},\ }\bibfield  {title}
  {\bibinfo {title} {{Standard Model Tested with Neutrinos}},\ }\href
  {https://doi.org/10.1103/dplq-dvc8} {\bibfield  {journal} {\bibinfo
  {journal} {Phys. Rev. Lett.}\ }\textbf {\bibinfo {volume} {135}},\ \bibinfo
  {pages} {231803} (\bibinfo {year} {2025}{\natexlab{a}})},\ \Eprint
  {https://arxiv.org/abs/2504.05272} {arXiv:2504.05272 [hep-ph]} \BibitemShut
  {NoStop}%
\bibitem [{\citenamefont {Barranco}\ \emph {et~al.}(2005)\citenamefont
  {Barranco}, \citenamefont {Miranda},\ and\ \citenamefont
  {Rashba}}]{Barranco:2005yy}%
  \BibitemOpen
  \bibfield  {author} {\bibinfo {author} {\bibfnamefont {J.}~\bibnamefont
  {Barranco}}, \bibinfo {author} {\bibfnamefont {O.~G.}\ \bibnamefont
  {Miranda}},\ and\ \bibinfo {author} {\bibfnamefont {T.~I.}\ \bibnamefont
  {Rashba}},\ }\bibfield  {title} {\bibinfo {title} {{Probing new physics with
  coherent neutrino scattering off nuclei}},\ }\href
  {https://doi.org/10.1088/1126-6708/2005/12/021} {\bibfield  {journal}
  {\bibinfo  {journal} {JHEP}\ }\textbf {\bibinfo {volume} {12}},\ \bibinfo
  {pages} {021}},\ \Eprint {https://arxiv.org/abs/hep-ph/0508299}
  {arXiv:hep-ph/0508299} \BibitemShut {NoStop}%
\bibitem [{\citenamefont {Giunti}(2020)}]{Giunti:2019xpr}%
  \BibitemOpen
  \bibfield  {author} {\bibinfo {author} {\bibfnamefont {C.}~\bibnamefont
  {Giunti}},\ }\bibfield  {title} {\bibinfo {title} {{General COHERENT
  constraints on neutrino nonstandard interactions}},\ }\href
  {https://doi.org/10.1103/PhysRevD.101.035039} {\bibfield  {journal} {\bibinfo
   {journal} {Phys. Rev. D}\ }\textbf {\bibinfo {volume} {101}},\ \bibinfo
  {pages} {035039} (\bibinfo {year} {2020})},\ \Eprint
  {https://arxiv.org/abs/1909.00466} {arXiv:1909.00466 [hep-ph]} \BibitemShut
  {NoStop}%
\bibitem [{\citenamefont {Coloma}\ \emph {et~al.}(2023)\citenamefont {Coloma},
  \citenamefont {Gonzalez-Garcia}, \citenamefont {Maltoni}, \citenamefont
  {Pinheiro},\ and\ \citenamefont {Urrea}}]{Coloma:2023ixt}%
  \BibitemOpen
  \bibfield  {author} {\bibinfo {author} {\bibfnamefont {P.}~\bibnamefont
  {Coloma}}, \bibinfo {author} {\bibfnamefont {M.~C.}\ \bibnamefont
  {Gonzalez-Garcia}}, \bibinfo {author} {\bibfnamefont {M.}~\bibnamefont
  {Maltoni}}, \bibinfo {author} {\bibfnamefont {J.~P.}\ \bibnamefont
  {Pinheiro}},\ and\ \bibinfo {author} {\bibfnamefont {S.}~\bibnamefont
  {Urrea}},\ }\bibfield  {title} {\bibinfo {title} {{Global constraints on
  non-standard neutrino interactions with quarks and electrons}},\ }\href
  {https://doi.org/10.1007/JHEP08(2023)032} {\bibfield  {journal} {\bibinfo
  {journal} {JHEP}\ }\textbf {\bibinfo {volume} {08}},\ \bibinfo {pages}
  {032}},\ \Eprint {https://arxiv.org/abs/2305.07698} {arXiv:2305.07698
  [hep-ph]} \BibitemShut {NoStop}%
\bibitem [{\citenamefont {Amaral}\ \emph {et~al.}(2023)\citenamefont {Amaral},
  \citenamefont {Cerdeno}, \citenamefont {Cheek},\ and\ \citenamefont
  {Foldenauer}}]{Amaral:2023tbs}%
  \BibitemOpen
  \bibfield  {author} {\bibinfo {author} {\bibfnamefont {D.~W.~P.}\
  \bibnamefont {Amaral}}, \bibinfo {author} {\bibfnamefont {D.}~\bibnamefont
  {Cerdeno}}, \bibinfo {author} {\bibfnamefont {A.}~\bibnamefont {Cheek}},\
  and\ \bibinfo {author} {\bibfnamefont {P.}~\bibnamefont {Foldenauer}},\
  }\bibfield  {title} {\bibinfo {title} {{A direct detection view of the
  neutrino NSI landscape}},\ }\href {https://doi.org/10.1007/JHEP07(2023)071}
  {\bibfield  {journal} {\bibinfo  {journal} {JHEP}\ }\textbf {\bibinfo
  {volume} {07}},\ \bibinfo {pages} {071}},\ \Eprint
  {https://arxiv.org/abs/2302.12846} {arXiv:2302.12846 [hep-ph]} \BibitemShut
  {NoStop}%
\bibitem [{\citenamefont {Alloni}\ \emph {et~al.}(2026)\citenamefont {Alloni}
  \emph {et~al.}}]{RES-NOVA:2026fii}%
  \BibitemOpen
  \bibfield  {author} {\bibinfo {author} {\bibfnamefont {D.}~\bibnamefont
  {Alloni}} \emph {et~al.} (\bibinfo {collaboration} {RES-NOVA}),\ }\href@noop
  {} {\bibinfo {title} {{Neutrino NSI in archaeological Pb}}} (\bibinfo {year}
  {2026}),\ \Eprint {https://arxiv.org/abs/2602.23419} {arXiv:2602.23419
  [hep-ph]} \BibitemShut {NoStop}%
\bibitem [{\citenamefont {Cadeddu}\ \emph {et~al.}(2021)\citenamefont
  {Cadeddu}, \citenamefont {Cargioli}, \citenamefont {Dordei}, \citenamefont
  {Giunti}, \citenamefont {Li}, \citenamefont {Picciau},\ and\ \citenamefont
  {Zhang}}]{Cadeddu:2020nbr}%
  \BibitemOpen
  \bibfield  {author} {\bibinfo {author} {\bibfnamefont {M.}~\bibnamefont
  {Cadeddu}}, \bibinfo {author} {\bibfnamefont {N.}~\bibnamefont {Cargioli}},
  \bibinfo {author} {\bibfnamefont {F.}~\bibnamefont {Dordei}}, \bibinfo
  {author} {\bibfnamefont {C.}~\bibnamefont {Giunti}}, \bibinfo {author}
  {\bibfnamefont {Y.~F.}\ \bibnamefont {Li}}, \bibinfo {author} {\bibfnamefont
  {E.}~\bibnamefont {Picciau}},\ and\ \bibinfo {author} {\bibfnamefont {Y.~Y.}\
  \bibnamefont {Zhang}},\ }\bibfield  {title} {\bibinfo {title} {{Constraints
  on light vector mediators through coherent elastic neutrino nucleus
  scattering data from COHERENT}},\ }\href
  {https://doi.org/10.1007/JHEP01(2021)116} {\bibfield  {journal} {\bibinfo
  {journal} {JHEP}\ }\textbf {\bibinfo {volume} {01}},\ \bibinfo {pages}
  {116}},\ \Eprint {https://arxiv.org/abs/2008.05022} {arXiv:2008.05022
  [hep-ph]} \BibitemShut {NoStop}%
\bibitem [{\citenamefont {Atzori~Corona}\ \emph
  {et~al.}(2022{\natexlab{b}})\citenamefont {Atzori~Corona}, \citenamefont
  {Cadeddu}, \citenamefont {Cargioli}, \citenamefont {Dordei}, \citenamefont
  {Giunti}, \citenamefont {Li}, \citenamefont {Picciau}, \citenamefont
  {Ternes},\ and\ \citenamefont {Zhang}}]{AtzoriCorona:2022moj}%
  \BibitemOpen
  \bibfield  {author} {\bibinfo {author} {\bibfnamefont {M.}~\bibnamefont
  {Atzori~Corona}}, \bibinfo {author} {\bibfnamefont {M.}~\bibnamefont
  {Cadeddu}}, \bibinfo {author} {\bibfnamefont {N.}~\bibnamefont {Cargioli}},
  \bibinfo {author} {\bibfnamefont {F.}~\bibnamefont {Dordei}}, \bibinfo
  {author} {\bibfnamefont {C.}~\bibnamefont {Giunti}}, \bibinfo {author}
  {\bibfnamefont {Y.~F.}\ \bibnamefont {Li}}, \bibinfo {author} {\bibfnamefont
  {E.}~\bibnamefont {Picciau}}, \bibinfo {author} {\bibfnamefont {C.~A.}\
  \bibnamefont {Ternes}},\ and\ \bibinfo {author} {\bibfnamefont {Y.~Y.}\
  \bibnamefont {Zhang}},\ }\bibfield  {title} {\bibinfo {title} {{Probing light
  mediators and (g \ensuremath{-} 2)$_{\mu}$ through detection of coherent
  elastic neutrino nucleus scattering at COHERENT}},\ }\href
  {https://doi.org/10.1007/JHEP05(2022)109} {\bibfield  {journal} {\bibinfo
  {journal} {JHEP}\ }\textbf {\bibinfo {volume} {05}},\ \bibinfo {pages}
  {109}},\ \Eprint {https://arxiv.org/abs/2202.11002} {arXiv:2202.11002
  [hep-ph]} \BibitemShut {NoStop}%
\bibitem [{\citenamefont {Allanach}\ \emph {et~al.}(2019)\citenamefont
  {Allanach}, \citenamefont {Davighi},\ and\ \citenamefont
  {Melville}}]{Allanach:2018vjg}%
  \BibitemOpen
  \bibfield  {author} {\bibinfo {author} {\bibfnamefont {B.~C.}\ \bibnamefont
  {Allanach}}, \bibinfo {author} {\bibfnamefont {J.}~\bibnamefont {Davighi}},\
  and\ \bibinfo {author} {\bibfnamefont {S.}~\bibnamefont {Melville}},\
  }\bibfield  {title} {\bibinfo {title} {{An Anomaly-free Atlas: charting the
  space of flavour-dependent gauged $U(1)$ extensions of the Standard Model}},\
  }\href {https://doi.org/10.1007/JHEP02(2019)082} {\bibfield  {journal}
  {\bibinfo  {journal} {JHEP}\ }\textbf {\bibinfo {volume} {02}},\ \bibinfo
  {pages} {082}},\ \bibinfo {note} {[Erratum: JHEP 08, 064 (2019)]},\ \Eprint
  {https://arxiv.org/abs/1812.04602} {arXiv:1812.04602 [hep-ph]} \BibitemShut
  {NoStop}%
\bibitem [{\citenamefont {De~Romeri}\ \emph {et~al.}(2023)\citenamefont
  {De~Romeri}, \citenamefont {Miranda}, \citenamefont {Papoulias},
  \citenamefont {Sanchez~Garcia}, \citenamefont {T\'ortola},\ and\
  \citenamefont {Valle}}]{DeRomeri:2022twg}%
  \BibitemOpen
  \bibfield  {author} {\bibinfo {author} {\bibfnamefont {V.}~\bibnamefont
  {De~Romeri}}, \bibinfo {author} {\bibfnamefont {O.~G.}\ \bibnamefont
  {Miranda}}, \bibinfo {author} {\bibfnamefont {D.~K.}\ \bibnamefont
  {Papoulias}}, \bibinfo {author} {\bibfnamefont {G.}~\bibnamefont
  {Sanchez~Garcia}}, \bibinfo {author} {\bibfnamefont {M.}~\bibnamefont
  {T\'ortola}},\ and\ \bibinfo {author} {\bibfnamefont {J.~W.~F.}\ \bibnamefont
  {Valle}},\ }\bibfield  {title} {\bibinfo {title} {{Physics implications of a
  combined analysis of COHERENT CsI and LAr data}},\ }\href
  {https://doi.org/10.1007/JHEP04(2023)035} {\bibfield  {journal} {\bibinfo
  {journal} {JHEP}\ }\textbf {\bibinfo {volume} {04}},\ \bibinfo {pages}
  {035}},\ \Eprint {https://arxiv.org/abs/2211.11905} {arXiv:2211.11905
  [hep-ph]} \BibitemShut {NoStop}%
\bibitem [{\citenamefont {Demirci}\ and\ \citenamefont
  {Mustamin}(2024)}]{Demirci:2023tui}%
  \BibitemOpen
  \bibfield  {author} {\bibinfo {author} {\bibfnamefont {M.}~\bibnamefont
  {Demirci}}\ and\ \bibinfo {author} {\bibfnamefont {M.~F.}\ \bibnamefont
  {Mustamin}},\ }\bibfield  {title} {\bibinfo {title} {{Solar neutrino
  constraints on light mediators through coherent elastic neutrino-nucleus
  scattering}},\ }\href {https://doi.org/10.1103/PhysRevD.109.015021}
  {\bibfield  {journal} {\bibinfo  {journal} {Phys. Rev. D}\ }\textbf {\bibinfo
  {volume} {109}},\ \bibinfo {pages} {015021} (\bibinfo {year} {2024})},\
  \Eprint {https://arxiv.org/abs/2312.17502} {arXiv:2312.17502 [hep-ph]}
  \BibitemShut {NoStop}%
\bibitem [{\citenamefont {Lindner}\ \emph {et~al.}(2017)\citenamefont
  {Lindner}, \citenamefont {Rodejohann},\ and\ \citenamefont
  {Xu}}]{Lindner:2016wff}%
  \BibitemOpen
  \bibfield  {author} {\bibinfo {author} {\bibfnamefont {M.}~\bibnamefont
  {Lindner}}, \bibinfo {author} {\bibfnamefont {W.}~\bibnamefont
  {Rodejohann}},\ and\ \bibinfo {author} {\bibfnamefont {X.-J.}\ \bibnamefont
  {Xu}},\ }\bibfield  {title} {\bibinfo {title} {{Coherent Neutrino-Nucleus
  Scattering and new Neutrino Interactions}},\ }\href
  {https://doi.org/10.1007/JHEP03(2017)097} {\bibfield  {journal} {\bibinfo
  {journal} {JHEP}\ }\textbf {\bibinfo {volume} {03}},\ \bibinfo {pages}
  {097}},\ \Eprint {https://arxiv.org/abs/1612.04150} {arXiv:1612.04150
  [hep-ph]} \BibitemShut {NoStop}%
\bibitem [{\citenamefont {Giunti}\ \emph {et~al.}(2016)\citenamefont {Giunti},
  \citenamefont {Kouzakov}, \citenamefont {Li}, \citenamefont {Lokhov},
  \citenamefont {Studenikin},\ and\ \citenamefont {Zhou}}]{Giunti:2015gga}%
  \BibitemOpen
  \bibfield  {author} {\bibinfo {author} {\bibfnamefont {C.}~\bibnamefont
  {Giunti}}, \bibinfo {author} {\bibfnamefont {K.~A.}\ \bibnamefont
  {Kouzakov}}, \bibinfo {author} {\bibfnamefont {Y.-F.}\ \bibnamefont {Li}},
  \bibinfo {author} {\bibfnamefont {A.~V.}\ \bibnamefont {Lokhov}}, \bibinfo
  {author} {\bibfnamefont {A.~I.}\ \bibnamefont {Studenikin}},\ and\ \bibinfo
  {author} {\bibfnamefont {S.}~\bibnamefont {Zhou}},\ }\bibfield  {title}
  {\bibinfo {title} {{Electromagnetic neutrinos in laboratory experiments and
  astrophysics}},\ }\href {https://doi.org/10.1002/andp.201500211} {\bibfield
  {journal} {\bibinfo  {journal} {Annalen Phys.}\ }\textbf {\bibinfo {volume}
  {528}},\ \bibinfo {pages} {198} (\bibinfo {year} {2016})},\ \Eprint
  {https://arxiv.org/abs/1506.05387} {arXiv:1506.05387 [hep-ph]} \BibitemShut
  {NoStop}%
\bibitem [{\citenamefont {Giunti}\ \emph {et~al.}(2025)\citenamefont {Giunti},
  \citenamefont {Kouzakov}, \citenamefont {Li},\ and\ \citenamefont
  {Studenikin}}]{Giunti:2024gec}%
  \BibitemOpen
  \bibfield  {author} {\bibinfo {author} {\bibfnamefont {C.}~\bibnamefont
  {Giunti}}, \bibinfo {author} {\bibfnamefont {K.}~\bibnamefont {Kouzakov}},
  \bibinfo {author} {\bibfnamefont {Y.-F.}\ \bibnamefont {Li}},\ and\ \bibinfo
  {author} {\bibfnamefont {A.}~\bibnamefont {Studenikin}},\ }\bibfield  {title}
  {\bibinfo {title} {{Neutrino Electromagnetic Properties}},\ }\href
  {https://doi.org/10.1146/annurev-nucl-102122-023242} {\bibfield  {journal}
  {\bibinfo  {journal} {Ann. Rev. Nucl. Part. Sci.}\ }\textbf {\bibinfo
  {volume} {75}},\ \bibinfo {pages} {1} (\bibinfo {year} {2025})},\ \Eprint
  {https://arxiv.org/abs/2411.03122} {arXiv:2411.03122 [hep-ph]} \BibitemShut
  {NoStop}%
\bibitem [{\citenamefont {Bonet}\ \emph {et~al.}(2022)\citenamefont {Bonet}
  \emph {et~al.}}]{CONUS:2022qbb}%
  \BibitemOpen
  \bibfield  {author} {\bibinfo {author} {\bibfnamefont {H.}~\bibnamefont
  {Bonet}} \emph {et~al.} (\bibinfo {collaboration} {CONUS}),\ }\bibfield
  {title} {\bibinfo {title} {{First upper limits on neutrino electromagnetic
  properties from the CONUS experiment}},\ }\href
  {https://doi.org/10.1140/epjc/s10052-022-10722-1} {\bibfield  {journal}
  {\bibinfo  {journal} {Eur. Phys. J. C}\ }\textbf {\bibinfo {volume} {82}},\
  \bibinfo {pages} {813} (\bibinfo {year} {2022})},\ \Eprint
  {https://arxiv.org/abs/2201.12257} {arXiv:2201.12257 [hep-ex]} \BibitemShut
  {NoStop}%
\bibitem [{\citenamefont {Beda}\ \emph {et~al.}(2010)\citenamefont {Beda},
  \citenamefont {Demidova}, \citenamefont {Starostin}, \citenamefont
  {Brudanin}, \citenamefont {Egorov}, \citenamefont {Medvedev}, \citenamefont
  {Shirchenko},\ and\ \citenamefont {Vylov}}]{Beda:2009kx}%
  \BibitemOpen
  \bibfield  {author} {\bibinfo {author} {\bibfnamefont {A.~G.}\ \bibnamefont
  {Beda}}, \bibinfo {author} {\bibfnamefont {E.~V.}\ \bibnamefont {Demidova}},
  \bibinfo {author} {\bibfnamefont {A.~S.}\ \bibnamefont {Starostin}}, \bibinfo
  {author} {\bibfnamefont {V.~B.}\ \bibnamefont {Brudanin}}, \bibinfo {author}
  {\bibfnamefont {V.~G.}\ \bibnamefont {Egorov}}, \bibinfo {author}
  {\bibfnamefont {D.~V.}\ \bibnamefont {Medvedev}}, \bibinfo {author}
  {\bibfnamefont {M.~V.}\ \bibnamefont {Shirchenko}},\ and\ \bibinfo {author}
  {\bibfnamefont {T.}~\bibnamefont {Vylov}},\ }\bibfield  {title} {\bibinfo
  {title} {{GEMMA experiment: Three years of the search for the neutrino
  magnetic moment}},\ }\href {https://doi.org/10.1134/S1547477110060063}
  {\bibfield  {journal} {\bibinfo  {journal} {Phys. Part. Nucl. Lett.}\
  }\textbf {\bibinfo {volume} {7}},\ \bibinfo {pages} {406} (\bibinfo {year}
  {2010})},\ \Eprint {https://arxiv.org/abs/0906.1926} {arXiv:0906.1926
  [hep-ex]} \BibitemShut {NoStop}%
\bibitem [{\citenamefont {Atzori~Corona}\ \emph
  {et~al.}(2025{\natexlab{b}})\citenamefont {Atzori~Corona}, \citenamefont
  {Cadeddu}, \citenamefont {Cargioli}, \citenamefont {Dordei},\ and\
  \citenamefont {Giunti}}]{AtzoriCorona:2025ygn}%
  \BibitemOpen
  \bibfield  {author} {\bibinfo {author} {\bibfnamefont {M.}~\bibnamefont
  {Atzori~Corona}}, \bibinfo {author} {\bibfnamefont {M.}~\bibnamefont
  {Cadeddu}}, \bibinfo {author} {\bibfnamefont {N.}~\bibnamefont {Cargioli}},
  \bibinfo {author} {\bibfnamefont {F.}~\bibnamefont {Dordei}},\ and\ \bibinfo
  {author} {\bibfnamefont {C.}~\bibnamefont {Giunti}},\ }\bibfield  {title}
  {\bibinfo {title} {{Reactor antineutrinos CE{\ensuremath{\nu}}NS on
  germanium: CONUS+ and TEXONO as a new gateway to SM and BSM physics}},\
  }\href {https://doi.org/10.1103/n563-8v8d} {\bibfield  {journal} {\bibinfo
  {journal} {Phys. Rev. D}\ }\textbf {\bibinfo {volume} {112}},\ \bibinfo
  {pages} {015007} (\bibinfo {year} {2025}{\natexlab{b}})},\ \Eprint
  {https://arxiv.org/abs/2501.18550} {arXiv:2501.18550 [hep-ph]} \BibitemShut
  {NoStop}%
\bibitem [{\citenamefont {Coloma}\ \emph {et~al.}(2022)\citenamefont {Coloma},
  \citenamefont {Esteban}, \citenamefont {Gonzalez-Garcia}, \citenamefont
  {Larizgoitia}, \citenamefont {Monrabal},\ and\ \citenamefont
  {Palomares-Ruiz}}]{Coloma:2022avw}%
  \BibitemOpen
  \bibfield  {author} {\bibinfo {author} {\bibfnamefont {P.}~\bibnamefont
  {Coloma}}, \bibinfo {author} {\bibfnamefont {I.}~\bibnamefont {Esteban}},
  \bibinfo {author} {\bibfnamefont {M.~C.}\ \bibnamefont {Gonzalez-Garcia}},
  \bibinfo {author} {\bibfnamefont {L.}~\bibnamefont {Larizgoitia}}, \bibinfo
  {author} {\bibfnamefont {F.}~\bibnamefont {Monrabal}},\ and\ \bibinfo
  {author} {\bibfnamefont {S.}~\bibnamefont {Palomares-Ruiz}},\ }\bibfield
  {title} {\bibinfo {title} {{Bounds on new physics with data of the Dresden-II
  reactor experiment and COHERENT}},\ }\href
  {https://doi.org/10.1007/JHEP05(2022)037} {\bibfield  {journal} {\bibinfo
  {journal} {JHEP}\ }\textbf {\bibinfo {volume} {05}},\ \bibinfo {pages}
  {037}},\ \Eprint {https://arxiv.org/abs/2202.10829} {arXiv:2202.10829
  [hep-ph]} \BibitemShut {NoStop}%
\bibitem [{\citenamefont {Li}\ \emph {et~al.}(2003)\citenamefont {Li} \emph
  {et~al.}}]{TEXONO:2002pra}%
  \BibitemOpen
  \bibfield  {author} {\bibinfo {author} {\bibfnamefont {H.~B.}\ \bibnamefont
  {Li}} \emph {et~al.} (\bibinfo {collaboration} {TEXONO}),\ }\bibfield
  {title} {\bibinfo {title} {{Limit on the electron neutrino magnetic moment
  from the Kuo-Sheng reactor neutrino experiment}},\ }\href
  {https://doi.org/10.1103/PhysRevLett.90.131802} {\bibfield  {journal}
  {\bibinfo  {journal} {Phys. Rev. Lett.}\ }\textbf {\bibinfo {volume} {90}},\
  \bibinfo {pages} {131802} (\bibinfo {year} {2003})},\ \Eprint
  {https://arxiv.org/abs/hep-ex/0212003} {arXiv:hep-ex/0212003} \BibitemShut
  {NoStop}%
\bibitem [{\citenamefont {Aprile}\ \emph {et~al.}(2022)\citenamefont {Aprile}
  \emph {et~al.}}]{XENON:2022ltv}%
  \BibitemOpen
  \bibfield  {author} {\bibinfo {author} {\bibfnamefont {E.}~\bibnamefont
  {Aprile}} \emph {et~al.} (\bibinfo {collaboration} {XENON}),\ }\bibfield
  {title} {\bibinfo {title} {{Search for New Physics in Electronic Recoil Data
  from XENONnT}},\ }\href {https://doi.org/10.1103/PhysRevLett.129.161805}
  {\bibfield  {journal} {\bibinfo  {journal} {Phys. Rev. Lett.}\ }\textbf
  {\bibinfo {volume} {129}},\ \bibinfo {pages} {161805} (\bibinfo {year}
  {2022})},\ \Eprint {https://arxiv.org/abs/2207.11330} {arXiv:2207.11330
  [hep-ex]} \BibitemShut {NoStop}%
\bibitem [{\citenamefont {Aalbers}\ \emph {et~al.}(2023)\citenamefont {Aalbers}
  \emph {et~al.}}]{LZ:2023poo}%
  \BibitemOpen
  \bibfield  {author} {\bibinfo {author} {\bibfnamefont {J.}~\bibnamefont
  {Aalbers}} \emph {et~al.} (\bibinfo {collaboration} {LZ}),\ }\bibfield
  {title} {\bibinfo {title} {{Search for new physics in low-energy electron
  recoils from the first LZ exposure}},\ }\href
  {https://doi.org/10.1103/PhysRevD.108.072006} {\bibfield  {journal} {\bibinfo
   {journal} {Phys. Rev. D}\ }\textbf {\bibinfo {volume} {108}},\ \bibinfo
  {pages} {072006} (\bibinfo {year} {2023})},\ \Eprint
  {https://arxiv.org/abs/2307.15753} {arXiv:2307.15753 [hep-ex]} \BibitemShut
  {NoStop}%
\bibitem [{\citenamefont {Kouzakov}\ and\ \citenamefont
  {Studenikin}(2017)}]{Kouzakov:2017hbc}%
  \BibitemOpen
  \bibfield  {author} {\bibinfo {author} {\bibfnamefont {K.~A.}\ \bibnamefont
  {Kouzakov}}\ and\ \bibinfo {author} {\bibfnamefont {A.~I.}\ \bibnamefont
  {Studenikin}},\ }\bibfield  {title} {\bibinfo {title} {{Electromagnetic
  properties of massive neutrinos in low-energy elastic neutrino-electron
  scattering}},\ }\href {https://doi.org/10.1103/PhysRevD.95.055013} {\bibfield
   {journal} {\bibinfo  {journal} {Phys. Rev. D}\ }\textbf {\bibinfo {volume}
  {95}},\ \bibinfo {pages} {055013} (\bibinfo {year} {2017})},\ \bibinfo {note}
  {[Erratum: Phys.Rev.D 96, 099904 (2017)]},\ \Eprint
  {https://arxiv.org/abs/1703.00401} {arXiv:1703.00401 [hep-ph]} \BibitemShut
  {NoStop}%
\bibitem [{\citenamefont {Abele}\ \emph
  {et~al.}(2025{\natexlab{b}})\citenamefont {Abele} \emph
  {et~al.}}]{Abele:2025rrl}%
  \BibitemOpen
  \bibfield  {author} {\bibinfo {author} {\bibfnamefont {H.}~\bibnamefont
  {Abele}} \emph {et~al.},\ }\bibfield  {title} {\bibinfo {title} {{Sub-keV
  Electron Recoil Calibration for Macroscopic Cryogenic Calorimeters Using a
  Novel X-ray Fluorescence Source}},\ }\href
  {https://doi.org/10.1007/s10909-025-03330-2} {\bibfield  {journal} {\bibinfo
  {journal} {J. Low Temp. Phys.}\ }\textbf {\bibinfo {volume} {221}},\ \bibinfo
  {pages} {265} (\bibinfo {year} {2025}{\natexlab{b}})},\ \Eprint
  {https://arxiv.org/abs/2505.17686} {arXiv:2505.17686 [physics.ins-det]}
  \BibitemShut {NoStop}%
\bibitem [{\citenamefont {Di~Domizio}\ \emph {et~al.}(2011)\citenamefont
  {Di~Domizio}, \citenamefont {Orio},\ and\ \citenamefont
  {Vignati}}]{DiDomizio:2010ph}%
  \BibitemOpen
  \bibfield  {author} {\bibinfo {author} {\bibfnamefont {S.}~\bibnamefont
  {Di~Domizio}}, \bibinfo {author} {\bibfnamefont {F.}~\bibnamefont {Orio}},\
  and\ \bibinfo {author} {\bibfnamefont {M.}~\bibnamefont {Vignati}},\
  }\bibfield  {title} {\bibinfo {title} {{Lowering the energy threshold of
  large-mass bolometric detectors}},\ }\href
  {https://doi.org/10.1088/1748-0221/6/02/P02007} {\bibfield  {journal}
  {\bibinfo  {journal} {JINST}\ }\textbf {\bibinfo {volume} {6}},\ \bibinfo
  {pages} {P02007}},\ \Eprint {https://arxiv.org/abs/1012.1263}
  {arXiv:1012.1263 [astro-ph.IM]} \BibitemShut {NoStop}%
\bibitem [{\citenamefont {Baxter}\ \emph {et~al.}(2025)\citenamefont {Baxter},
  \citenamefont {Essig}, \citenamefont {Hochberg}, \citenamefont {Kaznacheeva},
  \citenamefont {von Krosigk}, \citenamefont {Reindl}, \citenamefont {Romani},\
  and\ \citenamefont {Wagner}}]{Baxter:2025odk}%
  \BibitemOpen
  \bibfield  {author} {\bibinfo {author} {\bibfnamefont {D.}~\bibnamefont
  {Baxter}}, \bibinfo {author} {\bibfnamefont {R.}~\bibnamefont {Essig}},
  \bibinfo {author} {\bibfnamefont {Y.}~\bibnamefont {Hochberg}}, \bibinfo
  {author} {\bibfnamefont {M.}~\bibnamefont {Kaznacheeva}}, \bibinfo {author}
  {\bibfnamefont {B.}~\bibnamefont {von Krosigk}}, \bibinfo {author}
  {\bibfnamefont {F.}~\bibnamefont {Reindl}}, \bibinfo {author} {\bibfnamefont
  {R.~K.}\ \bibnamefont {Romani}},\ and\ \bibinfo {author} {\bibfnamefont
  {F.}~\bibnamefont {Wagner}},\ }\bibfield  {title} {\bibinfo {title}
  {{Low-Energy Backgrounds in Solid-State Phonon and Charge Detectors}},\
  }\href {https://doi.org/10.1146/annurev-nucl-121423-100849} {\bibfield
  {journal} {\bibinfo  {journal} {Ann. Rev. Nucl. Part. Sci.}\ }\textbf
  {\bibinfo {volume} {75}},\ \bibinfo {pages} {301} (\bibinfo {year} {2025})},\
  \Eprint {https://arxiv.org/abs/2503.08859} {arXiv:2503.08859
  [physics.ins-det]} \BibitemShut {NoStop}%
\bibitem [{\citenamefont {Cappelli}\ \emph {et~al.}(2026)\citenamefont
  {Cappelli} \emph {et~al.}}]{NUCLEUS:2026utx}%
  \BibitemOpen
  \bibfield  {author} {\bibinfo {author} {\bibfnamefont {M.}~\bibnamefont
  {Cappelli}} \emph {et~al.} (\bibinfo {collaboration} {NUCLEUS}),\ }\href@noop
  {} {\bibinfo {title} {{Sensitivity enhancement techniques for cryogenic
  calorimeters in the NUCLEUS experiment}}} (\bibinfo {year} {2026}),\ \Eprint
  {https://arxiv.org/abs/2603.28276} {arXiv:2603.28276 [physics.ins-det]}
  \BibitemShut {NoStop}%
\bibitem [{\citenamefont {Abrah{\~a}o}\ \emph {et~al.}(2021)\citenamefont
  {Abrah{\~a}o} \emph {et~al.}}]{DoubleChooz:2020vtr}%
  \BibitemOpen
  \bibfield  {author} {\bibinfo {author} {\bibfnamefont {T.}~\bibnamefont
  {Abrah{\~a}o}} \emph {et~al.} (\bibinfo {collaboration} {Double Chooz}),\
  }\bibfield  {title} {\bibinfo {title} {{Reactor rate modulation oscillation
  analysis with two detectors in Double Chooz}},\ }\href
  {https://doi.org/10.1007/JHEP01(2021)190} {\bibfield  {journal} {\bibinfo
  {journal} {JHEP}\ }\textbf {\bibinfo {volume} {01}},\ \bibinfo {pages}
  {190}},\ \Eprint {https://arxiv.org/abs/2007.13431} {arXiv:2007.13431
  [hep-ex]} \BibitemShut {NoStop}%
\bibitem [{\citenamefont {de~Kerret}\ \emph {et~al.}(2022)\citenamefont
  {de~Kerret} \emph {et~al.}}]{DoubleChooz:2022ukr}%
  \BibitemOpen
  \bibfield  {author} {\bibinfo {author} {\bibfnamefont {H.}~\bibnamefont
  {de~Kerret}} \emph {et~al.} (\bibinfo {collaboration} {Double Chooz}),\
  }\bibfield  {title} {\bibinfo {title} {{The Double Chooz antineutrino
  detectors}},\ }\href {https://doi.org/10.1140/epjc/s10052-022-10726-x}
  {\bibfield  {journal} {\bibinfo  {journal} {Eur. Phys. J. C}\ }\textbf
  {\bibinfo {volume} {82}},\ \bibinfo {pages} {804} (\bibinfo {year} {2022})},\
  \Eprint {https://arxiv.org/abs/2201.13285} {arXiv:2201.13285
  [physics.ins-det]} \BibitemShut {NoStop}%
\bibitem [{rte()}]{rte-data}%
  \BibitemOpen
  \href@noop {} {}\bibinfo {note}
  {{https://www.services-rte.com/en/view-data-published-by-rte.html}}\BibitemShut
  {NoStop}%
\bibitem [{\citenamefont {Cowan}\ \emph {et~al.}(2011)\citenamefont {Cowan},
  \citenamefont {Cranmer}, \citenamefont {Gross},\ and\ \citenamefont
  {Vitells}}]{Cowan:2010js}%
  \BibitemOpen
  \bibfield  {author} {\bibinfo {author} {\bibfnamefont {G.}~\bibnamefont
  {Cowan}}, \bibinfo {author} {\bibfnamefont {K.}~\bibnamefont {Cranmer}},
  \bibinfo {author} {\bibfnamefont {E.}~\bibnamefont {Gross}},\ and\ \bibinfo
  {author} {\bibfnamefont {O.}~\bibnamefont {Vitells}},\ }\bibfield  {title}
  {\bibinfo {title} {{Asymptotic formulae for likelihood-based tests of new
  physics}},\ }\href {https://doi.org/10.1140/epjc/s10052-011-1554-0}
  {\bibfield  {journal} {\bibinfo  {journal} {Eur. Phys. J. C}\ }\textbf
  {\bibinfo {volume} {71}},\ \bibinfo {pages} {1554} (\bibinfo {year}
  {2011})},\ \bibinfo {note} {[Erratum: Eur.Phys.J.C 73, 2501 (2013)]},\
  \Eprint {https://arxiv.org/abs/1007.1727} {arXiv:1007.1727 [physics.data-an]}
  \BibitemShut {NoStop}%
\bibitem [{\citenamefont {Alp{\'\i}zar-Venegas}\ \emph
  {et~al.}(2025)\citenamefont {Alp{\'\i}zar-Venegas}, \citenamefont {Flores},
  \citenamefont {Peinado},\ and\ \citenamefont
  {V{\'a}zquez-J{\'a}uregui}}]{Alpizar-Venegas:2025wor}%
  \BibitemOpen
  \bibfield  {author} {\bibinfo {author} {\bibfnamefont {M.}~\bibnamefont
  {Alp{\'\i}zar-Venegas}}, \bibinfo {author} {\bibfnamefont {L.~J.}\
  \bibnamefont {Flores}}, \bibinfo {author} {\bibfnamefont {E.}~\bibnamefont
  {Peinado}},\ and\ \bibinfo {author} {\bibfnamefont {E.}~\bibnamefont
  {V{\'a}zquez-J{\'a}uregui}},\ }\bibfield  {title} {\bibinfo {title}
  {{Exploring the standard model and beyond from the evidence of
  CE{\ensuremath{\nu}}NS with reactor antineutrinos in CONUS+}},\ }\href
  {https://doi.org/10.1103/PhysRevD.111.053001} {\bibfield  {journal} {\bibinfo
   {journal} {Phys. Rev. D}\ }\textbf {\bibinfo {volume} {111}},\ \bibinfo
  {pages} {053001} (\bibinfo {year} {2025})},\ \Eprint
  {https://arxiv.org/abs/2501.10355} {arXiv:2501.10355 [hep-ph]} \BibitemShut
  {NoStop}%
\bibitem [{\citenamefont {Akimov}\ \emph {et~al.}(2022)\citenamefont {Akimov}
  \emph {et~al.}}]{COHERENT:2021xmm}%
  \BibitemOpen
  \bibfield  {author} {\bibinfo {author} {\bibfnamefont {D.}~\bibnamefont
  {Akimov}} \emph {et~al.} (\bibinfo {collaboration} {COHERENT}),\ }\bibfield
  {title} {\bibinfo {title} {{Measurement of the Coherent Elastic
  Neutrino-Nucleus Scattering Cross Section on CsI by COHERENT}},\ }\href
  {https://doi.org/10.1103/PhysRevLett.129.081801} {\bibfield  {journal}
  {\bibinfo  {journal} {Phys. Rev. Lett.}\ }\textbf {\bibinfo {volume} {129}},\
  \bibinfo {pages} {081801} (\bibinfo {year} {2022})},\ \Eprint
  {https://arxiv.org/abs/2110.07730} {arXiv:2110.07730 [hep-ex]} \BibitemShut
  {NoStop}%
\bibitem [{\citenamefont {Cadeddu}\ \emph
  {et~al.}(2020{\natexlab{b}})\citenamefont {Cadeddu}, \citenamefont {Dordei},
  \citenamefont {Giunti}, \citenamefont {Li}, \citenamefont {Picciau},\ and\
  \citenamefont {Zhang}}]{Cadeddu:2020lky}%
  \BibitemOpen
  \bibfield  {author} {\bibinfo {author} {\bibfnamefont {M.}~\bibnamefont
  {Cadeddu}}, \bibinfo {author} {\bibfnamefont {F.}~\bibnamefont {Dordei}},
  \bibinfo {author} {\bibfnamefont {C.}~\bibnamefont {Giunti}}, \bibinfo
  {author} {\bibfnamefont {Y.~F.}\ \bibnamefont {Li}}, \bibinfo {author}
  {\bibfnamefont {E.}~\bibnamefont {Picciau}},\ and\ \bibinfo {author}
  {\bibfnamefont {Y.~Y.}\ \bibnamefont {Zhang}},\ }\bibfield  {title} {\bibinfo
  {title} {{Physics results from the first COHERENT observation of coherent
  elastic neutrino-nucleus scattering in argon and their combination with
  cesium-iodide data}},\ }\href {https://doi.org/10.1103/PhysRevD.102.015030}
  {\bibfield  {journal} {\bibinfo  {journal} {Phys. Rev. D}\ }\textbf {\bibinfo
  {volume} {102}},\ \bibinfo {pages} {015030} (\bibinfo {year}
  {2020}{\natexlab{b}})},\ \Eprint {https://arxiv.org/abs/2005.01645}
  {arXiv:2005.01645 [hep-ph]} \BibitemShut {NoStop}%
\bibitem [{\citenamefont {Atzori~Corona}\ \emph
  {et~al.}(2025{\natexlab{c}})\citenamefont {Atzori~Corona}, \citenamefont
  {Cadeddu}, \citenamefont {Cargioli}, \citenamefont {Co'}, \citenamefont
  {Dordei},\ and\ \citenamefont {Giunti}}]{AtzoriCorona:2025xgj}%
  \BibitemOpen
  \bibfield  {author} {\bibinfo {author} {\bibfnamefont {M.}~\bibnamefont
  {Atzori~Corona}}, \bibinfo {author} {\bibfnamefont {M.}~\bibnamefont
  {Cadeddu}}, \bibinfo {author} {\bibfnamefont {N.}~\bibnamefont {Cargioli}},
  \bibinfo {author} {\bibfnamefont {G.}~\bibnamefont {Co'}}, \bibinfo {author}
  {\bibfnamefont {F.}~\bibnamefont {Dordei}},\ and\ \bibinfo {author}
  {\bibfnamefont {C.}~\bibnamefont {Giunti}},\ }\bibfield  {title} {\bibinfo
  {title} {{Joint analysis of reactor and accelerator CE{\ensuremath{\nu}}NS
  data on germanium: implications for the standard model and nuclear
  physics}},\ }\href {https://doi.org/10.1016/j.physletb.2025.139856}
  {\bibfield  {journal} {\bibinfo  {journal} {Phys. Lett. B}\ }\textbf
  {\bibinfo {volume} {869}},\ \bibinfo {pages} {139856} (\bibinfo {year}
  {2025}{\natexlab{c}})},\ \Eprint {https://arxiv.org/abs/2506.13555}
  {arXiv:2506.13555 [hep-ph]} \BibitemShut {NoStop}%
\bibitem [{\citenamefont {De~Romeri}\ \emph
  {et~al.}(2025{\natexlab{a}})\citenamefont {De~Romeri}, \citenamefont
  {Papoulias},\ and\ \citenamefont {Ternes}}]{DeRomeri:2024iaw}%
  \BibitemOpen
  \bibfield  {author} {\bibinfo {author} {\bibfnamefont {V.}~\bibnamefont
  {De~Romeri}}, \bibinfo {author} {\bibfnamefont {D.~K.}\ \bibnamefont
  {Papoulias}},\ and\ \bibinfo {author} {\bibfnamefont {C.~A.}\ \bibnamefont
  {Ternes}},\ }\bibfield  {title} {\bibinfo {title} {{Bounds on new neutrino
  interactions from the first CE{\ensuremath{\nu}}NS data at direct detection
  experiments}},\ }\href {https://doi.org/10.1088/1475-7516/2025/05/012}
  {\bibfield  {journal} {\bibinfo  {journal} {JCAP}\ }\textbf {\bibinfo
  {volume} {05}},\ \bibinfo {pages} {012}},\ \Eprint
  {https://arxiv.org/abs/2411.11749} {arXiv:2411.11749 [hep-ph]} \BibitemShut
  {NoStop}%
\bibitem [{\citenamefont {Aguilar-Arevalo}\ \emph {et~al.}(2020)\citenamefont
  {Aguilar-Arevalo} \emph {et~al.}}]{CONNIE:2019xid}%
  \BibitemOpen
  \bibfield  {author} {\bibinfo {author} {\bibfnamefont {A.}~\bibnamefont
  {Aguilar-Arevalo}} \emph {et~al.} (\bibinfo {collaboration} {CONNIE}),\
  }\bibfield  {title} {\bibinfo {title} {{Search for light mediators in the
  low-energy data of the CONNIE reactor neutrino experiment}},\ }\href
  {https://doi.org/10.1007/JHEP04(2020)054} {\bibfield  {journal} {\bibinfo
  {journal} {JHEP}\ }\textbf {\bibinfo {volume} {04}},\ \bibinfo {pages}
  {054}},\ \Eprint {https://arxiv.org/abs/1910.04951} {arXiv:1910.04951
  [hep-ex]} \BibitemShut {NoStop}%
\bibitem [{\citenamefont {De~Romeri}\ \emph
  {et~al.}(2025{\natexlab{b}})\citenamefont {De~Romeri}, \citenamefont
  {Papoulias},\ and\ \citenamefont {Sanchez~Garcia}}]{DeRomeri:2025csu}%
  \BibitemOpen
  \bibfield  {author} {\bibinfo {author} {\bibfnamefont {V.}~\bibnamefont
  {De~Romeri}}, \bibinfo {author} {\bibfnamefont {D.~K.}\ \bibnamefont
  {Papoulias}},\ and\ \bibinfo {author} {\bibfnamefont {G.}~\bibnamefont
  {Sanchez~Garcia}},\ }\bibfield  {title} {\bibinfo {title} {{Implications of
  the first CONUS+ measurement of coherent elastic neutrino-nucleus
  scattering}},\ }\href {https://doi.org/10.1103/PhysRevD.111.075025}
  {\bibfield  {journal} {\bibinfo  {journal} {Phys. Rev. D}\ }\textbf {\bibinfo
  {volume} {111}},\ \bibinfo {pages} {075025} (\bibinfo {year}
  {2025}{\natexlab{b}})},\ \Eprint {https://arxiv.org/abs/2501.17843}
  {arXiv:2501.17843 [hep-ph]} \BibitemShut {NoStop}%
\bibitem [{\citenamefont {A.}\ \emph {et~al.}(2023)\citenamefont {A.},
  \citenamefont {Majumdar}, \citenamefont {Papoulias}, \citenamefont
  {Prajapati},\ and\ \citenamefont {Srivastava}}]{A:2022acy}%
  \BibitemOpen
  \bibfield  {author} {\bibinfo {author} {\bibfnamefont {S.~K.}\ \bibnamefont
  {A.}}, \bibinfo {author} {\bibfnamefont {A.}~\bibnamefont {Majumdar}},
  \bibinfo {author} {\bibfnamefont {D.~K.}\ \bibnamefont {Papoulias}}, \bibinfo
  {author} {\bibfnamefont {H.}~\bibnamefont {Prajapati}},\ and\ \bibinfo
  {author} {\bibfnamefont {R.}~\bibnamefont {Srivastava}},\ }\bibfield  {title}
  {\bibinfo {title} {{Implications of first LZ and XENONnT results: A
  comparative study of neutrino properties and light mediators}},\ }\href
  {https://doi.org/10.1016/j.physletb.2023.137742} {\bibfield  {journal}
  {\bibinfo  {journal} {Phys. Lett. B}\ }\textbf {\bibinfo {volume} {839}},\
  \bibinfo {pages} {137742} (\bibinfo {year} {2023})},\ \Eprint
  {https://arxiv.org/abs/2208.06415} {arXiv:2208.06415 [hep-ph]} \BibitemShut
  {NoStop}%
\bibitem [{\citenamefont {Blanco-Mas}\ \emph {et~al.}(2025)\citenamefont
  {Blanco-Mas}, \citenamefont {Coloma}, \citenamefont {Herrera}, \citenamefont
  {Huber}, \citenamefont {Kopp}, \citenamefont {Shoemaker},\ and\ \citenamefont
  {Tabrizi}}]{Blanco-Mas:2024ale}%
  \BibitemOpen
  \bibfield  {author} {\bibinfo {author} {\bibfnamefont {P.}~\bibnamefont
  {Blanco-Mas}}, \bibinfo {author} {\bibfnamefont {P.}~\bibnamefont {Coloma}},
  \bibinfo {author} {\bibfnamefont {G.}~\bibnamefont {Herrera}}, \bibinfo
  {author} {\bibfnamefont {P.}~\bibnamefont {Huber}}, \bibinfo {author}
  {\bibfnamefont {J.}~\bibnamefont {Kopp}}, \bibinfo {author} {\bibfnamefont
  {I.~M.}\ \bibnamefont {Shoemaker}},\ and\ \bibinfo {author} {\bibfnamefont
  {Z.}~\bibnamefont {Tabrizi}},\ }\bibfield  {title} {\bibinfo {title}
  {{Clarity through the neutrino fog: constraining new forces in dark matter
  detectors}},\ }\href {https://doi.org/10.1007/JHEP08(2025)043} {\bibfield
  {journal} {\bibinfo  {journal} {JHEP}\ }\textbf {\bibinfo {volume} {08}},\
  \bibinfo {pages} {043}},\ \Eprint {https://arxiv.org/abs/2411.14206}
  {arXiv:2411.14206 [hep-ph]} \BibitemShut {NoStop}%
\bibitem [{\citenamefont {Karada{\u{g}}}\ \emph {et~al.}(2025)\citenamefont
  {Karada{\u{g}}} \emph {et~al.}}]{TEXONO:2025sub}%
  \BibitemOpen
  \bibfield  {author} {\bibinfo {author} {\bibfnamefont {S.}~\bibnamefont
  {Karada{\u{g}}}} \emph {et~al.} (\bibinfo {collaboration} {TEXONO}),\
  }\bibfield  {title} {\bibinfo {title} {{Constraints on new physics with light
  mediators and generalized neutrino interactions via coherent elastic neutrino
  nucleus scattering}},\ }\href {https://doi.org/10.1103/63xf-t6fz} {\bibfield
  {journal} {\bibinfo  {journal} {Phys. Rev. D}\ }\textbf {\bibinfo {volume}
  {112}},\ \bibinfo {pages} {035038} (\bibinfo {year} {2025})},\ \Eprint
  {https://arxiv.org/abs/2502.20007} {arXiv:2502.20007 [hep-ex]} \BibitemShut
  {NoStop}%
\bibitem [{\citenamefont {Aristizabal~Sierra}\ \emph
  {et~al.}(2025)\citenamefont {Aristizabal~Sierra}, \citenamefont {Mishra},\
  and\ \citenamefont {Strigari}}]{AristizabalSierra:2024nwf}%
  \BibitemOpen
  \bibfield  {author} {\bibinfo {author} {\bibfnamefont {D.}~\bibnamefont
  {Aristizabal~Sierra}}, \bibinfo {author} {\bibfnamefont {N.}~\bibnamefont
  {Mishra}},\ and\ \bibinfo {author} {\bibfnamefont {L.}~\bibnamefont
  {Strigari}},\ }\bibfield  {title} {\bibinfo {title} {{Implications of first
  neutrino-induced nuclear recoil measurements in direct detection experiments:
  Probing nonstandard interaction via CE{\ensuremath{\nu}}NS}},\ }\href
  {https://doi.org/10.1103/PhysRevD.111.055007} {\bibfield  {journal} {\bibinfo
   {journal} {Phys. Rev. D}\ }\textbf {\bibinfo {volume} {111}},\ \bibinfo
  {pages} {055007} (\bibinfo {year} {2025})},\ \Eprint
  {https://arxiv.org/abs/2409.02003} {arXiv:2409.02003 [hep-ph]} \BibitemShut
  {NoStop}%
\bibitem [{\citenamefont {Daraktchieva}\ \emph {et~al.}(2005)\citenamefont
  {Daraktchieva} \emph {et~al.}}]{MUNU:2005xnz}%
  \BibitemOpen
  \bibfield  {author} {\bibinfo {author} {\bibfnamefont {Z.}~\bibnamefont
  {Daraktchieva}} \emph {et~al.} (\bibinfo {collaboration} {MUNU}),\ }\bibfield
   {title} {\bibinfo {title} {{Final results on the neutrino magnetic moment
  from the MUNU experiment}},\ }\href
  {https://doi.org/10.1016/j.physletb.2005.04.030} {\bibfield  {journal}
  {\bibinfo  {journal} {Phys. Lett. B}\ }\textbf {\bibinfo {volume} {615}},\
  \bibinfo {pages} {153} (\bibinfo {year} {2005})},\ \Eprint
  {https://arxiv.org/abs/hep-ex/0502037} {arXiv:hep-ex/0502037} \BibitemShut
  {NoStop}%
\bibitem [{\citenamefont {Beda}\ \emph {et~al.}(2012)\citenamefont {Beda},
  \citenamefont {Brudanin}, \citenamefont {Egorov}, \citenamefont {Medvedev},
  \citenamefont {Pogosov}, \citenamefont {Shirchenko},\ and\ \citenamefont
  {Starostin}}]{Beda:2012zz}%
  \BibitemOpen
  \bibfield  {author} {\bibinfo {author} {\bibfnamefont {A.~G.}\ \bibnamefont
  {Beda}}, \bibinfo {author} {\bibfnamefont {V.~B.}\ \bibnamefont {Brudanin}},
  \bibinfo {author} {\bibfnamefont {V.~G.}\ \bibnamefont {Egorov}}, \bibinfo
  {author} {\bibfnamefont {D.~V.}\ \bibnamefont {Medvedev}}, \bibinfo {author}
  {\bibfnamefont {V.~S.}\ \bibnamefont {Pogosov}}, \bibinfo {author}
  {\bibfnamefont {M.~V.}\ \bibnamefont {Shirchenko}},\ and\ \bibinfo {author}
  {\bibfnamefont {A.~S.}\ \bibnamefont {Starostin}},\ }\bibfield  {title}
  {\bibinfo {title} {{The results of search for the neutrino magnetic moment in
  GEMMA experiment}},\ }\href {https://doi.org/10.1155/2012/350150} {\bibfield
  {journal} {\bibinfo  {journal} {Adv. High Energy Phys.}\ }\textbf {\bibinfo
  {volume} {2012}},\ \bibinfo {pages} {350150} (\bibinfo {year}
  {2012})}\BibitemShut {NoStop}%
\bibitem [{\citenamefont {Wong}\ \emph {et~al.}(2007)\citenamefont {Wong} \emph
  {et~al.}}]{TEXONO:2006xds}%
  \BibitemOpen
  \bibfield  {author} {\bibinfo {author} {\bibfnamefont {H.~T.}\ \bibnamefont
  {Wong}} \emph {et~al.} (\bibinfo {collaboration} {TEXONO}),\ }\bibfield
  {title} {\bibinfo {title} {{A Search of Neutrino Magnetic Moments with a
  High-Purity Germanium Detector at the Kuo-Sheng Nuclear Power Station}},\
  }\href {https://doi.org/10.1103/PhysRevD.75.012001} {\bibfield  {journal}
  {\bibinfo  {journal} {Phys. Rev. D}\ }\textbf {\bibinfo {volume} {75}},\
  \bibinfo {pages} {012001} (\bibinfo {year} {2007})},\ \Eprint
  {https://arxiv.org/abs/hep-ex/0605006} {arXiv:hep-ex/0605006} \BibitemShut
  {NoStop}%
\bibitem [{\citenamefont {Ternes}\ and\ \citenamefont
  {T{\'o}rtola}(2025)}]{Ternes:2025lqh}%
  \BibitemOpen
  \bibfield  {author} {\bibinfo {author} {\bibfnamefont {C.~A.}\ \bibnamefont
  {Ternes}}\ and\ \bibinfo {author} {\bibfnamefont {M.}~\bibnamefont
  {T{\'o}rtola}},\ }\bibfield  {title} {\bibinfo {title} {{Neutrino magnetic
  moments: effective versus fundamental parameters}},\ }\href
  {https://doi.org/10.1016/j.nuclphysb.2025.117107} {\bibfield  {journal}
  {\bibinfo  {journal} {Nucl. Phys. B}\ }\textbf {\bibinfo {volume} {1019}},\
  \bibinfo {pages} {117107} (\bibinfo {year} {2025})},\ \Eprint
  {https://arxiv.org/abs/2505.02633} {arXiv:2505.02633 [hep-ph]} \BibitemShut
  {NoStop}%
\bibitem [{\citenamefont {Aprile}\ \emph {et~al.}(2023)\citenamefont {Aprile}
  \emph {et~al.}}]{XENON:2023cxc}%
  \BibitemOpen
  \bibfield  {author} {\bibinfo {author} {\bibfnamefont {E.}~\bibnamefont
  {Aprile}} \emph {et~al.} (\bibinfo {collaboration} {XENON}),\ }\bibfield
  {title} {\bibinfo {title} {{First Dark Matter Search with Nuclear Recoils
  from the XENONnT Experiment}},\ }\href
  {https://doi.org/10.1103/PhysRevLett.131.041003} {\bibfield  {journal}
  {\bibinfo  {journal} {Phys. Rev. Lett.}\ }\textbf {\bibinfo {volume} {131}},\
  \bibinfo {pages} {041003} (\bibinfo {year} {2023})},\ \Eprint
  {https://arxiv.org/abs/2303.14729} {arXiv:2303.14729 [hep-ex]} \BibitemShut
  {NoStop}%
\bibitem [{\citenamefont {Akerib}\ \emph
  {et~al.}(2025{\natexlab{b}})\citenamefont {Akerib} \emph
  {et~al.}}]{LZ:2025zpw}%
  \BibitemOpen
  \bibfield  {author} {\bibinfo {author} {\bibfnamefont {D.~S.}\ \bibnamefont
  {Akerib}} \emph {et~al.} (\bibinfo {collaboration} {LZ}),\ }\href@noop {}
  {\bibinfo {title} {{Search for New Physics via Low-Energy Electron Recoils
  with a 4.2 Tonne{\texttimes} Year Exposure from the LZ Experiment}}}
  (\bibinfo {year} {2025}{\natexlab{b}}),\ \Eprint
  {https://arxiv.org/abs/2511.17350} {arXiv:2511.17350 [hep-ex]} \BibitemShut
  {NoStop}%
\bibitem [{\citenamefont {Agostini}\ \emph {et~al.}(2017)\citenamefont
  {Agostini} \emph {et~al.}}]{Borexino:2017fbd}%
  \BibitemOpen
  \bibfield  {author} {\bibinfo {author} {\bibfnamefont {M.}~\bibnamefont
  {Agostini}} \emph {et~al.} (\bibinfo {collaboration} {Borexino}),\ }\bibfield
   {title} {\bibinfo {title} {{Limiting neutrino magnetic moments with Borexino
  Phase-II solar neutrino data}},\ }\href
  {https://doi.org/10.1103/PhysRevD.96.091103} {\bibfield  {journal} {\bibinfo
  {journal} {Phys. Rev. D}\ }\textbf {\bibinfo {volume} {96}},\ \bibinfo
  {pages} {091103} (\bibinfo {year} {2017})},\ \Eprint
  {https://arxiv.org/abs/1707.09355} {arXiv:1707.09355 [hep-ex]} \BibitemShut
  {NoStop}%
\bibitem [{\citenamefont {Chattaraj}\ \emph {et~al.}(2025)\citenamefont
  {Chattaraj}, \citenamefont {Majumdar},\ and\ \citenamefont
  {Srivastava}}]{Chattaraj:2025fvx}%
  \BibitemOpen
  \bibfield  {author} {\bibinfo {author} {\bibfnamefont {A.}~\bibnamefont
  {Chattaraj}}, \bibinfo {author} {\bibfnamefont {A.}~\bibnamefont
  {Majumdar}},\ and\ \bibinfo {author} {\bibfnamefont {R.}~\bibnamefont
  {Srivastava}},\ }\bibfield  {title} {\bibinfo {title} {{Probing standard
  model and beyond with reactor CE{\ensuremath{\nu}}NS data of CONUS+
  experiment}},\ }\href {https://doi.org/10.1016/j.physletb.2025.139438}
  {\bibfield  {journal} {\bibinfo  {journal} {Phys. Lett. B}\ }\textbf
  {\bibinfo {volume} {864}},\ \bibinfo {pages} {139438} (\bibinfo {year}
  {2025})},\ \Eprint {https://arxiv.org/abs/2501.12441} {arXiv:2501.12441
  [hep-ph]} \BibitemShut {NoStop}%
\bibitem [{\citenamefont {Wilks}(1938)}]{Wilks:1938dza}%
  \BibitemOpen
  \bibfield  {author} {\bibinfo {author} {\bibfnamefont {S.~S.}\ \bibnamefont
  {Wilks}},\ }\bibfield  {title} {\bibinfo {title} {{The Large-Sample
  Distribution of the Likelihood Ratio for Testing Composite Hypotheses}},\
  }\href {https://doi.org/10.1214/aoms/1177732360} {\bibfield  {journal}
  {\bibinfo  {journal} {Annals Math. Statist.}\ }\textbf {\bibinfo {volume}
  {9}},\ \bibinfo {pages} {60} (\bibinfo {year} {1938})}\BibitemShut {NoStop}%
\bibitem [{\citenamefont {Abele}\ \emph {et~al.}(2023)\citenamefont {Abele}
  \emph {et~al.}}]{CRAB:2022rcm}%
  \BibitemOpen
  \bibfield  {author} {\bibinfo {author} {\bibfnamefont {H.}~\bibnamefont
  {Abele}} \emph {et~al.} (\bibinfo {collaboration} {CRAB, NUCLEUS}),\
  }\bibfield  {title} {\bibinfo {title} {{Observation of a Nuclear Recoil Peak
  at the 100~eV Scale Induced by Neutron Capture}},\ }\href
  {https://doi.org/10.1103/PhysRevLett.130.211802} {\bibfield  {journal}
  {\bibinfo  {journal} {Phys. Rev. Lett.}\ }\textbf {\bibinfo {volume} {130}},\
  \bibinfo {pages} {211802} (\bibinfo {year} {2023})},\ \Eprint
  {https://arxiv.org/abs/2211.03631} {arXiv:2211.03631 [nucl-ex]} \BibitemShut
  {NoStop}%
\bibitem [{\citenamefont {Abele}\ \emph
  {et~al.}(2025{\natexlab{c}})\citenamefont {Abele} \emph
  {et~al.}}]{Abele:2025vrj}%
  \BibitemOpen
  \bibfield  {author} {\bibinfo {author} {\bibfnamefont {H.}~\bibnamefont
  {Abele}} \emph {et~al.},\ }\bibfield  {title} {\bibinfo {title} {{The CRAB
  facility at the TU Wien TRIGA reactor: status and related physics program}},\
  }\href {https://doi.org/10.1140/epjc/s10052-025-14809-3} {\bibfield
  {journal} {\bibinfo  {journal} {Eur. Phys. J. C}\ }\textbf {\bibinfo {volume}
  {85}},\ \bibinfo {pages} {1188} (\bibinfo {year} {2025}{\natexlab{c}})},\
  \Eprint {https://arxiv.org/abs/2505.15227} {arXiv:2505.15227
  [physics.ins-det]} \BibitemShut {NoStop}%
\bibitem [{\citenamefont {Li}\ \emph {et~al.}(2025)\citenamefont {Li},
  \citenamefont {Herrera},\ and\ \citenamefont {Huber}}]{Li:2025pfw}%
  \BibitemOpen
  \bibfield  {author} {\bibinfo {author} {\bibfnamefont {Y.}~\bibnamefont
  {Li}}, \bibinfo {author} {\bibfnamefont {G.}~\bibnamefont {Herrera}},\ and\
  \bibinfo {author} {\bibfnamefont {P.}~\bibnamefont {Huber}},\ }\bibfield
  {title} {\bibinfo {title} {{New physics versus quenching factors in Coherent
  Neutrino Scattering}},\ }\href {https://doi.org/10.1007/JHEP11(2025)022}
  {\bibfield  {journal} {\bibinfo  {journal} {JHEP}\ }\textbf {\bibinfo
  {volume} {11}},\ \bibinfo {pages} {022}},\ \Eprint
  {https://arxiv.org/abs/2502.12308} {arXiv:2502.12308 [hep-ph]} \BibitemShut
  {NoStop}%
\bibitem [{\citenamefont {Abe}\ \emph {et~al.}(2014)\citenamefont {Abe} \emph
  {et~al.}}]{DoubleChooz:2014kuw}%
  \BibitemOpen
  \bibfield  {author} {\bibinfo {author} {\bibfnamefont {Y.}~\bibnamefont
  {Abe}} \emph {et~al.} (\bibinfo {collaboration} {Double Chooz}),\ }\bibfield
  {title} {\bibinfo {title} {{Improved measurements of the neutrino mixing
  angle $\theta_{13}$ with the Double Chooz detector}},\ }\href
  {https://doi.org/10.1007/JHEP02(2015)074} {\bibfield  {journal} {\bibinfo
  {journal} {JHEP}\ }\textbf {\bibinfo {volume} {10}},\ \bibinfo {pages}
  {086}},\ \bibinfo {note} {[Erratum: JHEP 02, 074 (2015)]},\ \Eprint
  {https://arxiv.org/abs/1406.7763} {arXiv:1406.7763 [hep-ex]} \BibitemShut
  {NoStop}%
\end{thebibliography}%

\end{document}